\documentclass[aps,prb,twocolumn,showpacs,amsmath]{revtex4}
\usepackage{graphicx}
\usepackage{bm}
\usepackage{color}
\usepackage[normalem]{ulem}	

\begin{document}

\title{Geometry-dependent critical currents in superconducting nanocircuits}

\author{John R.\ Clem}
\affiliation{%
   Ames Laboratory and Department of Physics and Astronomy, \\
   Iowa State University, Ames, Iowa, 50011--3160 USA}

\author{Karl K. Berggren}
\affiliation{%
	Massachusetts Institute
of Technology, Cambridge, MA 02139 USA and \\
Kavli Institute of Nanoscience, Delft University of Technology, Lorentzweg 1, 2628CJ Delft,
The Netherlands}

\date{\today}

\begin{abstract} 
In this paper we calculate the critical currents in thin superconducting strips with sharp right-angle turns, 180-degree turnarounds, and more complicated geometries, where all the line widths are much smaller than the Pearl length $\Lambda = 2 \lambda^2/d$.  We define the critical current as the current that reduces the Gibbs free-energy barrier to zero.  We show that current crowding, which occurs whenever the current rounds a sharp turn, tends to reduce the critical current, but we also show that when the radius of curvature is less than the coherence length this effect is partially compensated by a radius-of-curvature effect.  We propose several patterns with rounded corners to avoid critical-current reduction due to current crowding. These results are relevant to superconducting nanowire single-photon detectors, where they suggest a means of improving the bias conditions and reducing dark counts. These results also have relevance to normal-metal nanocircuits, as these patterns can reduce the electrical resistance, electromigration,   and hot spots caused by nonuniform heating.

\end{abstract}


\maketitle

\section{Introduction\label{intro}}

When an electrical  current travels through a 180-degree hairpin
turn or around a sharp corner in a thin film, it tends
to concentrate on the inner boundary of the curve. For
normal metals, this effect is known as current crowding, and results in an additional resistance
over what would be expected from na\"ive geometrical
arguments.\cite{Hagedorn63} 
The situation with a superconductor
is analogous to that in normal metals, except that  the 
effect manifests itself primarily as a reduced critical current
for the pattern.
A superconducting
wire will become resistive as soon as the critical current for the inner
corner is exceeded, even if the total current is lower than the critical current of the
connecting straight-line segments. 

Current crowding has an important implication for
measurements of critical currents in thin films because
such measurements typically include at least one
sharp corner. For example, in two-point measurements, narrow,
straight segments connect to  larger contacts at the end, forming
sharp inner corners on either side. Even the ubiquitous
Kelvin Bridge resistor structure
typically includes
sharp corners where the voltage leads contact the
center current-carrying lead. These measurements are
used extensively in materials and device characterization
for nanoscale structures, but in practice these measurement methods can in general underestimate
the  critical current.

Nearly 50 years ago,  Hagedorn and Hall\cite{Hagedorn63} extended earlier work by Cockroft\cite{Cockroft28} to consider a strip conductor with a right-angle bend.  They calculated both the resistance for a normal-metal strip and the current-density distribution for a superconducting strip.  Using conformal transformations, the authors showed that the current density increases on the inside corner of a bend, and derived  the mathematical form for an optimally  rounded inner boundary for which the critical current for the bend would be the same as for a straight strip.  The authors' resistance result and its extension to other geometries\cite{Hall68} became widely used in the field of integrated-circuit layout, but their critical-current calculations were apparently forgotten. Despite extensive work on the critical current of superconducting materials over the intervening period,\cite{DewHughes01} to our knowledge Hagedorn and Hall's predictions have never been tested experimentally.

Although Hagedorn and Hall correctly predicted the presence of current crowding, their treatment did not permit a quantitative estimate of its effect on the critical current, and was limited to considering only simple 90$^\circ$ corner goemetries. 
A new aspect emphasized in our paper, not considered in Ref.\ \onlinecite{Hagedorn63}, is that the critical current of a superconducting strip in the absence of thermal excitations is the current at which a nucleating vortex surmounts the Gibbs free-energy barrier at the wire edge and then is driven entirely across the strip.  The resulting voltage along the strip is proportional to the rate at which these processes occur.  Our approach permits an estimate of the critical current of a variety of thin-film patterns. To calculate the critical current, one needs to know both the current distribution and the vortex's interaction energy with the edge of the strip. An interesting and important consequence of this treatment is that the critical current of a right-angle bend is finite, even though the current density at the sharp inner corner calculated using the London equation, as done in the Hagedorn paper, would diverge. In this paper, we perform theoretical calculations of the critical current using conformal transformations going beyond Hagedorn and Hall's calculations for simple right-angle bends\cite{Hagedorn63} to include many more pattern geometries.

Thermal excitation of vortices over the Gibbs free-energy barrier can produce random voltage pulses due to vortex motion across the strip.\cite{Bartolf10}  The resulting time-averaged voltage gives rise to a strongly temperature-dependent finite resistance of the superconductor at currents below the critical current that would occur in the absence of thermal excitations. Voltage pulses like these have been proposed as the origin of so-called dark counts in  superconducting nanowire single-photon detectors (SNSPDs).\cite{Goltsman01,Engel06,Bell07,Kitaygorsky07,Bulaevskii11,Bulaevskii11a,Yamashita11}  Our results should provide guidance regarding the extent to which thermally activated voltages depend upon the geometry of the underlying nanocircuitry.

These results may have important implications for SNSPD performance. SNSPDs consist of thin films patterned to form a continuous wire arranged in a boustrophedonic pattern on the substrate. Their performance depends sensitively on the bias current:  Typically, detectors increase in sensitivity with increased bias current until the critical current is exceeded. Commonly, somewhere between a few and 100\% of fabricated devices suffer from an effect known as ``constriction,''\cite{Kerman07} in which the critical current is anomalously suppressed. This suppression is established by comparing the measured critical current to one predicted by measuring the device inductance and fitting to theoretical expectations (which can be used to extract a ``true'' critical current). At present, the cause of constrictions in SNSPDs is not understood. There are no unambiguous demonstrations in which a constriction was identified. It seems from our work that sharp corners, hairpin turns, contact inner corners, and line asperities (both widenings and narrowings) could all result in constriction.
Our results, as well as the earlier results of Hagedorn and Hall, are of direct relevance not only to superconducting wires, but also to conventional current flow in patterned thin films. Sharp corners and current crowding are known to cause problems of electromigration and the formation of hot spots in the pattern. Such effects can limit the performance and reliability of integrated circuits. The results shown here specify how to optimally design a variety of patterns without current crowding, thus minimizing the undesirable impact of this effect.

In Sec.\ \ref{uniformK}, we show why the current density is very nearly uniform in straight, narrow, thin superconducting strips. In Sec.\ \ref{procedure}, we define the critical current as the current that lowers to zero the Gibbs free energy barrier against vortex entry into the strip. In Sec.\ \ref{Kcstripsection}, we present a general procedure for calculating the critical current and, as an example, apply it to the case of a long, straight, narrow, thin superconducting strip.  In Sec.\ \ref{curve}, we show how current crowding, which occurs whenever the current bends around a curve, tends to reduce the critical current.   In Sec.\ \ref{rhocsec}, we show how the current-crowding effect is ameliorated by a radius-of-curvature effect. In Sec.\ \ref{180general}, we calculate the extent to which the critical current is reduced in 180-degree turnarounds.  We show that sharp inner corners reduce the critical current, and we show how the inner curve can be optimally rounded so that there is no critical-current reduction.  In Sec.\  \ref{rightangle}, we calculate the extent to which the critical current is reduced in 90-degree turns.  We show that sharp inner corners reduce the critical current, and we show how the inner curve can be optimally rounded, as noted in Ref.\ \onlinecite{Hagedorn63}, so that there is no critical-current reduction.  In Sec.\ \ref{Tsec}, we show how current crowding at the sharp corners of sidebar contacts in the shape of T intersections reduce the critical current, and in Sec.\ \ref{WideEndsec}, we show how  a similar critical-current reduction occurs because of current crowding at the sharp corner where a long strip makes electrical contact with a wide contact pad. In both Secs.\ \ref{Tsec} and \ref{WideEndsec}, we also suggest new patterns with rounded corners that should avoid critical-current reduction.
In Sec.\ \ref{EdgeDefects}, we examine the role played by edge defects in reducing the critical current.
In Sec.\ \ref{BarrierClimbing}, we show how thermal excitation over the Gibbs free-energy barrier leads to broadening of the switching-current distribution observed when the critical current is measured repeatedly.
  In Sec.\ \ref{ExpSec}, we compare our calculations with previously unexplained experimental results by Yang et al.\cite{Yang09}  In Sec.\ \ref{Discussion}, we list some limitations of our results and discuss how these might be improved, and we examine some practical consequences.  Appendix A contains a calculation of the self-energy of a Pearl vortex in a sector of arbitrary angle $\alpha$.

\section{Nearly uniform current density in a straight, narrow, thin superconducting strip\label{uniformK}}

We focus our attention on the properties of a thin superconducting strip of thickness $d$ much less than the London penetration depth $\lambda$, width $W$ much less than the Pearl length\cite{Pearl64} $\Lambda = 2 \lambda^2/d$, and Ginzburg-Landau coherence length $\xi$ much less than the width.  In particular, let us examine the current density in  an infinitely long thin superconducting strip of uniform width $W$ in the $xy$ plane, centered on the $x$ axis.  Suppose a current $I$ flows in the $x$ direction.  Since the film thickness obeys $d \ll \lambda$, the current density $\bm j$ is very nearly independent of $z$ across the film thickness, and therefore it is convenient to consider only the sheet current (the integral of $\bm j$ across the thickness) $\bm K = \bm j d$. When $W \ll \Lambda$, $\bm K$ is  very nearly independent of $y$ ($|y| < W/2$).  

According to the London equation,\cite{London61} the sheet-current density in the  strip obeys $\bm K = -(2/\mu_0 \Lambda)[\bm A+(\phi_0/2\pi)\nabla \gamma],$
where $\bm A$ is the vector potential ($\bm B = \nabla \times \bm A$) and $\gamma$ is the phase of the order parameter.  When the strip contains no vortices but carries a current $I = \int_{-W/2}^{W/2} K_x(y) dy$ in the $x$ direction, the gauge of the vector potential can be chosen to absorb the constant $(\phi_0/2\pi)\nabla \gamma$, such that $K_x(y)=-2A_x(y)/\mu_0\Lambda$, which  is an even function and the self-field $B_{z}(y) = -dA_{x}/dy$ is an odd function of $y$.
That $K_x$ is very nearly independent of $y$ when $W \ll \Lambda$ can be shown using perturbation theory by expanding $K_x = -2A_x/\mu_0 \Lambda = K_{x0} + K_{x1}+K_{x2}+...$ and $A_x = A_{x0} + A_{x1}+A_{x2}+...$, where $K_{x0} = I/W$, $K_{xn}=-2A_{xn}(y)/\mu_0\Lambda$ is proportional to $(W/\Lambda)^n$, and $\int_{-W/2}^{W/2} K_{xn}(y)dy=-2\int_{-W/2}^{W/2}A_{xn}(y)dy/\mu_0\Lambda=0$ for $n\ge 1$.  
To obtain the first-order correction term $K_{x1}$, we use the Biot-Savart law with $K_{x0}$ to  obtain
\begin{equation}
B_{z1}(y) = \frac{\mu_0 K_{x0}}{2\pi}\ln \Big|\frac{y+W/2}{y-W/2}\Big|
\end{equation}
and then integrate $B_{z1}(y) = -dA_{x1}/dy$ with respect to $y$ for $|y| \le W/2$ to obtain $A_{x1}(y)$.  
The result is $K_{x1}(y)=-2A_{x1}(y)/\mu_0\Lambda=(W/\Lambda)f(2y/W) K_{x0},$
where\begin{equation}
f(u)=\frac{1}{2\pi}[1-\ln4+(1+u)\ln(1+u)
+(1-u)\ln(1-u)].
\end{equation}
The function $f(2y/W)$ varies smoothly from a minimum of $f(0) = -0.061$ at the middle of the strip to $f(\pm 1) = 0.159$ at the edges of the strip.  Thus for the experiments reported in this paper, for which $W <$ 100 nm and $\Lambda \approx$  100 $\mu$m, the current density  $K_x(y)$ vs $y$ in a long, straight superconducting strip is flat to better than  0.02\%.

In general, the current flow in the strip obeys $\nabla \cdot \bm K = 0$ and (via the London equation) $\nabla \times \bm K = -2{\bm B}/\mu_0 \Lambda$.  However, as we have seen above, when $W \ll \Lambda$, the effect of $\bm B$ on the current flow is negligible, and to excellent approximation the current flow can be obtained from $\nabla \cdot \bm K = 0$ and $\nabla \times \bm K = 0$, as in thin nonsuperconducting strips.  As has been demonstrated in Refs.\ \onlinecite{Hagedorn63}-\onlinecite{Hall68}, the current distributions in nonsuperconducting thin films with complicated geometries can be  obtained using  complex-field methods.\cite{Churchill48}   Here we express the $x$ and $y$ components of the sheet-current density ${\bm K}(x,y) = \hat x K_x(x,y) + \hat y K_y(x,y)$ as the real and imaginary parts of a complex sheet-current density ${\cal K}(\zeta) = K_x(x,y) -i K_y(x,y)$, where ${\cal K}(\zeta)$ is an analytic function of the complex variable $\zeta = x + iy$.  The Cauchy-Riemann conditions obeyed by analytic functions correspond to the physical conditions that $\nabla \cdot \bm K = 0$ and  $\nabla \times \bm K=0$.  

\section{Procedure for estimating the critical current of narrow superconducting strips\label{procedure}}

Within the context of the Ginzburg-Landau (GL) theory,\cite{deGennes66,StJames69,Tinkham96}
the upper limit to the dc supercurrent that can be carried before a voltage appears along the length of the sample is the GL critical current density, $J_{GL}=\phi_0/3^{3/2}\pi\mu_0 \xi \lambda^2$, critical sheet-current density, $K_{GL}=2\phi_0/3^{3/2}\pi\mu_0 \xi \Lambda$, or critical current, $I_{GL}= 2\phi_0 W/3^{3/2}\pi\mu_0 \xi \Lambda$.  However, since $J_{GL}$ was calculated theoretically assuming that both $d$ and $W$ were much smaller than both $\lambda$ and $\xi$, these results cannot be applied to the case of interest here, $\xi \ll W$.  

In this paper we use a general but approximate method for estimating the critical current at which a voltage first appears along the sample length, the voltage being produced by the nucleation of a vortex at an edge or corner and its subsequent transit across the width.  The method must be capable of calculating the critical current for samples of many different geometries, including width variations and turnarounds. We will start by writing down the Gibbs free energy $G(\bm r_v)$, where $\bm r_v$ is the two-dimensional coordinate of a vortex, $G(\bm r_v) = E_{self}(\bm r_v)-W_I(\bm r_v)$. $E_{self}(\bm r_v)$ is the self-energy of the vortex accounting for all its interactions with the sample edges, including image vortices, and  $W_I(\bm r_v)$ is the work done by the sources of the current in moving the vortex away from the sample edge to the position $\bm r_v$.  

Neglecting core contributions, we use the London model of a vortex to calculate the self-energy $E_{self}$ as the area integral of the kinetic-energy density per unit area $\mu_0 \Lambda K^2/4$ outside the vortex core, taken to have   radius $\xi$, the Ginzburg-Landau coherence length.\cite{Bulaevskii11,Bulaevskii11a,Kogan94,Maksimova98,Kuit08}  The divergence theorem can be used to obtain $E_{self}(\bm r_v)= \phi_0 I_{circ}(\bm r_v)/2$, where $I_{circ}(\bm r_v)$ is the net self-generated supercurrent circulating around the vortex core when the vortex is at $\bm r_v$.   

When the vortex is placed in a sample carrying a current $I$ distributed as the sheet-current density $\bm K_I(\bm r)$, the work $W_I(\bm r_v)$ is the line integral of the Lorentz force $\bm F_L(\bm r) = \bm K_I(\bm r) \times \phi_0 \hat z$ from the vortex's point of entry to the position $\bm r_v$.  Thus $W_I(\bm r_v) = \phi_0 \Delta I(\bm r_v),$ where $\Delta I(\bm r_v)$ is the portion of the current $I$ that flows between the vortex's point of entry and the position $\bm r_v$. In contrast to the approach used in Refs.\ \onlinecite{Bulaevskii11} and \onlinecite{Bulaevskii11a}, we neglect any suppression of the magnitude of the GL order parameter by the current density.

We wish to be able to calculate the critical current for a variety of sample geometries.  In each case we will find that for small currents $I$ there is a Gibbs free-energy barrier  at $\bm r_b$ that prevents the nucleation of a vortex.  At the top of the  barrier, where $\bm r_v = \bm r_b$, $\nabla G(\bm r_v) = 0$, which corresponds to a balance of forces: The Lorentz force, which tends to repel the vortex away from the sample edge, is balanced by the image force, which tends to attract the vortex back to the edge.  We will define the critical current $I_c$ as that value of the current for which the height of the barrier is reduced to zero and $\bm r_b$ moves to the position $\bm r_c$, where $G(\bm r_c) = 0$.  In the following calculations we will find for each case that $\bm r_c$ is only a short distance, of the order of $\xi$, from the vortex entry point.  Since we consider $\xi \ll W$, the functions needed to calculate $G(\bm r_v)$ can all be obtained using power-law expansions.

\section{Critical current density of a long, straight, narrow, thin superconducting strip\label{Kcstripsection}}

In this section, we present a general procedure for calculating the critical current in  all geometries but apply it as an example to the simplest case, a long, thin strip of uniform width $W$, here assumed to occupy the space $0 <y < W$.\cite{footnote1} 
In general, we will first use conformal mapping to find the applied sheet-current distribution and to calculate $W_I(\bm r_v)$ for vortex positions $\bm r_v$ close to the nucleation point.  Next we will use the same mapping to find the current distribution around the nucleating vortex and to calculate $E_{self}(\bm r_v)$.  We will then examine the Gibbs free energy $G(\bm r_v)$ for $\bm r_v$ at or near the barrier, and we will define the critical current as the current that reduces the Gibbs free-energy barrier to zero.

The conformal mapping,\cite{Churchill48a}
\begin{eqnarray}
\zeta'(w)&=&\frac{d\zeta(w)}{dw}= \frac{2W}{\pi(w^2-1)},
\label{zetaprimestraight}\\
\zeta(w)&=& \frac{W}{\pi}\ln\Big(\frac{w-1}{w+1}\Big),
\label{zetastraight}\\
w(\zeta)&=& -\coth\big(\frac{\pi\zeta}{2W}\Big),
\label{wstraight}
\end{eqnarray}
maps points in the upper half $w$-plane $v \ge 0$ ($w = u + iv$) into the strip $0 \le y \le W$  in the $\zeta$-plane ($\zeta= x + i y$).

The complex potential 
\begin{equation}
{\cal G}_w(w)=\frac{I}{\pi}\ln\Big(\frac{w-1}{w+1}\Big)
\label{calGw}
\end{equation}
describes the flow of current $I$ in the upper half $w$-plane from a source at $w = 1$ to a drain at $w = -1$.  In the $\zeta$-plane the same complex potential is
\begin{equation}
{\cal G}_\zeta(\zeta)={\cal G}_w(w(\zeta))=\frac{I}{\pi}\ln\Big(\frac{w(\zeta)-1}{w(\zeta)+1}\Big),
\label{calGzeta}
\end{equation}
and the corresponding complex sheet current is ${\cal K}_\zeta(\zeta)=d{\cal G}_\zeta(\zeta)/d\zeta = K_x(x,y)-iK_y(x,y)$. The streamlines of the applied sheet-current density $\bm K = \hat x K_x + \hat y K_y$  are obtained as contours of the stream function $S(x,y) = \Im{\cal G}_\zeta(x+iy)$, the imaginary part of ${\cal G}_\zeta(\zeta)$. In general, $K_x(x,y) = \partial S(x,y)/\partial y$ and $K_y(x,y) = -\partial S(x,y)/\partial x$.  

In the example for which $w(\zeta)$ is given by Eq.\ (\ref{wstraight}), we have, with $I = K_I W$, ${\cal G}_\zeta(\zeta)=K_I \zeta$, ${\cal K}_\zeta(\zeta) = K_I$, $S(x,y) = K_I y,$ $K_x = K_I$, and $K_y = 0$.

The complex potential describing the sheet-current flowing around a Pearl vortex\cite{Pearl64} at $w = w_v$ in the upper half $w$-plane, subject to the boundary condition that at $v = 0$ there be no current flow perpendicular to the $u$ axis, can be obtained simply by the method of images, so long as the vortex's distance from the edge of the film is much less than the Pearl length $\Lambda$.\cite{Kogan94}  The self-generated sheet-current density circulating at distances $\rho \ll \Lambda$  around the axis of a Pearl vortex  has magnitude $K = \phi_0/\pi\mu_0\Lambda \rho$.\cite{Pearl64}  We therefore have 
\begin{equation}
{\cal G}_{vw}(w_v;w)=\frac{i\phi_0}{\pi \mu_0 \Lambda} \ln \Big(\frac{w-w_v^*}{w-w_v}\Big),
\label{Gvw}
\end{equation}
where the term in the numerator arises from the negative image at $w = w^*_v$ needed to satisfy the boundary condition at $v = 0$.  
The complex potential describing the sheet-current flow circulating around a vortex at $\zeta_v = x_v + i y_v$ in the $\zeta$ plane can be obtained by starting from Eq.\ (\ref{Gvw}) and using  $w(\zeta)$ from Eq.\ (\ref{wstraight}) to obtain
\begin{equation}
{\cal G}_{v\zeta}(\zeta_v;\zeta)=\frac{i\phi_0}{\pi \mu_0 \Lambda} \ln \Big(\frac{w(\zeta)-w^*(\zeta_v)}{w(\zeta)-w(\zeta_v)}\Big).
\label{Gvzeta}
\end{equation}
This function automatically accounts for the infinite sets of positive and negative image vortices needed to satisfy the boundary condition that the sheet-current density be parallel to the edges at $ y = 0$ and $y = W$.  The imaginary part of ${\cal G}_{v\zeta}$ is the stream function $S_v(x_v,y_v;x,y) = \Im{\cal G}_{v\zeta}(\zeta_v;\zeta)$, which rises to its largest value as $\zeta \to \zeta_v$ and $w(\zeta)\to w(\zeta_v)$; it is zero on the boundaries $y = 0$ and $y = W$.   

The current circulating around the vortex can be obtained from Eq.\ (\ref{Gvzeta}) by evaluating the stream function at a cutoff radius equal to the Ginzburg-Landau coherence length, i.e., a distance $\xi$ from the vortex position  $\zeta_v$, where we make the approximation that $\xi \ll |\zeta_v|$. In the numerator of the argument of the logarithm we replace $w(\zeta)$ by $w(\zeta_v)$, and in the denominator we replace $w(\zeta)-w(\zeta_v)$ by $\xi dw(\zeta_v)/d\zeta_v=\xi/\zeta'(w(\zeta_v))$.   When the vortex is at $(x,y) = (x_v,y_v)$, the circulating current is
\begin{eqnarray}
\!\!\!\!\!\!\!\!I_{circ}(x_v,y_v) &\!=& \!\!S_v(x_v,y_v;x_v+\xi,y_v) \nonumber 
\\
&=&\!\!\!\!\frac{\phi_0}{\pi \mu_0 \Lambda}\! \ln \!\Big[\frac{|w(\zeta_v)\!-\!w^*(\zeta_v)||\zeta'(w(\zeta_v))|}{\xi}\!\Big].
\label{Icirc}
\end{eqnarray}

We can use the stream function $S(x,y)$ to calculate the work $W_I(\bm r_v) = \phi_0 \Delta I(\bm r_v)$ done by the source of the current in moving the vortex from its point of entry $(x,y) = (x_{en},y_{en})$ to the point  $(x,y) = (x_v,y_v)$,
\begin{equation}
\Delta I(x_v,y_v) = S(x_{en},y_{en})-S(x_v,y_v).
\label{DeltaI}  
\end{equation}
Note that $\Delta I$ also can be obtained by integrating the sheet-current density  $\bm K_I(\bm r)$ that passes between $(x_{en},y_{en})$ and $(x_v,y_v)$.

When $\xi \ll W$, the distance between the point of entry and the point where the Gibbs free energy is zero is only a little larger than $\xi$ for all the cases considered here.  We therefore will use expansions of $w(\zeta)$ valid at small distances from the vortex entry point to evaluate Eqs.\ (\ref{calGzeta})-(\ref{DeltaI}) for all cases considered in this paper.  

As an example, let us now apply the above general procedure to evaluate the critical current for a long straight strip of width $W$.  If the vortex entry point is at $(x_{en},y_{en}) = (0,W)$, which corresponds to $\zeta = iW$ and $w=0$, and the vortex position of interest is $(x_v,y_v) = (0,W-\delta)$, we can use Eq.\ (\ref{zetaprimestraight})  to obtain the approximation $w = i \pi \delta/2W$, which is valid for $\delta \ll W$.  Using this $w$ in place of $w(\zeta)$ in Eqs.\ (\ref{calGzeta}) and (\ref{Gvzeta}) and evaluating   Eqs.\ (\ref{Icirc}) and (\ref{DeltaI}) by expanding them to lowest order in $\delta/W$ yields 

\begin{equation}
G = \frac{\phi_0^2}{2\pi\mu_0\Lambda}\ln\Big(\frac{2\delta}{\xi}\Big)-\phi_0 K_I\delta.
\label{Gdelta}
\end{equation}

The first term on the right-hand side of Eq.\ (\ref{Gdelta}) is the self-energy accounting for the vortex's interaction with its nearest negative image vortex at a distance $2\delta$, and the second term is the negative of the work done by the source of the current in moving the vortex a distance $\delta$ in from the edge.

The free energy barrier occurs at $\delta = \delta_b$.  Setting $\partial G/\partial \delta = 0$ there, we obtain
\begin{equation}
\delta_b = \frac{\phi_0}{2\pi \mu_0 \Lambda K_I},
\end{equation}
which is equivalent to the force balance between the repulsive Lorentz force $\phi_0 K_I$ and the attractive force of the image vortex $\phi_0^2/2\pi\mu_0 \Lambda s$. 
Setting $G=0$ at $\delta_b$ yields  $\delta_b = \delta_c= e\xi/2 = 1.36\xi$ and the critical sheet current $K_I = K_c$,
\begin{equation}
K_c = \frac{\phi_0}{e \pi\mu_0\xi\Lambda},
\label{Kcstrip}
\end{equation}
where $e$ is Euler's number, 2.718....  Despite the approximations made in deriving Eq.\ (\ref{Kcstrip}), this result for $K_c$ is numerically close to $K_{GL}$. The barrier height for $K_I < K_c$ is
\begin{equation}
G_b =  \frac{\phi_0^2}{2\pi\mu_0\Lambda}\ln\Big(\frac{K_c}{K_I}\Big).
\label{Gbstrip}
\end{equation}

In later sections we have followed the above procedure, employing complex fields and conformal mapping, to calculate the current flow and the critical sheet current in strips with various geometries, including turns and turnarounds.  The main difference between these cases is the mathematical form of the conformal mappings, which replace Eqs.\ (\ref{zetaprimestraight})-(\ref{wstraight}).

\section{Current crowding \label{curve}}

\begin{figure}
\includegraphics[width=4cm]{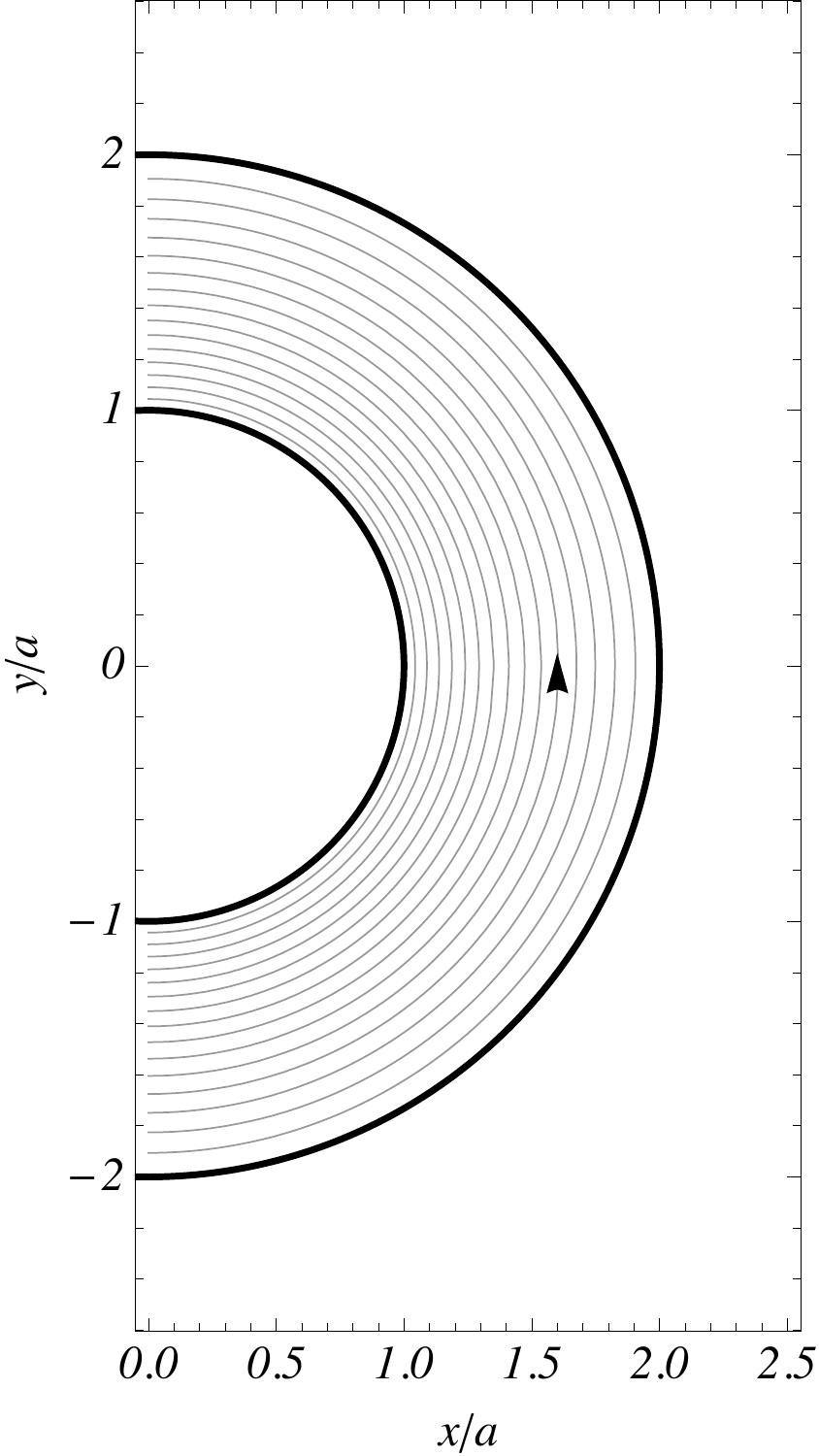}
\caption{Current flow in a strip carrying average sheet-current density $K_I$ around a circular arc of inner radius $a$ and outer radius $b=2a$, showing current crowding near the inner radius. }
\label{arcplot}
\end{figure} 

When a superconducting strip of constant width $W$ follows a curving path, current crowding occurs near the inner radius of a bend.  As a result, the  critical current of the strip is reduced because vortices more readily nucleate where the edge has the minimum radius of curvature. 
Here we examine this effect in two limits, $\xi \ll W$ and $\xi \gg W$.

\subsection{London-model calculation for $\xi \ll W$}
 
This critical-current reduction   can be understood using the London model for the geometry shown in Fig.\ \ref{arcplot}.  When  a sheet-current $\bm K(\rho)= K_\phi(\rho) \hat\phi$ flows around an annulus of inner and outer radii $a$ and $b = a+W$, where $\rho$ is the radial coordinate and $W \ll \Lambda$, the solution  obeying 
 $\nabla \cdot \bm K = 0$ and  $\nabla \times \bm K=0$ is $K_\phi(\rho) = (b-a)K_I/\ln(b/a)\rho,$ where $K_I$ is the average sheet-current density.  The current density at the inner radius, whose radius of curvature is  $\rho_c= a$, has the enhanced value, $K_\phi(a) = (b-a)K_I/\ln(b/a)a.$  As a function of the distance $\delta$ of a nucleating vortex from the point $(x,y) = (a,0)$, the self-energy of the vortex, including its interaction with the nearest negative image, is the same as the first term in Eq.\  (\ref{Gdelta}) except for correction terms of order $\delta/a$, which are negligible in critical-current calculations  when $\xi \ll a$.  Thus when $\delta \ll (b-a)$, the Gibbs free energy becomes 
\begin{equation}
G= \frac{\phi_0^2}{2\pi\mu_0\Lambda}\ln\Big(\frac{2\delta}{\xi}\Big)-\phi_0 K_\phi(a)\delta.
\label{Gdeltacurve}
\end{equation}
Following the steps that led to Eq.\ (\ref{Kcstrip}), we obtain 
\begin{equation}
\delta_b = \frac{\phi_0}{2\pi \mu_0 \Lambda K_\phi(a)}, \;{\rm and}\;
\delta_c=e\xi/2=1.36\xi,
\end{equation}
which tells us that when $\xi \ll \rho_c = a$, the critical current is reached when 
$K_\phi(a)=\phi_0/e \pi\mu_0\xi\Lambda$.  However, since  $K_I< K_\phi(a)$ and $K_c$ is the critical value of $K_I$, we have
\begin{equation}
K_c = \frac{\phi_0}{e \pi\mu_0\xi\Lambda}R,\;{\rm where }\;
R=\frac{a\ln(b/a)}{(b-a)}
\label{Rarc}
\end{equation}
is the reduction factor due to current crowding at the curving inner boundary, relative to the $K_c$ for a long straight strip [Eq.\ (\ref{Kcstrip})]. For example, $R= 0.69$ when $(b-a)=W =a$. Note that $R \to 1$ in the limit of infinite radius of curvature ($\rho_c = a\to \infty$) for fixed $W =(b-a)$.  The barrier height for $K_I < K_c$ is
\begin{equation}
G_b =  \frac{\phi_0^2}{2\pi\mu_0\Lambda}\ln\Big(\frac{K_c}{K_I}\Big).
\label{Gbarc}
\end{equation}

\subsection{Ginzburg-Landau calculation for $\xi \gg W$}

The calculation of the critical-current reduction factor $R$ of Eq.\ (\ref{Rarc}) assumed that $\xi \ll a$, but the amount of the reduction is not as great when $\xi \approx a$ or when  $\xi \gg a$.  For the latter case it is straightforward to calculate the critical current as an extension of the Ginzburg-Landau (GL) calculation\cite{deGennes66,StJames69,Tinkham96} that yields the GL critical current density of a long, straight strip, $J_{GL}=\phi_0/3^{3/2}\pi\mu_0 \xi \lambda^2$, critical sheet-current density, $K_{GL}=2\phi_0/3^{3/2}\pi\mu_0 \xi \Lambda$, or critical current, $I_{GL}= 2\phi_0 W/3^{3/2}\pi\mu_0 \xi \Lambda$, which assumes that that both $d$ and $W$ are much smaller than both $\lambda$ and $\xi$.  For the geometry shown in Fig.\ \ref{arcplot}, the GL sheet-current density is\cite{deGennes66,StJames69,Tinkham96} $\bm K = -(2f^2/\mu_0 \Lambda)[\bm A+(\phi_0/2\pi)\nabla \gamma],$ where $f$ is the magnitude of the normalized order parameter. Since we can neglect the self-field, we may choose a gauge such that the current around the arc is determined by the gradient of the phase of the order parameter, $\nabla \gamma = -\hat \phi k/\rho$, where $k$ is a dimensionless constant.  The sheet-current density becomes $\bm K = \hat \phi K_\phi$, where
\begin{equation}
K_\phi(\rho) = \Big(\frac{\phi_0}{\pi\mu_0\Lambda\xi}\Big)f^2v_s,
\end{equation}
 $v_s = \rho_0/\rho$ is the magnitude of the superfluid velocity in GL dimensionless units,\cite{deGennes66,StJames69,Tinkham96} and $\rho_0 = k\xi$ is a measure of the total current around the arc.  When $\xi \gg W = (b-a)$, the first Ginzburg-Landau equation yields $f^2 = 1-v_s^2,$ so that the radial dependence of $K_\phi(\rho)$ is given by 
\begin{equation}
K_\phi(\rho) =  \Big(\frac{\phi_0}{\pi\mu_0\Lambda\xi}\Big)\Big(\frac{\rho_0}{\rho}\Big)\Big[1-(\frac{\rho_0}{\rho}\Big)^2\Big].
\end{equation}
The integral of $K_\phi$ over the width $W = b-a$ yields the total current $I$ and the average sheet-current density $K_I = I/W$,
\begin{equation}
K_I =  \Big(\frac{\phi_0}{\pi\mu_0\Lambda\xi}\Big)\Big[\frac{\rho_0 \ln(b/a)}{b-a}-\frac{\rho_0^3(b+a)}{2a^2b^2}\Big].
\end{equation}
$K_I$ is maximized when $\rho_0 = \rho_{0max}$, where 
\begin{equation}
\rho_{0max}=ab\Big[\frac{2\ln(b/a)}{3(b^2-a^2)}\Big]^{1/2}.
\end{equation}
The corresponding maximum value of $K_I$ is the arc's Ginzburg-Landau critical sheet-current density,
\begin{equation}
K_{cGL}=K_{GL}R_{GL},
\end{equation}
where $K_{GL}=2\phi_0/3^{3/2}\pi\mu_0 \xi \Lambda$, and 
\begin{equation}
R_{GL}=\frac{ab}{\sqrt{(a+b)/2}}\Big[\frac{\ln(b/a)}{b-a}\Big]^{3/2}
\end{equation}
is the GL reduction factor due to the inhomogeneous current density around the arc.  For $(b-a)=W =a$, the case shown in Fig.\ \ref{arcplot}, $R_{GL}=0.94$.  Expanding for small values of $(b-a)/(b+a)$ yields $R_{GL}\approx 1-[(b-a)/(b+a)]^2/2.$

\section{Radius-of-curvature effect\label{rhocsec}}

As we have seen above, when the superconducting strip is curved, current crowding occurs at the point of minimum radius of curvature $\rho_c$, and this reduces the critical sheet-current density of the strip below the value given in Eq.\ (\ref{Kcstrip}) when $\xi \ll \rho_c$.  However, for values of $\rho_c$ of the order of $\xi$ or smaller, the critical-current reduction due to current crowding is ameliorated by a radius-of-curvature effect. In strips fabricated with sharp corners, one might at first expect the critical current to vanish, because the sheet-current density diverges at the sharp inner corner.  However, for such cases the radius-of-curvature effect partially compensates for the current-crowding effect and leads to a critical current that is reduced by a factor proportional to $(\xi/a)^n$, where $a$ is a characteristic linear dimension of the strip and $n$ is a geometry-dependent fractional exponent. 

\subsection{Rounded 180-degree turnaround\label{round180}}

\begin{figure}
\includegraphics[width=6cm]{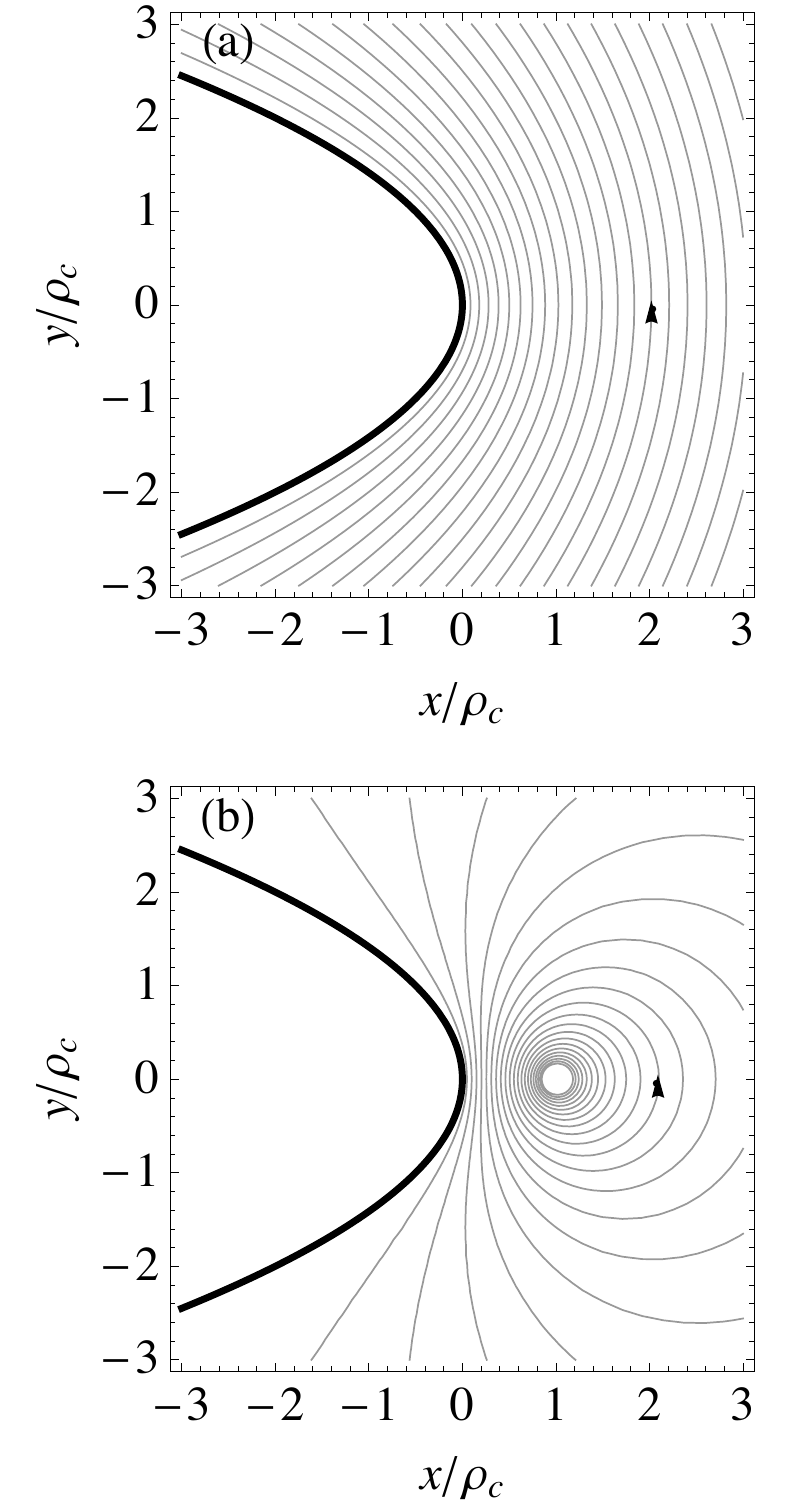}
\caption{(a) Current crowding around a parabolic bend of radius of curvature $\rho_c$, shown by the contour plot of the stream function $S(x,y)= \Im{\cal G}_\zeta(x+iy)$ [Eq.\ (\ref{calGparabolic})], which has the value $S=0$ along the parabolic boundary, $x = -y^2/2\rho_c$.  The contours correspond to streamlines of the sheet-current density $\bm K$, and the arrow shows the current direction.  (b) Current flow generated by a vortex interacting with the parabolic boundary, shown by the contour plot of the stream function $S_v(x_v,y_v;x,y)$, which has the value $S_v=0$ for $(x,y)$  along the boundary. The contours, shown here for $(x_v,y_v) = (\rho_c,0)$, correspond to streamlines of the vortex-generated sheet-current density $\bm K_v$, and the arrow shows the direction of the current.}
\label{parabolicplot}
\end{figure}

To analyze this radius-of-curvature effect near the inside corner of a rounded 180-degree turnaround, we use the conformal mapping,\cite{Kober57p39}
\begin{eqnarray}
\zeta(w)&=&-\rho_c w^2/2-i\rho_c w,\label{zetaofwpara}\\
w(\zeta)&=&i(\sqrt{2\zeta/\rho_c + 1}-1)\label{wofzetapara},
\end{eqnarray} 
which maps points in the upper half $w$ plane into points in the $\zeta$ plane ($\zeta = x + i y)$ to the right of the parabola $x = -y^2/2\rho_c.$  This parabola has radius of curvature $\rho_c$ at the origin.   See Fig.\ \ref{parabolicplot}.

The complex potential describing uniform current flow from right to left  in the $w$ plane is ${\cal G}_w = -K_0 \rho_c w$, and in the $\zeta$ plane the corresponding complex potential 
\begin{equation}
{\cal G}_\zeta(\zeta) = -K_0 \rho_c w(\zeta)
\label{calGparabolic}
\end{equation}
 describes the current crowding around the point $(x,y) = (0,0)$.  The imaginary part yields the stream function $S(x,y) = \Im{\cal G}_\zeta(\zeta)$, shown in Fig.\ \ref{parabolicplot}(a).  The complex current density is
\begin{equation}
{\cal K}(\zeta) \!= \!\frac{d{\cal G}_\zeta(\zeta)}{d\zeta} \!= \!\frac{-i K_0}{\sqrt{2\zeta/\rho_c \!+ \!1}} \!= \!K_x(x,y)\!-\!i K_y(x,y). 
\label{calKparabolic} 
\end{equation}
The constant prefactors in Eq.\ (\ref{calGparabolic}) are chosen such that the current density at the origin is $\bm K(0,0) = \hat y K_0$.

The critical current is reached when a vortex can be nucleated from the parabolic tip at $(x,y) = (0,0)$.  To calculate it, we first need the work term $W_I(\bm r_v) = \phi_0 \Delta I(\bm r_v)=\phi_0 [S(0,0)-S(x_v,y_v)]$, which can be obtained from Eq.\  (\ref{calGparabolic}).
For  $\bm r_v=(\delta,0)$ or $\zeta_v = \delta$, 
\begin{equation}
\Delta I(\delta) = K_0 \rho_c(\sqrt{2\delta/\rho_c + 1}-1).
\label{DeltaIparabolic}
\end{equation}

The self-energy $E_{self}(\bm r_v)= \phi_0 I_{circ}(\bm r_v)/2$ can be obtained as follows.  The complex potential describing the sheet-current flowing around a vortex in the $\zeta$ plane of Fig.\ \ref{parabolicplot}(b) is given by Eq.\ (\ref{Gvzeta}) but with $w(\zeta)$ given by Eq.\ (\ref{wofzetapara}).  We are interested only in the case when the vortex is on the $x$ axis at $\bm r_v=(\delta,0)$ or $\zeta_v = \delta$. The imaginary part is the stream function $S_v(x_v,y_v;x,y) = \Im{\cal G}_{v\zeta}(\zeta_v;\zeta)$, shown in Fig.\ \ref{parabolicplot}(b). 
When $\xi \ll a$, $I_{circ}(\bm r_v) = S_v(\delta,0;\delta+\xi,0)$ can be evaluated as described in Sec.\ \ref{Kcstripsection}.
The result is 
\begin{equation}
I_{circ}(\delta)\! = \!\frac{\phi_0}{\pi\mu_0\Lambda}\!\ln\!\Big[\frac{2\rho_c}{\xi}\sqrt{\frac{2\delta}{\rho_c}\!+\!1}\Big(\sqrt{\frac{2\delta}{\rho_c}\!+\!1}-1\Big)\Big].  
\label{Icircparabolic}
\end{equation}

\begin{figure}
\includegraphics[width=7cm]{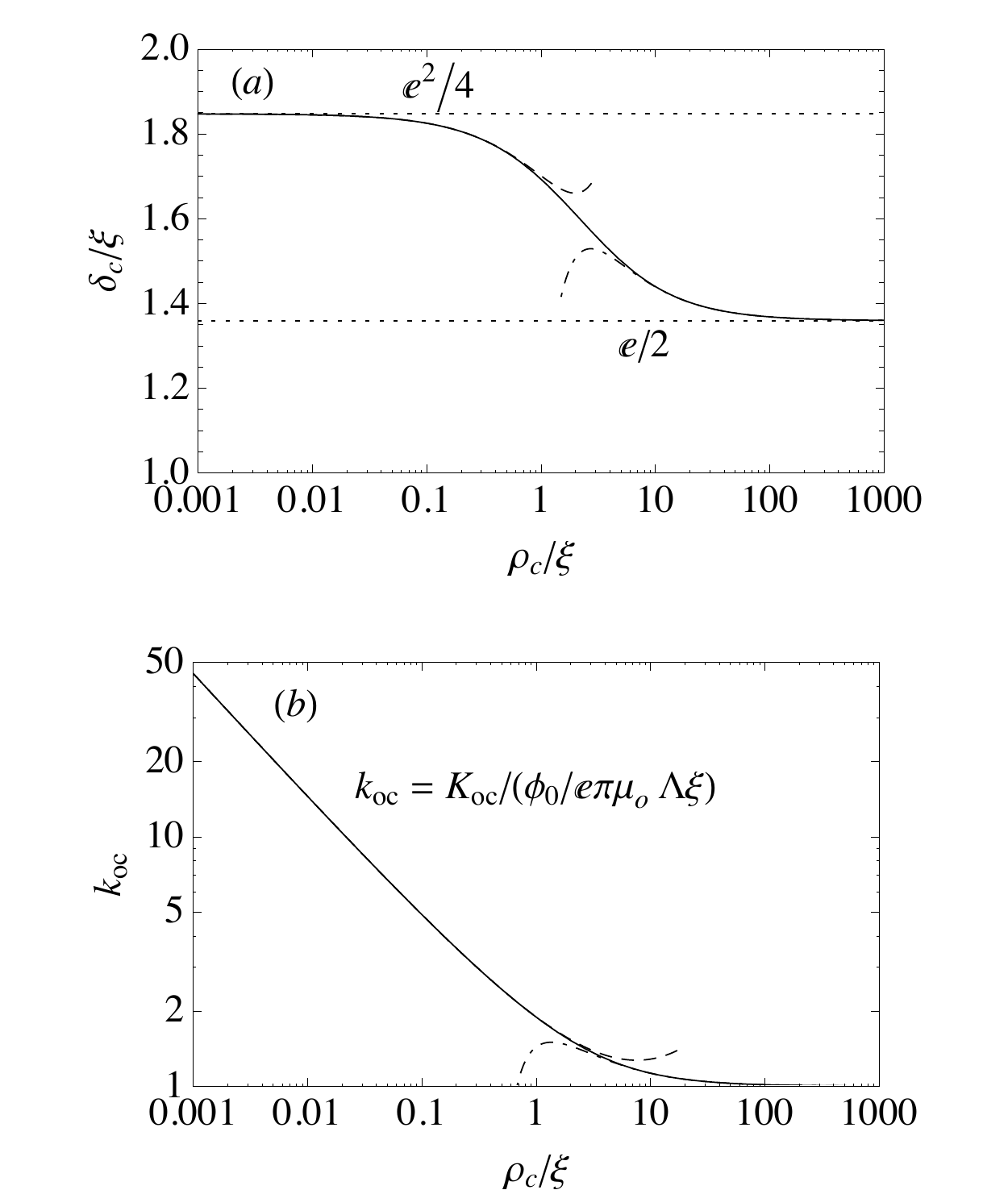}
\caption{Numerical results for (a) $\delta_c/\xi$ and (b) $K_{0c}$ (normalized to $\phi_0/e \pi\mu_0\xi\Lambda$) as functions of the ratio of the radius of curvature $\rho_c$ (see Fig.\ \ref{parabolicplot}) to the coherence length $\xi$. Expansions in powers of $\rho_c/\xi$ are shown as dashed curves for $\delta_c$ [Eq.\ (\ref{deltacsmrhoc})] and $K_{0c}$ [Eq.\ (\ref{K0csmrhoc})].  Expansions in powers of $\xi/\rho_c$ are shown as dot-dashed curves for $\delta_c$ [Eq.\ (\ref{deltaclgrhoc})] and $K_{0c}$ [Eq.\ (\ref{K0clgrhoc})].}
\label{deltacK0cparaplot}
\end{figure}

The Gibbs free energy is
\begin{eqnarray}
G &=& \frac{\phi_0^2}{2\pi \mu_0 \Lambda} \!\ln\!\Big[\frac{2\rho_c}{\xi}\sqrt{\frac{2\delta}{\rho_c}\!+\!1}\Big(\sqrt{\frac{2\delta}{\rho_c}\!+\!1}-1)\Big] \nonumber\\
&-&\phi_0 K_0 \rho_c\Big(\sqrt{\frac{2\delta}{\rho_c} + 1}-1\Big).
\label{Gparabolic}
\end{eqnarray}
The position of the barrier $\delta_b$ is the value of $\delta$ at the force-balance condition, $\partial G/\partial \delta=0$.  The critical sheet current density at $(x,y) = (0,0)$ is reached ($K_0 = K_{0c}$) when the barrier height $G$ is reduced to zero at $\delta_b = \delta_c$, and  
\begin{equation}
K_{0c}= \frac{\phi_0}{e \pi\mu_0\xi\Lambda} k_{0c},
\label{K0c}
\end{equation}
where numerical results for  $\delta_c/\xi$ and  $k_{0c}$ as  functions of $\rho_c/\xi$ are shown in Fig.\  \ref{deltacK0cparaplot}.  Analytic expansions (including only the first few terms in the series) for $\rho_c/\xi \ll 1$ are 
\begin{eqnarray}
\frac{\delta_c}{\xi}&=&\frac{e^{2}}{4}-\frac{1}{2}\Big(\frac{\rho_c}{2\xi}\Big)+\frac{2}{3e}\Big(\frac{\rho_c}{2\xi}\Big)^{\!\!3/2}\!\!+\frac{1}{2e^2}\Big(\frac{\rho_c}{2\xi}\Big)^{\!\!2} , \label{deltacsmrhoc}\\
k_{0c} \!\!\!&= &\!\!\!\Big(\frac{2\xi}{\rho_c}\Big)^{\!\!1/2}\!\!\!\!+\!\frac{1}{e}\!+\!\frac{1}{e^2}\Big(\frac{\rho_c}{2\xi}\Big)^{\!\!1/2}\!\!\!\!\!+\!\frac{2}{3e^3}\Big(\frac{\rho_c}{2\xi}\Big)\!,
\label{K0csmrhoc}
\end{eqnarray}
and  corresponding expansions for $\rho_c/\xi \gg 1$ are
\begin{eqnarray}
\frac{\delta_c}{\xi}&=&\frac{e}{2}+\frac{e^2}{4}\Big(\frac{\xi}{2\rho_c}\Big)-\frac{e^3}{4}\Big(\frac{\xi}{2\rho_c}\Big)^{2}, \label{deltaclgrhoc}\\
k_{0c} &= &1+e\Big(\frac{\xi}{2\rho_c}\Big)-\frac{e^2}{2}\Big(\frac{\xi}{2\rho_c}\Big)^2.
\label{K0clgrhoc}
\end{eqnarray}

Regardless of the size of $\rho_c$ relative to $\xi$, vortex nucleation  occurs when the barrier height is reduced to zero at a distance $\delta_c$ of the order of $\xi$ from the point of minimum radius of curvature [see Fig.\ \ref{deltacK0cparaplot}(a)].   When $\xi \ll \rho_c$, vortex nucleation occurs when the sheet-current density $K_{0}$ at this point exceeds $\phi_0/e \pi\mu_0\xi\Lambda$.  However, in the opposite limit ($\rho_c \ll \xi$), vortex nucleation does not occur until $K_{0}$ reaches much larger values [see Fig.\ \ref{deltacK0cparaplot}(b)].

\subsection{Rounded 90-degree turn\label{round90}}

\begin{figure}
\includegraphics[width = 5 cm]{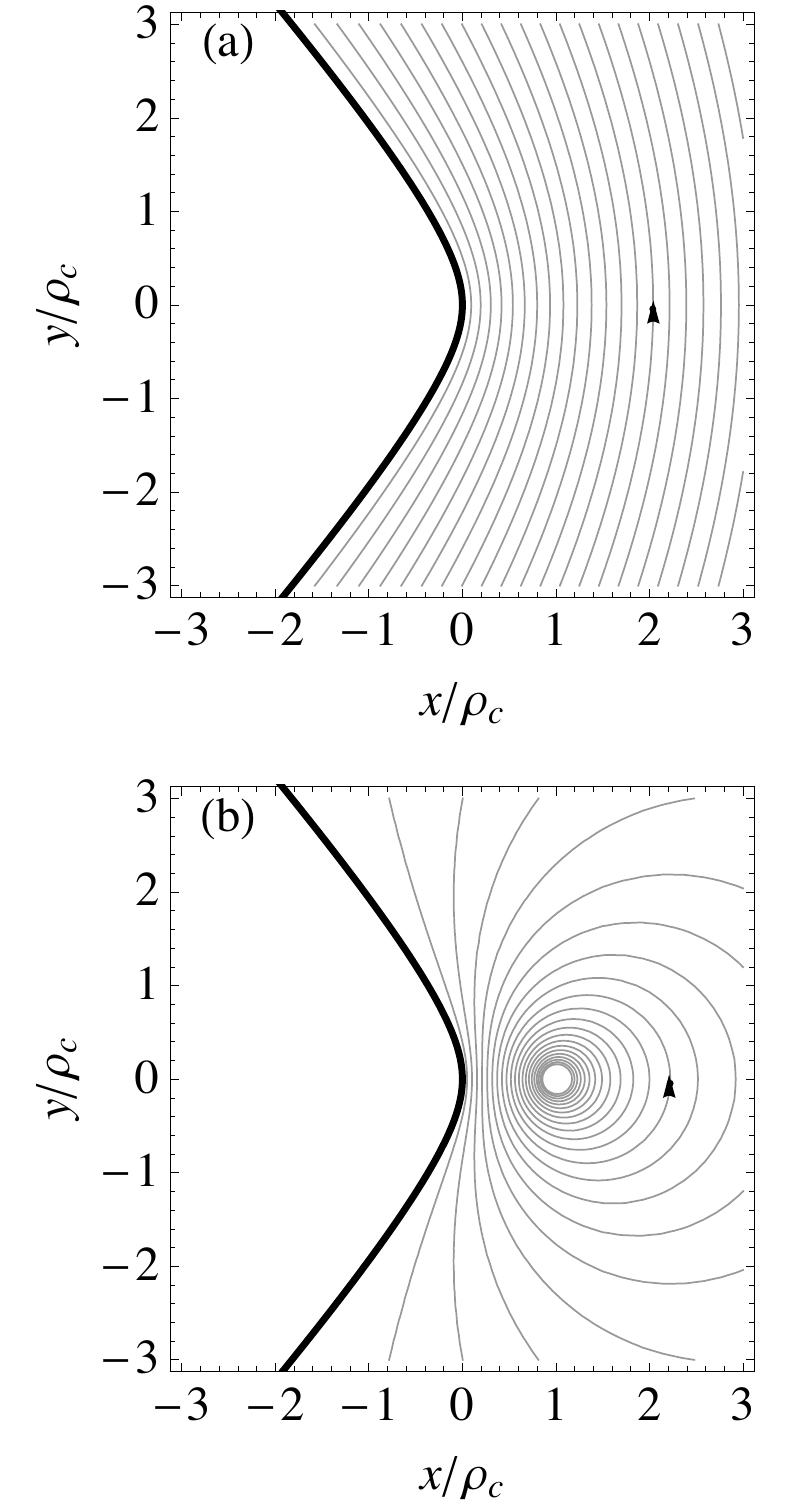}
\caption{(a) Current crowding around a generalized hyperbolic bend of radius of curvature $\rho_c$, shown by the contour plot of the stream function $S(x,y)= \Im{\cal G}_\zeta(x+iy)$ [Eq.\ (\ref{calGgenhyp})], which has the value $S=0$ along the  boundary.  The contours correspond to streamlines of the sheet-current density $\bm K$, and the arrow shows the current direction.  (b) Current flow generated by a vortex interacting with the  boundary, shown by the contour plot of the stream function $S_v(x_v,y_v;x,y)$, which has the value $S_v=0$ for $(x,y)$  along the boundary. The contours, shown here for $(x_v,y_v) = (\rho_c,0)$, correspond to streamlines of the vortex-generated sheet-current density $\bm K_v$, and the arrow shows the direction of the current.}
\label{genhypplot}
\end{figure}

To analyze the radius-of-curvature effect near the inside corner of a rounded 90-degree turn, we use a different  conformal mapping,\cite{Kober57p41}
\begin{eqnarray}
\zeta(w)&=&(\rho_c/3)[(1-iw)^{3/2}-1],\label{zetaofwhyp}\\
w(\zeta)&=&i[(3\zeta/\rho_c+1)^{2/3}-1]\label{wofzetahyp},
\end{eqnarray} 
which maps points in the upper half $w$ plane into points in the $\zeta$ plane ($\zeta = x + i y)$ to the right of a generalized hyperbola whose radius of curvature is $\rho_c$ at the origin.   See Fig.\ \ref{genhypplot}.

The complex potential describing uniform current flow from right to left  in the $w$ plane is ${\cal G}_w = -(K_0 \rho_c/2) w$.  In the $\zeta$ plane the corresponding complex potential 
\begin{equation}
{\cal G}_\zeta(\zeta) = -(K_0 \rho_c/2) w(\zeta)
\label{calGgenhyp}
\end{equation}
 describes the current crowding around the point $(x,y) = (0,0)$.  The imaginary part yields the stream function $S(x,y) = \Im{\cal G}_\zeta(\zeta)$, shown in Fig.\ \ref{genhypplot}(a).  The complex current density is
\begin{equation}
{\cal K}(\zeta) \!= \!\frac{d{\cal G}_\zeta(\zeta)}{d\zeta} \!= \!\frac{-i K_0}{(3\zeta/\rho_c \!+ \!1)^{1/3}} \!= \!K_x(x,y)\!-\!i K_y(x,y). 
\label{calKgenhyp} 
\end{equation}
The constant prefactors in Eq.\ (\ref{calGgenhyp}) are chosen such that the current density at the origin is $\bm K(0,0) = \hat y K_0$. 
The critical current is reached when a vortex can be nucleated from the tip at $(x,y) = (0,0)$.  To calculate it, we first need  the work term $W_I(\bm r_v) = \phi_0 \Delta I(\bm r_v)=\phi_0 [S(0,0)-S(x_v,y_v)]$, which can be obtained from Eq.\  (\ref{calGgenhyp}).
For  $\bm r_v=(\delta,0)$ or $\zeta_v = \delta$, 
\begin{equation}
\Delta I(\delta) = (K_0 \rho_c/2)[(3\delta/\rho_c + 1)^{2/3}-1).
\label{DeltaIgenhyp}
\end{equation}

The self-energy $E_{self}(\bm r_v)= \phi_0 I_{circ}(\bm r_v)/2$ can be obtained as follows.  The complex potential describing the sheet-current flowing around a vortex in the $\zeta$ plane of Fig.\ \ref{genhypplot}(b) is given by Eq.\ (\ref{Gvzeta}), but
where $w(\zeta)$ is given by Eq.\ (\ref{wofzetahyp}) and we are interested only in the case when the vortex is on the $x$ axis at $\bm r_v=(\delta,0)$ or $\zeta_v = \delta$. The imaginary part is the stream function $S_v(x_v,y_v;x,y) = \Im{\cal G}_{v\zeta}(\zeta_v;\zeta)$, shown in Fig.\ \ref{genhypplot}(b). 
When $\xi \ll a$, $I_{circ}(\bm r_v) = S_v(\delta,0;\delta+\xi,0)$  can be evaluated as described in Sec.\ \ref{Kcstripsection}.
The result is 
\begin{equation}
I_{circ}(\delta)\! = \!\frac{\phi_0}{\pi\mu_0\Lambda}\!\ln\!\Big\{\frac{\rho_c}{\xi}\Big(\frac{3\delta}{\rho_c}+\!1\Big)^{1/3}\Big[\Big(\frac{3\delta}{\rho_c}\!+\!1\Big)^{2/3}\!\!\!\!-1\Big)\Big]\Big\}.  
\label{Icircgenhyp}
\end{equation}

\begin{figure}
\includegraphics[width=7cm]{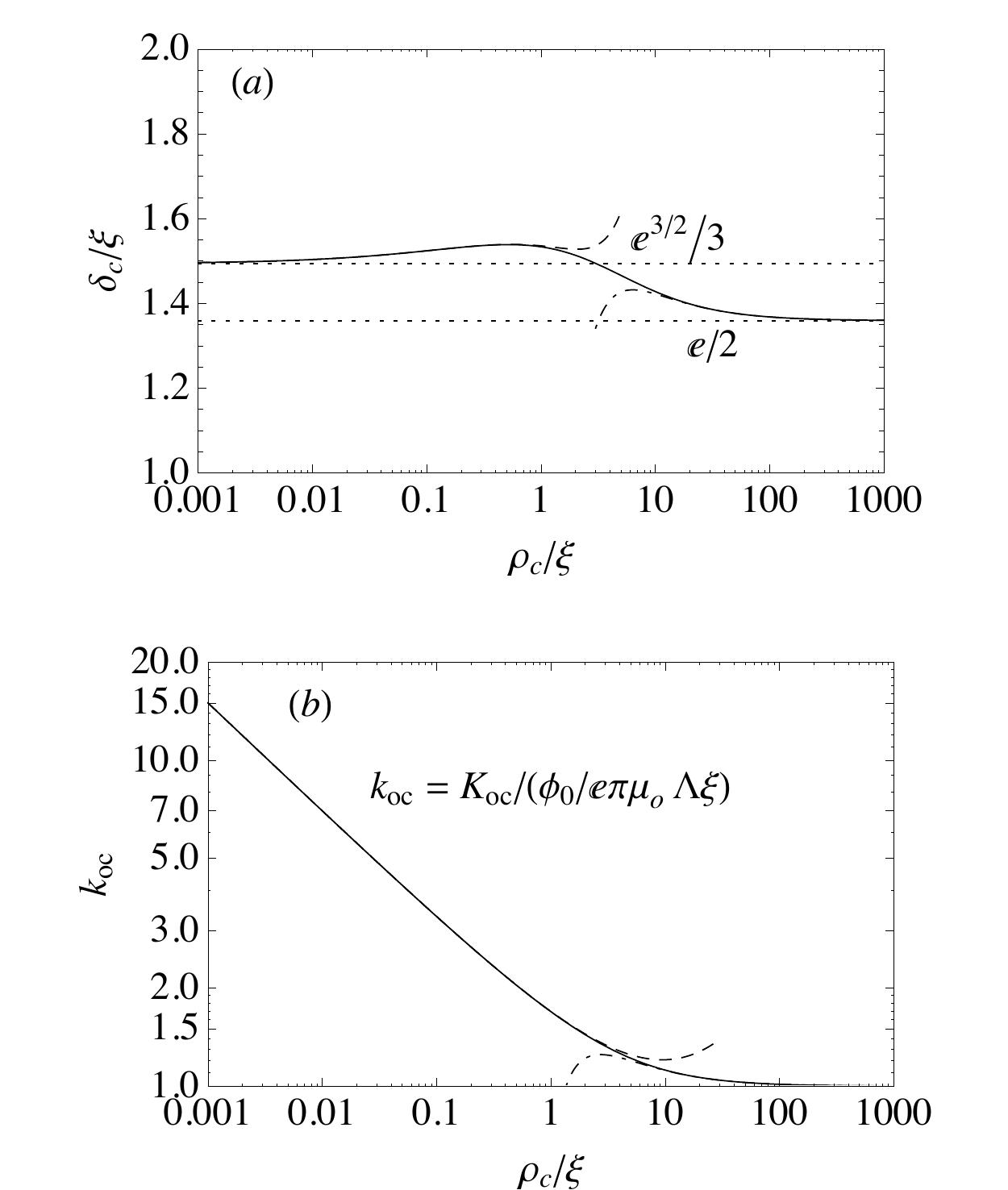}
\caption{Numerical results for (a) $\delta_c/\xi$ and (b) $K_{0c}$ (normalized to $\phi_0/e \pi\mu_0\xi\Lambda$) as functions of the ratio of the radius of curvature $\rho_c$ (see Fig.\ \ref{genhypplot}) to the coherence length $\xi$. Expansions in powers of $\rho_c/\xi$ are shown as dashed curves for $\delta_c$ [Eq.\ (\ref{deltacsmrhocgenhyp})] and $K_{0c}$ [Eq.\ (\ref{K0csmrhocgenhyp})].  Expansions in powers of $\xi/\rho_c$ are shown as dot-dashed curves for $\delta_c$ [Eq.\ (\ref{deltaclgrhocgenhyp})] and $K_{0c}$ [Eq.\ (\ref{K0clgrhocgenhyp})].}
\label{deltacK0cgenhypplot}
\end{figure}

The Gibbs free energy is
\begin{eqnarray}
G &=& \frac{\phi_0^2}{2\pi \mu_0 \Lambda} \!\ln\!\Big\{\frac{\rho_c}{\xi}\Big(\frac{3\delta}{\rho_c}\!+\!1\Big)^{1/3}\Big[\Big(\frac{3\delta}{\rho_c}\!+\!1\Big)^{2/3}\!\!\!\!\!-1\Big)\Big]\Big\} \nonumber\\
&-&\frac{\phi_0 K_0 \rho_c}{2}\Big[\Big(\frac{3\delta}{\rho_c}\!+\!1\Big)^{2/3}\!\!\!\!\!-1\Big)\Big].
\label{Ggenhyp}
\end{eqnarray}
The position of the barrier $\delta_b$ is the value of $\delta$ at the force-balance condition, $\partial G/\partial \delta=0$.  The critical sheet current density at $(x,y) = (0,0)$ is reached ($K_0 = K_{0c}$) when the barrier height $G$ is reduced to zero at $\delta_b = \delta_c$, and
\begin{equation}
K_{0c}= \frac{\phi_0}{e \pi\mu_0\xi\Lambda} k_{0c},
\label{K0cgenhyp}
\end{equation}
where numerical results for  $\delta_c/\xi$ and  $k_{0c}$ as  functions of $\rho_c/\xi$ are shown in Fig.\  \ref{deltacK0cgenhypplot}.  Analytic expansions (including only the first few terms in the series) for $\rho_c/\xi \ll 1$ are 
\begin{eqnarray}
\frac{\delta_c}{\xi}&=&\frac{e^{3/2}}{3}+\frac{e^{1/2}}{6}\Big(\frac{\rho_c}{\xi}\Big)^{2/3}-\frac{1}{3}\Big(\frac{\rho_c}{\xi}\Big)\nonumber \\ &+&\frac{11e^{-1/2}}{72}\Big(\frac{\rho_c}{\xi}\Big)^{-4/3}+\frac{5e^{-3/2}}{144}\Big(\frac{\rho_c}{\xi}\Big)^{2} , \label{deltacsmrhocgenhyp}\\
k_{0c} \!\!\!&= &\frac{3}{2}\Big(\frac{\xi}{\rho_c}\Big)^{\!\!1/3}\!\!\!\!+\!\frac{1}{2e}\Big(\frac{\rho_c}{\xi}\Big)^{\!\!1/3}\!+\!\frac{1}{12e^2}\Big(\frac{\rho_c}{\xi}\Big),
\label{K0csmrhocgenhyp}
\end{eqnarray}
and  corresponding expansions for $\rho_c/\xi \gg 1$ are
\begin{eqnarray}
\frac{\delta_c}{\xi}&=&\frac{e}{2}+\frac{e^2}{4}\Big(\frac{\xi}{2\rho_c}\Big)-\frac{e^3}{4}\Big(\frac{\xi}{2\rho_c}\Big)^{2}, \label{deltaclgrhocgenhyp}\\
K_{0c} &= &\frac{\phi_0}{e \pi\mu_0\xi\Lambda}\Big[1+e\Big(\frac{\xi}{2\rho_c}\Big)-\frac{e^2}{2}\Big(\frac{\xi}{2\rho_c}\Big)^2\Big].
\label{K0clgrhocgenhyp}
\end{eqnarray}

Regardless of the size of $\rho_c$ relative to $\xi$, vortex nucleation  occurs when the barrier height is reduced to zero at a distance $\delta_c$ of the order of $\xi$ from the point of minimum radius of curvature [see Fig.\ \ref{deltacK0cgenhypplot}(a)].   When $\xi \ll \rho_c$, vortex nucleation occurs when the sheet-current density $K_{0}$ at this point exceeds $\phi_0/e \pi\mu_0\xi\Lambda$.  However, in the opposite limit ($\rho_c \ll \xi$), vortex nucleation does not occur until $K_{0}$ reaches much larger values [see Fig.\ \ref{deltacK0cgenhypplot}(b)].

\section{180-degree Turnarounds\label{180general}}

In this section we address the extent to which the critical current is reduced by a 180-degree turnaround.  We consider the following specific examples: a sharp 180-degree turnaround,  an optimally rounded 180-degree turnaround,  a 180-degree turnaround intermediate between sharp and optimally rounded, a rounded 180-degree turnaround at the end of straight strips, and  a sharp rectangular 180-degree turnaround.  The latter geometry was used in the experiments of Yang et al,\cite{Yang09}  which will be discussed in more detail in Sec.\ \ref{ExpSec}.

\subsection{Sharp 180-degree turnaround\label{180sharp}}

\begin{figure}
\includegraphics[width = 6cm]{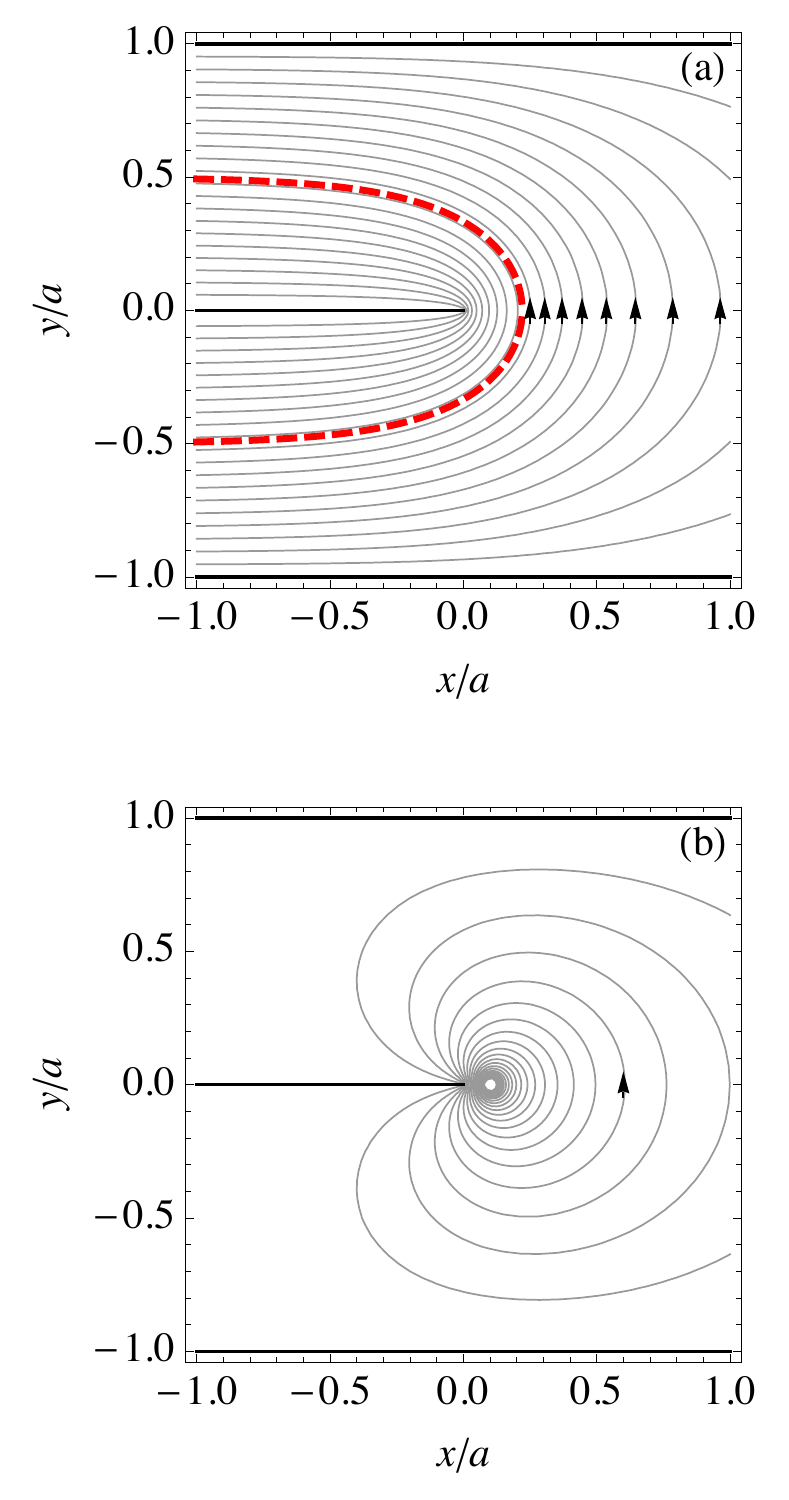}
\caption{(a) Current flow in a strip carrying current $K_I a$ around a 180-degree turn, shown by the contour plot of the stream function $S(x,y)= \Im{\cal G}_\zeta(x+iy)$, which has the values $S=0$ along the lines $y = \pm a$ at the outer boundaries and $S=K_I a$ on either side of the narrow gap along the line $y=0$ for $x < 0$.  The contours correspond to streamlines of the sheet-current density $\bm K$, and the arrows show the direction of the current.  The  dashed curve, which corresponds to $S = K_I a/2$, separates the current-crowding region close to $(x,y) = (0,0)$ from the current-expanding region  outside.  The magnitude of $\bm K$ is constant ($\bm K=K_I$) along the dashed curve. (b) Vortex-generated current flow, shown by the contour plot of the stream function $S_v(x_v,y_v;x,y)$, which has the values $S_v=0$ for $(x,y)$  along the boundaries. The contours, shown here for $(x_v,y_v) = (0.1a,0)$, correspond to streamlines of the vortex-generated sheet-current density $\bm K_v$, and the arrow shows the direction of the current.  }
\label{180fig}
\end{figure}

Consider the current flow in a strip of width $a$ with a sharp 180-degree turnaround as shown in Fig.\ \ref{180fig}. Alternatively, we can think of this as a wider strip of width $2a$, cut along the $x$ axis for $x < 0$.
The conformal mapping\cite{Kober57p94}
\begin{eqnarray}
\zeta'(w)&=&\frac{d\zeta(w)}{dw}= \frac{2a}{\pi}\frac{w}{w^2-1},
\label{zetaprime180}\\
\zeta(w)&=& (2a/\pi)(\ln\sqrt{w-1}+\ln\sqrt{w+1}-i\pi/2),
\end{eqnarray}
maps points in the upper half $w$-plane ($w = u + i v$)  into the strip $-a \le y \le a$ in the $\zeta$-plane  ($\zeta = x + i y$) as shown in Fig.\ \ref{180fig}.  The inverse mapping is given by 
\begin{equation}
w(\zeta) = \pm\sqrt{1-\exp(\pi\zeta/a)},
\label{wofzeta180}
\end{equation}
where the upper (lower) sign holds when $\Im\zeta \le 0$ ($\Im\zeta > 0$).

The complex potential is given by Eq.\ (\ref{calGzeta}) but with $w(\zeta)$ from Eq.\ (\ref{wofzeta180}) and $I = K_I a$, and the corresponding complex sheet current ${\cal K}_\zeta(\zeta)=d{\cal G}_\zeta(\zeta)/d\zeta = K_x(x,y)-iK_y(x,y)$ is
\begin{equation}
{\cal K}_\zeta(\zeta)=\frac{K_I}{w(\zeta)}.
\label{calK180}
\end{equation}
For  $-a < y < 0$, $K_x(x,y) \to K_I$ as $x \to -\infty$, and  for  $0 < y < a$, $K_x(x,y) \to -K_I$ as $x \to -\infty$.  The streamlines of the  sheet current $\bm K = \hat K_x + \hat y K_y$  are obtained as contours of the stream function $S(x,y) = \Im{\cal G}_\zeta(x+iy)$, the imaginary part of ${\cal G}_\zeta(\zeta)$, shown in Fig.\ \ref{180fig}(a).  

The critical current of the 180$^\circ$ turnaround is reached when a vortex can be nucleated at the sharp point in the center of Fig.\ \ref{180fig}(a) or (b) at $(x,y) = (0,0)$, which corresponds to $\zeta =0$ and $w = 0$.  To calculate the critical current, we examine the behavior when a vortex is at $\zeta_v = \delta$, where $\delta \ll a$.  Expanding Eq.\ (\ref{zetaprime180}), we obtain $w(\zeta_v) = i(\pi\delta/a)^{1/2}$ to lowest order, and from Eqs.\ (\ref{calGzeta}), (\ref{Icirc}), and (\ref{DeltaI}) we obtain 
the Gibbs free energy,
\begin{equation}
G = \frac{\phi_0^2}{2\pi \mu_0 \Lambda} \ln \Big(\frac{4\delta}{\xi}\Big)-2\phi_0 K_I \Big(\frac{a\delta}{\pi}\Big)^{1/2}.
\label{Gdelta180}
\end{equation}
Following the steps that led to Eq.\ (\ref{Kcstrip}), we obtain  with $K_I = K_c$,
\begin{eqnarray}
\delta_b^{1/2} &=& \frac{\phi_0}{2\mu_0 \Lambda K_I\sqrt{\pi a}},\;
\delta_c=e^{2}\xi/4=1.85\xi,\\
K_c &= &\frac{\phi_0}{e \pi\mu_0\xi\Lambda}R,\;{\rm where }\;
R=\Big(\frac{\pi\xi}{a}\Big)^{1/2}
\label{R180}
\end{eqnarray}
is the reduction factor due to current crowding near $(x,y)= (0,0)$. For example, $R= 0.56$ when $\xi = 0.1 a$ and $R= 0.18$ when $\xi = 0.01 a$. 
The barrier height for $K_I < K_c$ is
\begin{equation}
G_b =  \frac{\phi_0^2}{\pi\mu_0\Lambda}\ln\Big(\frac{K_c}{K_I}\Big).
\label{Gb180}
\end{equation}
Note that the prefactor is larger  than that in Eq.\ (\ref{Gbstrip}) by a factor of 2, which arises from the term proportional to $\delta^{1/2}$ in  Eq.\ (\ref{Gdelta180}).

\subsection{Optimally rounded 180-degree turnaround\label{180optimal}}

For the case of the 180$^\circ$ turn shown in Fig.\ \ref{180fig}, the complex sheet-current density is given by Eq.\ (\ref{calK180}).
Examination of Fig.\ \ref{180fig}(a) reveals that current crowding occurs, i.e., $K = |\bm K|$ increases along streamlines inside the  dashed curve as the current turns around the end of the gap, reaching a maximum at $(x,y) = (0,0)$.  On the other hand, $K$ decreases along streamlines outside the  dashed curve  in the current-turnaround region.  The dashed  curve, which corresponds to the contour for which $S=K_I a/2$, but which also can be obtained by setting $K = K_I$,  is given by either of the following equations:
\begin{eqnarray}
x_o(y) &=& (a/\pi)\ln[2\cos(\pi y/a)],\label{xo}\\
y_o(x)&=&\pm (a/\pi)\cos^{-1}[\exp(\pi x/a)/2]. \label{yo}
\end{eqnarray}
In the latter equation,  $x \le (a/\pi)\ln 2 = 0.221$, and the upper (lower) sign holds for positive (negative) values of $y$.  Note that $y$ in Eq.\ (\ref{yo}) rapidly approaches $\pm a/2$ for $x < -a$ as $\exp(\pi x/a)\to 0$. 

The above results can tell us the optimal film design that will prevent any significant reduction of the critical current due to current crowding at a 180-degree turn.  Consider a long superconducting strip of width $W$ and critical sheet current given by Eq.\ (\ref{Kcstrip}).  If we wish the direction of the current to change by 180$^\circ$ at the corner of a strip for which the outer boundaries are straight, as shown in Fig.\ \ref{180fig}(a), the inner boundary of the strip should be chosen to be the smooth curve given by Eq.\ (\ref{xo}) or (\ref{yo}) but with $a/2=W$.  The minimum radius of curvature of this curve is $\rho_c=2W/\pi = 0.637 W$, which occurs at $(x,y)=(x_o(0),0) = (2W\ln2/\pi,0)=(0.441W,0)$.  As discussed in Secs.\ \ref{curve} and \ref{rhocsec}, so long as $\xi \ll \rho_c$, the self-energy $E_{self}$ of a nucleating vortex for small $\delta$ is the same as in Eq.\ (\ref{Gdelta}) to excellent approximation.  Moreover, since the sheet-current density along the entire inner boundary is constant with the value $K_I$, the work term $W_I$  and hence the entire Gibbs free energy for small $\delta$ are very nearly the same as in Eq.\ (\ref{Gdelta}).  Therefore, the critical current for a strip with a 180-degree turnaround of the above-described design, i.e., the area between the dashed curve and the outer boundary in Fig.\ \ref{180fig}, should be the same as that of a very long strip of constant width $W$ [see Eq.\ (\ref{Kcstrip})], so long as $\xi \ll W$.  See also Sec.\ \ref{C180Optimal}, where the optimally rounded 180-degree turnaround is examined from a different perspective.

\subsection{180-degree turnaround intermediate between sharp and optimally rounded\label{180nonoptimal}}

Consider a long strip of width $W$ whose critical sheet current is given by Eq.\ (\ref{Kcstrip}) when $\xi \ll W$.  
In Sec.\ \ref{180sharp}, we showed that current crowding reduces this critical current  by the factor $R= (\pi\xi/W)^{1/2}$ at the sharp 180-degree turn as shown in Fig.\ \ref{180fig} when $a = W$. In Sec.\ \ref{180optimal}, we described an optimal 180-degree turnaround geometry that avoids current crowding, such that there is no reduction of the critical sheet current given by Eq.\ (\ref{Kcstrip}) so long as $\xi \ll W$.
However, the optimal 180-degree turnaround requires that the gap between strips be double the strip width $W$, so that in meander arrays the filling factor (fraction of surface area covered by the superconducting film) is only 1/3.  Device designers may wish to increase the filling factor at the price of reducing the critical current.  In this section we therefore show how to estimate the critical current in 180-degree turnarounds that are intermediate between the sharp and optimally rounded cases discussed in Secs.\ \ref{180sharp} and \ref{180optimal}.

Consider a 180-degree turnaround consisting of a superconducting strip whose shape is chosen such that the outer boundaries are the straight lines at $y = \pm a$ as shown in Fig.\ \ref{180fig}(a) and the inner boundary is the curve defined by the stream-function contour $S(x,y)  = \Im{\cal G}_\zeta(x+iy) = K_I W$.  Far to the left of the turnaround, the film has  a nearly constant width $W$ extending from $y =-a$ to $y = -(a-W)$ for $y < 0$, where it carries a sheet-current density $\bm K = \hat x K_I$, and it has a nearly constant width $W$ extending from $y =+a$ to $y = +(a-W)$ for $y < 0$, where it carries a sheet-current density $\bm K = -\hat x K_I$.  Analysis of the contour $S(x,y) = K_I W$ reveals that it intersects the $x$ axis at $(x,y) = (x_W,0)$, where 
\begin{equation}
x_W = (2a/\pi) \ln[1/\sin(\pi W/2a)],
\end{equation}
the radius of curvature is
\begin{equation}
\rho_c(x_W) = \frac{2a}{\pi} (1-e^{-\pi x_W/a})=\frac{2a}{\pi} \cos^2(\pi W/2a),
\label{rhocxW}
\end{equation}
and the sheet-current density is, from Eq.\ (\ref{calK180}), 
\begin{equation}
\bm K(x_W,0)= \hat y \frac{K_I}{e^{\pi x_W/a}-1} = \hat y K_I \tan(\pi W/2a).
\label{KyxW}
\end{equation}
\begin{figure}
\includegraphics[width=8cm]{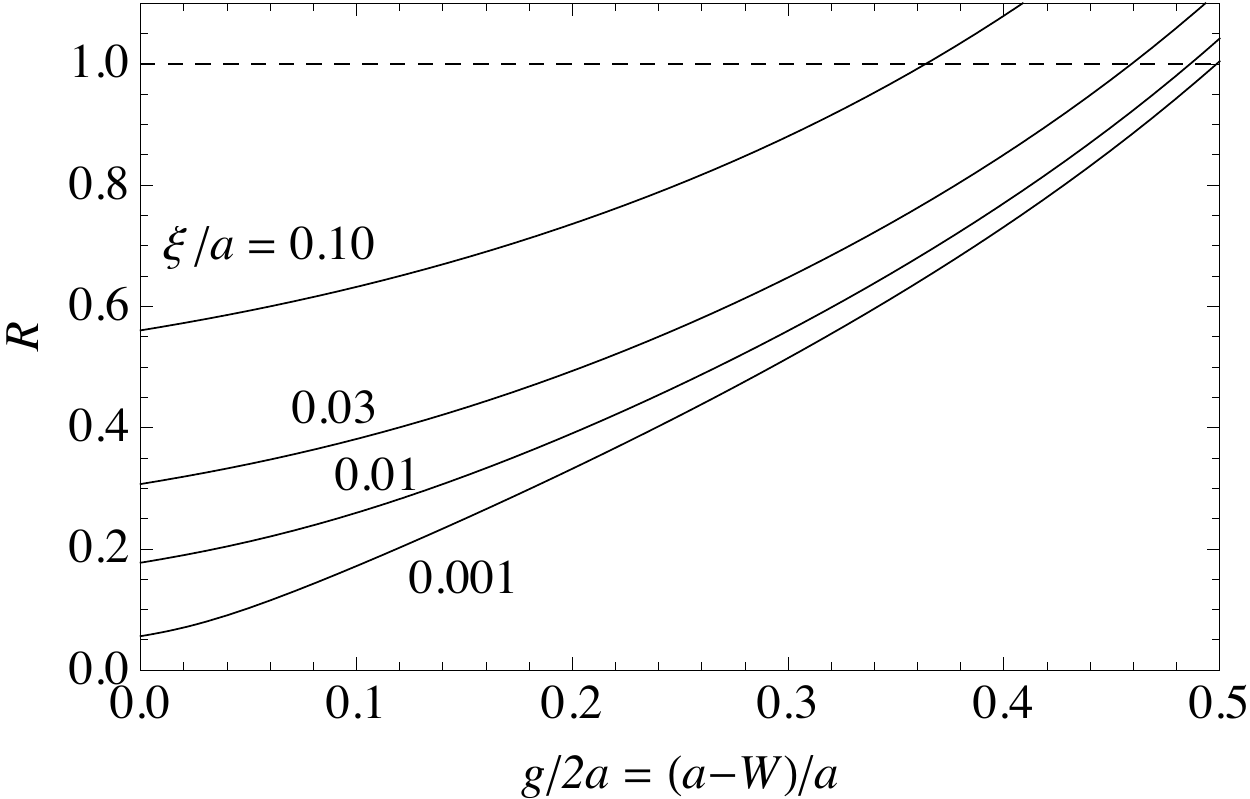}
\caption{Critical-current reduction factor $R$ [Eq.\ (\ref{Rrounded180})] vs $g/2a = (a-W)/a$ for several values of $\xi/a$.}
\label{Rvsgplot}
\end{figure}

We now make use of the results of Sec.\ \ref{rhocsec}, in which the critical sheet-current density $K_{0c}$ at the point of minimum radius of curvature $\rho_c$ is given by Eq.\ (\ref{K0c}), where $k_{0c}$ is the function of $\rho_c/\xi$ plotted in Fig.\ \ref{deltacK0cparaplot}(b).  Since $K_I = K_c$ at the critical current, we have  $K_{0c} = K_c\tan(\pi W/2a)$ from Eq.\ (\ref{KyxW}), such that 
\begin{equation}
K_c = \frac{\phi_0}{e \pi\mu_0\xi\Lambda}R,\;{\rm 
where}\; 
R=k_{0c}\cot(\frac{\pi W}{2a}),
\label{Rrounded180}
\end{equation}
and the argument of $k_{0c}$ [see Fig.\ \ref{deltacK0cparaplot}(b)] is $\rho_c/\xi=(2a/\pi\xi) \cos^2(\pi W/2a)$, obtained  from Eq.\ (\ref{rhocxW}).  Figure  \ref{Rvsgplot} shows calculated values of $R$ as a function of $g/2a$, the ratio of the width of the gap $g=2a-2W$ to the distance $2a$ between the two outer boundaries in Fig.\ \ref{180fig}.   $R$ decreases monotonically as $g$ decreases and reduces to Eq.\ (\ref{R180}) in the limit as $g \to 0$ and $W \to a$. 

As expected, for very small values of $\xi/a$ (see plot for $\xi/a = 0.001$), $R = 1$ when $g/2a = 1/2$ or $W = a/2$. 
For larger values of  $\xi/a$, on the other hand, the behavior of $R$ vs $g/2a$ becomes more  strongly affected by the radius-of-curvature effect (see Sec.\ \ref{rhocsec}), which  counteracts the current-crowding reduction of $K_c$.  The portions of the curves for which $R > 1$ tell us the values of $\xi/a$, $g$, and $W$ for which the critical current is limited by vortex nucleation not at the point of minimum radius of curvature, $(x,y) = (x_W,0)$, but rather somewhere along the straight portion of the strip, far from the bend.

\subsection{Rounded 180-degree turnaround at the end of straight strips\label{CockroftSection}}

We now use a method inspired by that of Cockroft\cite{Cockroft28} to calculate the current-crowding critical-current reduction factor $R$ for the case that straight strips with constant width $W$ and gap $2r'$ between them  are connected at their ends with a rounded corner, as shown in Fig.\ \ref{C180fig}. The results depend in detail upon the geometry chosen.  In Sec.\  \ref{C180General} we discuss the behavior for which the dimensions shown in Fig.\ \ref{C180fig}(a) obey $W' = W$, in Sec.\ \ref{beqa} we examine the behavior for $W' \to \infty$ and $r'/W < 1$, and in Sec.\ \ref{C180Optimal} we treat the limiting case for   $W' \to \infty$ and $r'/W = 1$, which produces a curving inner boundary of the turnaround corresponding to optimal rounding discussed in Sec.\ \ref{180optimal}.
\subsubsection{General case\label{C180General}}
\begin{figure}
\includegraphics[width = 7cm]{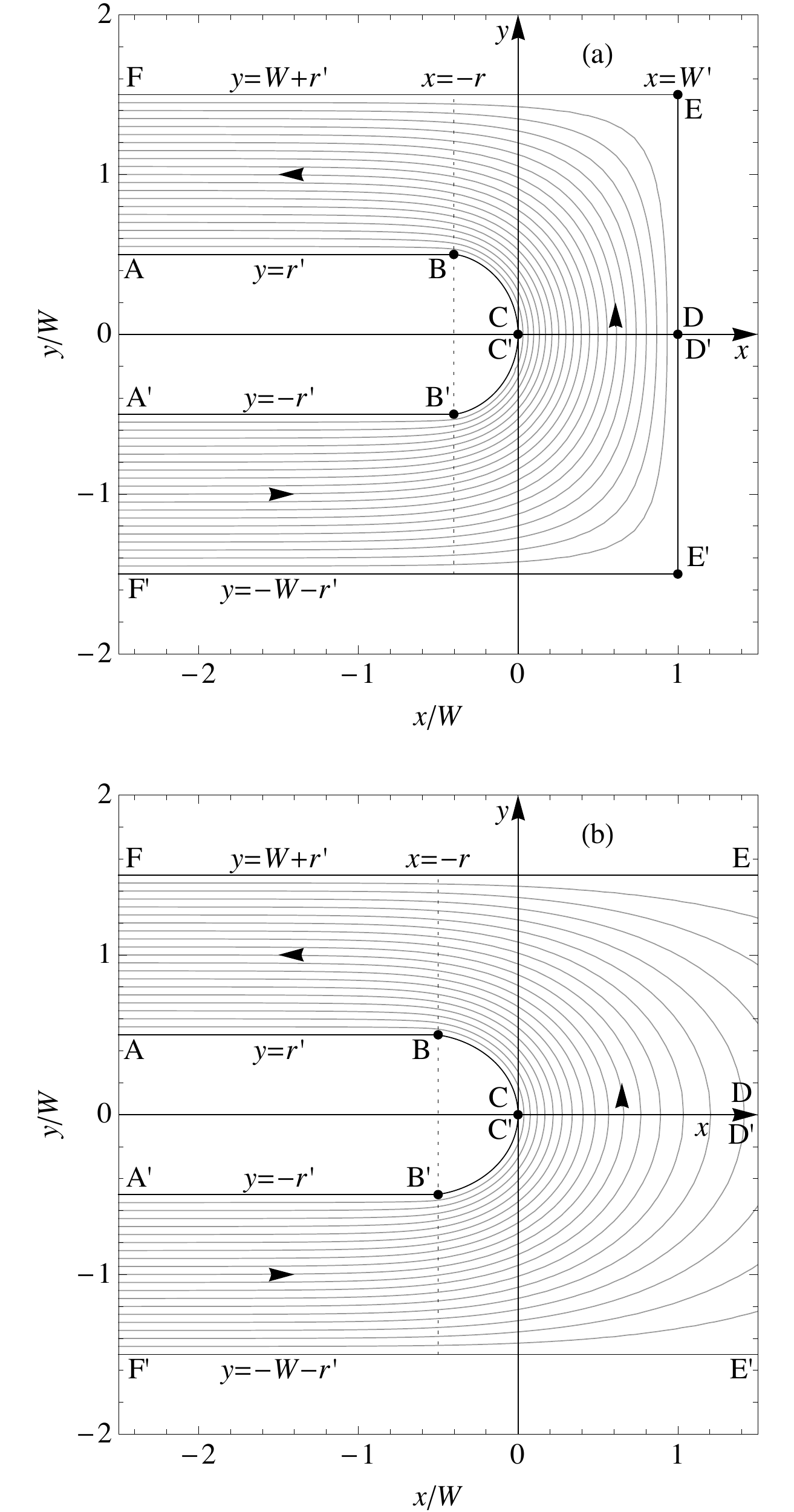}
\caption{(a) Current flow in a film carrying current $K_I W$ from a straight strip of width $W$ around a rounded 180-degree turn of length CD = $W'=W$ into another straight strip of width $W$, shown by the contour plot of the stream function $S(x,y)= \Im{\cal G}_\zeta(x+iy)$, which has the values $S=0$ along the outer boundaries of the film and $S=K_I W$ along the inner boundary.  Here $r'= 0.500 W$ and $r = 0.402W$. The contours correspond to streamlines of the sheet-current density $\bm K$, and the arrows show the direction of the current.   (b) Same as (a), except that the connection is infinite in length (CD = $\infty$), $r'=0.500W$, and $r = 0.525W$.  }
\label{C180fig}
\end{figure}

The conformal mapping\cite{Cockroft28}
\begin{eqnarray}
\zeta'(w)&=&\frac{d\zeta(w)}{dw}= -A\frac{\sqrt{w+1}+\sqrt{\frac{b+1}{b-1}}\sqrt{w-1}}{\sqrt{w-1}\sqrt{w-a}\sqrt{w-b}},
\label{zetaprimeC180}\\
\zeta(w)&=&  -A\Big\{\frac{2i}{\sqrt{(a+1)(b-1)}}[(a-1)\Pi(\phi_1,n_1,k_1) \nonumber \\
&&\;\;\;\;\;\;\;\;\;\;\;\;\;\;\;\;\;\;\;\;\;\;\;\;\;\;\;\;\;\;\;\;\;-(a+1)F(\phi_1,k_1)] \nonumber 
\label{zetaC180} \\
&+&2\sqrt{\frac{b+1}{b-1}}\ln\Big(\frac{\sqrt{a-w}+\sqrt{b-w}}{\sqrt{a-1}+\sqrt{b-1}}\Big)\Big\},\\
A&=&W/g(b),\label{AWgb}\\
g(b)&=&\pi\Big(1+\sqrt{\frac{b+1}{b-1}}\Big),\label{gb}\\
\phi_1&=&\sin^{-1}\sqrt{\frac{(a+1)(1-w)}{2(a-w)}},\\
n_1&=&\frac{2}{a+1},\;n_2 = \frac{a-b}{a+1},\\
k_1&=&\sqrt{\frac{2(b-a)}{(a+1)(b-1)}},
\end{eqnarray}
maps points in the upper half $w$-plane ($w = u + i v$)  into the area ABCDEF in the upper half $\zeta$-plane  ($\zeta = x + i y$) shown in Fig.\ \ref{C180fig}.  The inverse mapping $w(\zeta)$ can be obtained numerically. Here, $F(\phi,n,k)$ and $\Pi(\phi,n,k)$ are elliptic integrals of the first and third kind with parameter $n$ and modulus $k$,\cite{Gradshteyn00}
and $K(k) = F(\pi/2,k)$ and $\Pi(n,k)=\Pi(\pi/2,n,k)$ are the corresponding complete elliptic integrals.  
Special points in the $\zeta$ and $w$ planes (see Fig.\ \ref{C180fig}) are related as follows: A, $w = -\infty$; B, $w = -1$; C, $w = +1$; D, $w=a$; E, $w = b$; and F, $w = +\infty$, where $1 < a < b$.  The lengths CD = $W'$, $r$, and $r'$ shown in Fig.\ \ref{C180fig}(a) are determined as functions of $a$ and $b$ as follows:
\begin{eqnarray}
W' &=& A\Big\{\frac{2}{\sqrt{(a+1)(b-1)}}[(b+1)K(k_2)\label{Wpab}\nonumber \\
&&\;\;\;\;\;\;\;\;\;\;\;\;\;\;\;\;\;\;\;\;\;\;\;\;\;-(b-a)\Pi(n_3,k_2)] \nonumber \\
&&+2\sqrt{\frac{b+1}{b-1}}\ln\Big(\frac{\sqrt{a-1}+\sqrt{b-1}}{\sqrt{b-a}}\Big)\Big\},\\
n_3&=&\frac{a-1}{b-1}, \;k_2 = \sqrt{\frac{(a-1)(b+1)}{(a+1)(b-1)}},
\end{eqnarray}
\begin{eqnarray}
r &=& 2A\sqrt{\frac{b+1}{b-1}}\ln\Big(\frac{\sqrt{a+1}+\sqrt{b+1}}{\sqrt{a-1}+\sqrt{b-1}}\Big),\label{rab}\\
r' &=& \frac{2A}{\sqrt{(a+1)(b-1)}}[(1+a)K(k_1) \nonumber \\
&&\;\;\;\;\;\;\;\;\;\;\;\;\;\;\;\;\;\;\;\;\;\;\;\;+(1-a)\Pi(n_1,k_1)].\label{rpab}
\end{eqnarray}
Note that DE = $W +r'$.

The complex potential describing the current flow within the area ABCDEF shown in Fig.\ \ref{C180fig} is ${\cal G}_\zeta(\zeta)={\cal G}_w(w)$, where 
\begin{equation}
{\cal G}_w(w)=\frac{I}{\pi}\ln(2w-1-a+2\sqrt{w-1}\sqrt{w-a})
\label{calGzetaC180}
\end{equation}
and $w = w(\zeta)$.
The imaginary part of this is the stream function $S$, whose contours, shown in  Fig.\ \ref{C180fig}, are the streamlines of the sheet-current density $\bm K = \hat x K_x + \hat y K_y$.  The  corresponding complex sheet current, ${\cal K}_\zeta(\zeta)=d{\cal G}_\zeta(\zeta)/d\zeta = K_x(x,y)-iK_y(x,y)$, is
\begin{equation}
{\cal K}_\zeta(\zeta) = -\frac{IA}{\pi}\frac{\sqrt{w-b}}{\sqrt{w+1}+\sqrt{\frac{b+1}{b-1}}\sqrt{w-1}}.
\end{equation}
For $r' < y < W+r'$ and $x \ll -W$, which corresponds to $w \to \infty,$ we have to good approximation $K_x = -K_I = -IA/g(b)=-I/W$.  However, around the arc BC, which corresponds to $w = u$, where $-1 \le u \le 1$, we find that $K_{\rm BC} = |{\cal K}|$ is given by 
\begin{equation}
\frac{K_{\rm BC}}{K_I}=\frac{\sqrt{b+1}+\sqrt{b-1}}{\sqrt{2}}.
\end{equation}
Since $b$ is required to obey $b > 1$ except in limiting cases, this equation tells us that current-crowding  ($K_{\rm BC} > K_I$) almost always occurs.  As discussed in Secs.\ \ref{curve} and  \ref{rhocsec}, so long as $\xi \ll r'$, this current-crowding therefore leads to a critical sheet-current density of the form 
\begin{equation}
K_c = \frac{\phi_0}{e \pi\mu_0\xi\Lambda}R,\;{\rm where }\;
R=\frac{\sqrt{2}}{\sqrt{b+1}+\sqrt{b-1}}
\label{RBC}
\end{equation}
is the reduction factor due to current crowding at the curving inner boundary BC, relative to the $K_c$ for a long straight strip [Eq.\ (\ref{Kcstrip})].  However, to treat the case for which $r' \le \xi$, an approach similar to that in Sec.\ \ref{rhocsec} would be required.

Using Eqs.\ (\ref{AWgb}), (\ref{Wpab}), and (\ref{rpab}), one can obtain two equations from which the values of $b$ and $a$ can be obtained for desired values of the ratios $W'/W$ and $r'/W$. 
Figure \ref{C180fig}(a) was calculated for the following parameters:  $W = W' = 1$, $r' = 0.500$, $r = 0.402$, $A = 0.102$, $a = 1.216$, and $b = 1.572$.

Combining displaced mirror images of turnarounds like these in a two-dimensional layout (a ``boustrophedonic'' pattern) results in a fill factor $f= W/p$, where $p = W +2r'$ is the pattern period (pitch) and $W$ and $2r'$ are the strip and gap widths away from the turnarounds.  For example, Fig.\ \ref{C180fig} corresponds to $f = 1/2.$  The solid curves of Fig.\ \ref{C180Retcfig} show plots of $R$, $r'/W = (1-f)/2f$, $r'/r$, and $b-1$ vs $f= 1/(1+2r'/W)$ for $W'/W = 1$.  For $f \ll 1$, the following expansions have been used to plot the functions for $f \le 0.12$:
\begin{eqnarray}
\delta b&=&b-1=\frac{128}{\pi^2}\exp[-\frac{\pi(1-f)}{2f}],\\
R&=&1/(\sqrt{1+\delta b/2} + \sqrt{\delta b/2}),\\
r'/r&=&\ln \Big(\frac{128}{\pi^2 \delta b}\Big)/\ln\Big(\frac{8}{\delta b}\Big)\nonumber \\&=&\frac{1-f}{1-f[1+(4/\pi)\ln(4/\pi)]}.
\end{eqnarray}
While Fig.\ \ref{C180Retcfig} shows $R$ approaching zero for $f \to 1$ and  $r'/W \to 0,$ bear in mind that Eq.\ (\ref{RBC}) is valid only for $r' \gg \xi$.

\subsubsection{$W' \to \infty$ and $r'/W < 1$\label{beqa}}

\begin{figure}
\includegraphics[width = 6cm]{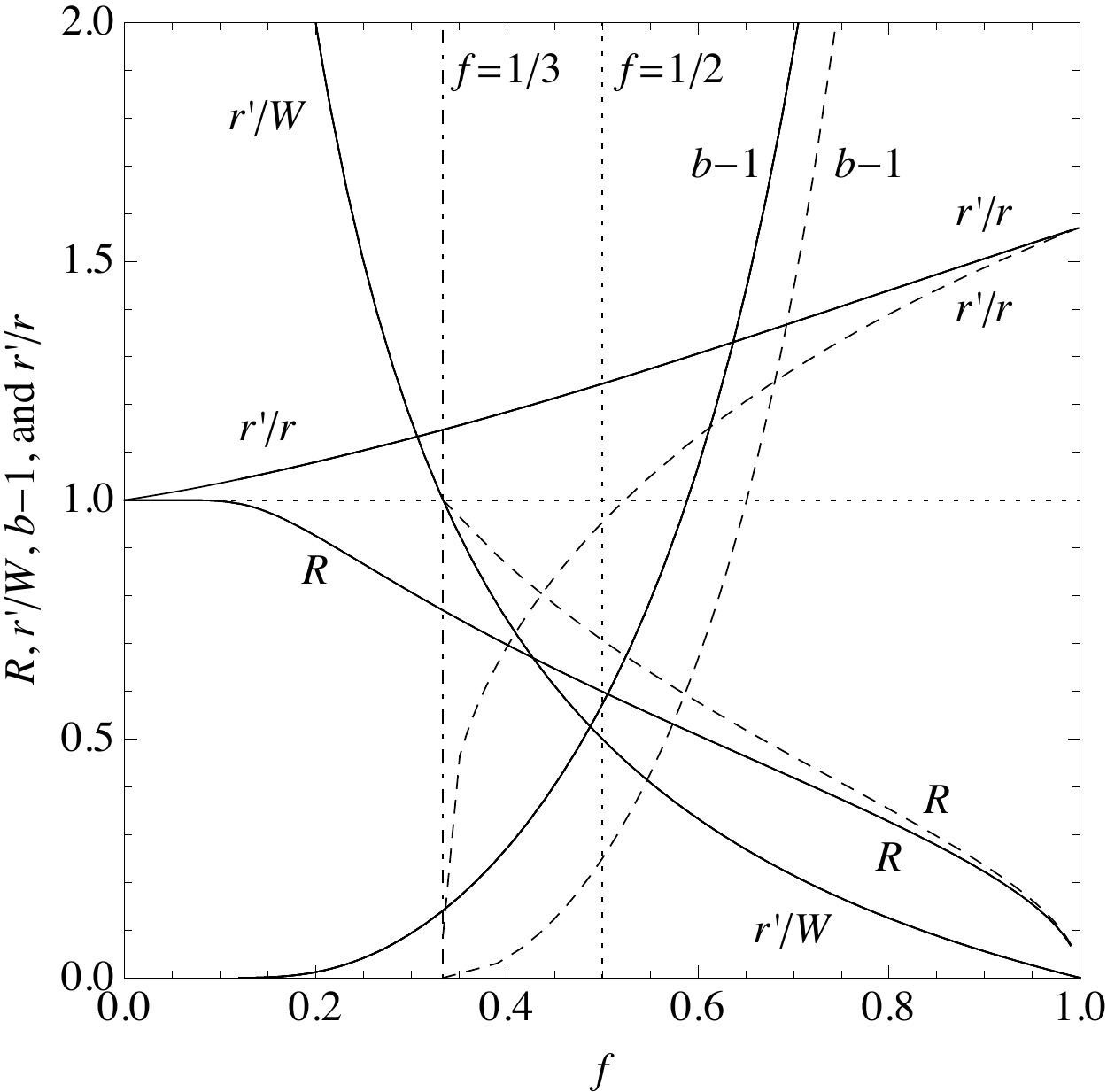}
\caption{Plots of the current-crowding critical-current reduction factor $R$, $\delta b = b-1$, and the ratios $r'/W$ and $r'/r$ vs the fill factor $f = 1/(1+2r'/W)$ for  $W' = W$ (solid curves) and $W' = \infty$ (dashed curves).  The vertical dotted line marks the parameters for Fig.\ \ref{C180fig}(a) ($f = 0.5,$ $r'/W = 0.5$, $R = 0.599$, $b-1 = 0.572$, and $r'/r = $1.243) and Fig.\ \ref{C180fig}(b) ($f = 0.5,$ $r'/W = 0.5$, $R = 0.707$, $b-1 = 0.250$, and $r'/r = $0.953). The vertical dot-dashed line marks the smallest $f$ for which solutions can be found for $W' = \infty$ and the value of $f = 1/3$ for which $r'/W = 1$,  $r/W \to \infty$, and the inner boundary takes the optimally rounded shape for a 180-degree turnaround so that $R = 1$.  }
\label{C180Retcfig}
\end{figure}

In the limit CD = $W' \to \infty$, $b \to a$, and Eqs.\ (\ref{zetaprimeC180})-(\ref{rpab}) simplify to
\begin{eqnarray}
\zeta'(w)&=&\frac{d\zeta(w)}{dw}= -A\frac{\sqrt{w+1}+\sqrt{\frac{a+1}{a-1}}\sqrt{w-1}}{\sqrt{w-1}(w-a)},
\label{zetaprimeC180infWp}\\
\zeta(w)&=&  -A\Big\{2i[\tan^{-1}\sqrt{\frac{1-w}{1+w}} -\sqrt{\frac{a+1}{a-1}}\phi_1]\nonumber \\
&+&2\sqrt{\frac{a+1}{a-1}}\ln\Big(\frac{\sqrt{a-w}}{\sqrt{a-1}}\Big)\Big\},
\label{zetaC180infWp}\\
A&=&W/g(a),\label{AWga}\\
g(a)&=&\pi\Big(1+\sqrt{\frac{a+1}{a-1}}\Big),\label{ga}\\
r &=& 2A\sqrt{\frac{a+1}{a-1}}\ln\Big(\frac{\sqrt{a+1}}{\sqrt{a-1}}\Big),\label{ra}\\
r' &=& A\pi\Big(\sqrt{\frac{a+1}{a-1}}-1\Big)\label{rpa}.
\end{eqnarray}
For $W' \to \infty$, Eqs.\ (\ref{AWga}) and (\ref{rpa}) can be used to obtain an equation from which $b=a$ can be determined as a function of $r'/W$, but only for $r'/W \le 1$.
Equations (\ref{calGzetaC180})-(\ref{RBC}) still apply, and 
Fig.\ \ref{C180fig} (b) was calculated for the following parameters:  $W = 1$, $W' \to \infty$, $r' = 0.500$, $r = 0.525$, $A = 1/4\pi,$ and $b = a = 1.25$.  The dashed curves of Fig.\ \ref{C180Retcfig} show plots of $R$, $r'/W = (1-f)/2f$, $r'/r$, and $b-1=a-1$ vs $f= 1/(1+2r'/W)$ for $W'/W = 1$.
While the dashed curve for $R$ in Fig.\ \ref{C180Retcfig} shows $R$ approaching zero for $f \to 1$ and  $r'/W \to 0,$ recall that Eq.\ (\ref{RBC}) is valid only for $r' \gg \xi$.  

\subsubsection{Optimal rounding when $W' \to \infty$ and $r/W \to \infty$\label{C180Optimal}}

\begin{figure}
\includegraphics[width = 6cm]{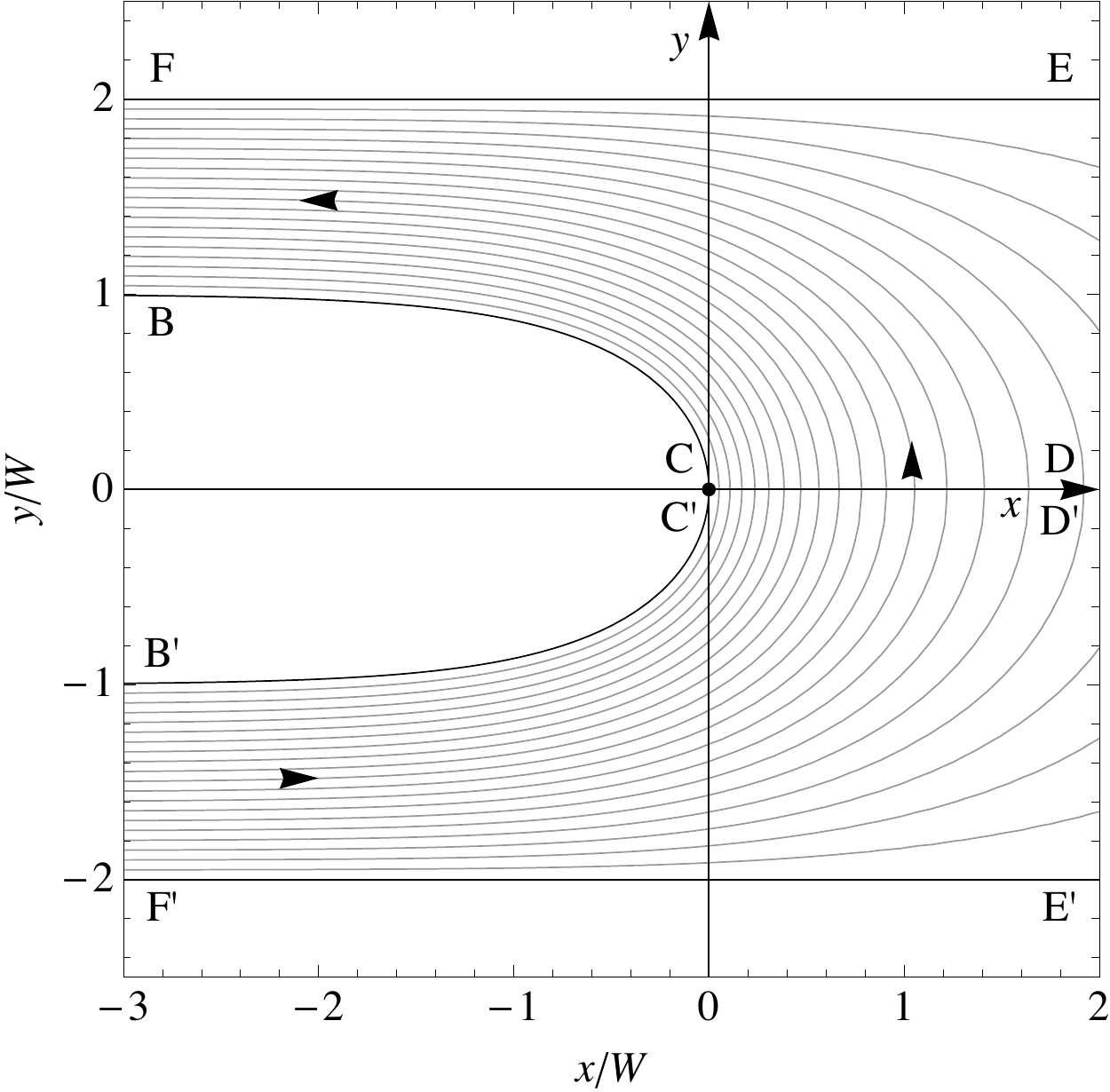}
\caption{Current flow around a 180-degree turnaround with an optimally rounded inner boundary, shown by the contour plot of the stream function $S(x,y)= \Im{\cal G}_\zeta(x+iy)$, which has the values $S=0$ along the outer boundaries of the film and $S=K_I W$ along the inner boundary.  The contours correspond to streamlines of the sheet-current density $\bm K$, and the arrows show the direction of the current.  Since there is no current crowding along the curve BCC$'$B$'$, the critical current is predicted to be the same as for a long, straight strip. }
\label{C180optfig}
\end{figure}

Note from the dashed curves in Fig.\ \ref{C180Retcfig} that in the limit $W' \to \infty$, we find $R = 1$, $r'/W =1$, $r/W \to \infty$, and $f = 1/3$ in the limit  $a = b \to 1$.  The dependence of $\zeta(w)$ then becomes dominated by values of $w$ very close to 1 and $a$, and it is appropriate to introduce the variable $\omega = (w-1)/(a-1)$ in Eqs.\ (\ref{zetaprimeC180infWp})-(\ref{rpa}) and to take the limit as $a \to 1$.  This leads to the conformal mapping 
\begin{eqnarray}
\zeta'(\omega)&=&\frac{d\zeta(\omega)}{d\omega}= -\frac{W(1+\sqrt{\omega})}{\sqrt{\omega}(\omega-1)},
\label{zetaprimeofomega180}\\
\zeta(\omega)&=&  -\frac{2W}{\pi}\ln(1-\sqrt{\omega}),
\label{zetaofomega180}\\
w(\zeta)&=&[1-\exp(-\pi\zeta/2W)]^2,
\end{eqnarray}
which maps points in the upper half $\omega$-plane  into the area BCDEF in the upper half $\zeta$-plane  ($\zeta = x + i y$) shown in Fig.\ \ref{C180optfig}. 
Special points in the $\zeta$ and $\omega$ planes (see Fig.\ \ref{C180optfig}) are related as follows:  B, $\omega = -\infty$; C, $\omega = 0$; D, $\omega = 1-\epsilon$; E, $\omega = 1+\epsilon$; and F, $w = +\infty$, where $\epsilon$ is a positive infinitessimal. The curves BC and B$'$C$'$ for $x < 0$ are given by 
\begin{equation}
y_{opt}(x) = \pm (2W/\pi)\cos^{-1}[\exp(\pi x/2W)]
\label{yopt}
\end{equation}
with the upper (lower) sign holding for BC (B$'$C$'$).  These curves correspond to the optimally rounded inner boundary of a 180-degree turnaround discussed in Sec.\ \ref{180optimal}.  Note that $y_{opt}(x)$ in Eq.\  (\ref{yopt}) is the same as $y_o(x)$ in Eq.\ (\ref{yo}) with 
$a =2W$ and the origin shifted along the $x$ axis by $\Delta x = (a/\pi)\ln2$.

The complex potential describing the current flow within the area BCDEF shown in Fig.\ \ref{C180optfig} is 
\begin{equation}
{\cal G}_\zeta(\zeta)=\frac{2I}{\pi}\ln[\sqrt{\omega(\zeta)}+\sqrt{\omega(\zeta)-1})].
\label{calGzetaC180opt}
\end{equation}
The imaginary part of this is the stream function $S$, whose contours, shown in  Fig.\ \ref{C180optfig}, are the streamlines of the sheet-current density $\bm K = \hat x K_x + \hat y K_y$.  The  corresponding complex sheet current, ${\cal K}_\zeta(\zeta)=d{\cal G}_\zeta(\zeta)/d\zeta = K_x(x,y)-iK_y(x,y)$, is
\begin{equation}
{\cal K}_\zeta(\zeta) = -\frac{I}{W}\frac{\sqrt{\omega(\zeta)-1}}{1+\sqrt{\omega(\zeta)}}.
\end{equation}
For $y_{opt} < y < 2W$ and $x \ll -W$, which corresponds to $\omega(\zeta) \to \infty,$ we obtain $K_x = -K_I$, where $K_I = I/W$.  Along the arc BC, which corresponds to $\omega = u$, where $-\infty \le u \le 0$, we find that $K_{\rm BC} = |{\cal K}| = K_I$.  In other words, there is no current crowding, and the critical current is predicted to be exactly the same as for a long straight strip [Eq.\ (\ref{Kcstrip})].
As discussed above, however, this conclusion is based on the assumption that the coherence length $\xi$ is much smaller than the radius of curvature at the origin $\rho_c = 2W/\pi$. To treat the case for which $\xi \ge \rho_c$, an approach similar to that in Sec.\ \ref{rhocsec} would be required.
  
\subsection{Sharp rectangular 180-degree turnaround\label{rect180section}}

\begin{figure}
\includegraphics[width = 6cm]{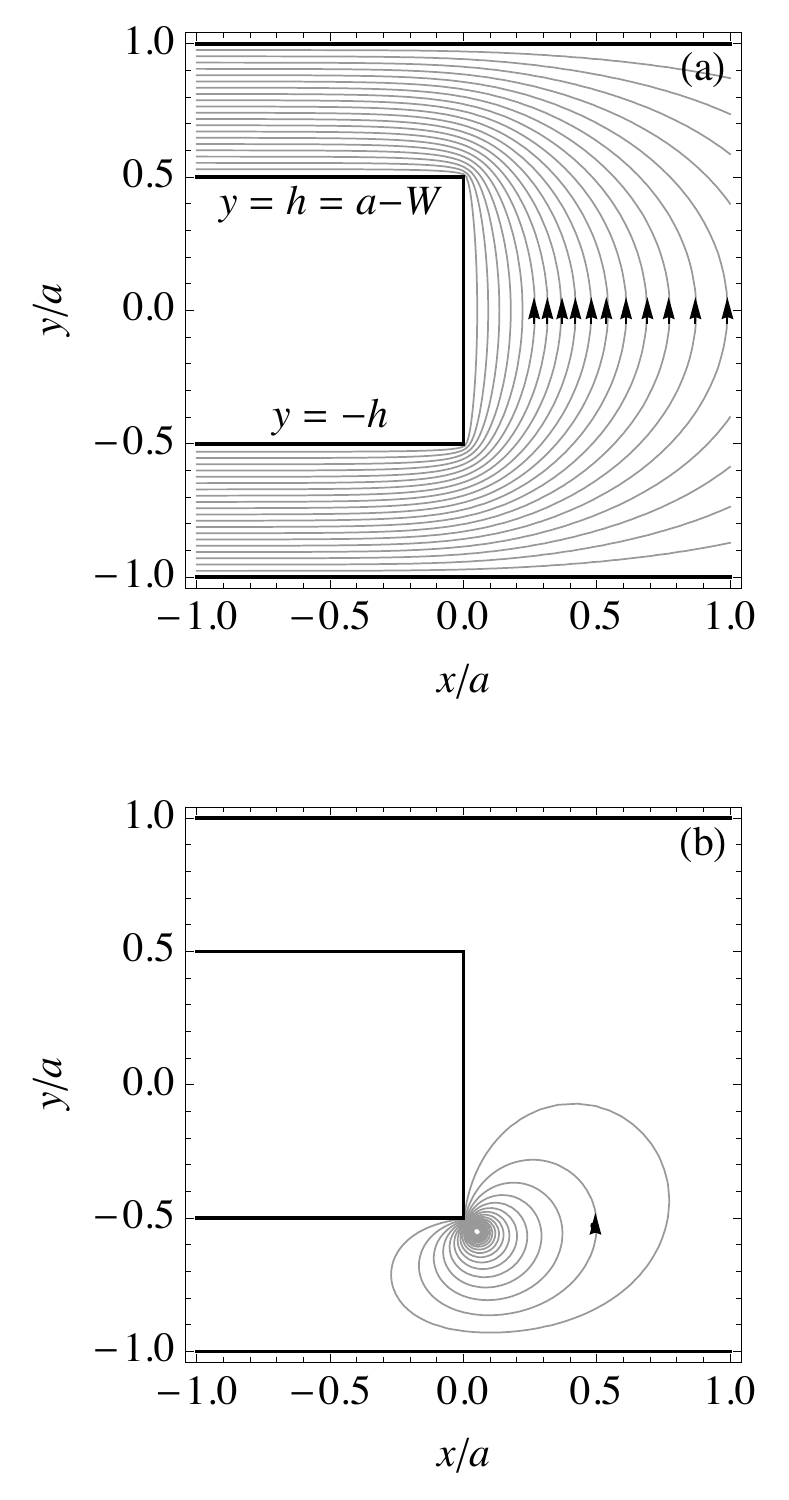}
\caption{(a) Current flow in a film carrying current $K_I W$ around a sharp rectangular 180-degree turn, shown by the contour plot of the stream function $S(x,y)= \Im{\cal G}_\zeta(x+iy)$, which has the values $S=0$ along the lines $y = \pm a$ at the outer boundary of the film and $S=K_I W$ along the inner boundary [$|y| = h = a-W$ for  $x < 0$ or $|y| \le h$ for $x = 0$].  (In this figure, $h = a/2 = W$, and $\alpha = \sqrt{3}/2$ = 0.866.) The contours correspond to streamlines of the sheet-current density $\bm K$, and the arrows show the direction of the current.   (b) Vortex-generated current flow, shown by the contour plot of the stream function $S_v(x_v,y_v;x,y)$, which has the values $S_v=0$ for $(x,y)$  along the boundaries. The contours, shown here for $(x_v,y_v) = (0.05a,-0.55a)$, correspond to streamlines of the vortex-generated sheet-current density $\bm K_v$, and the arrow shows the direction of the current.  }
\label{rect180fig}
\end{figure}

We next examine the current flow around a sharp rectangular turnaround shown in Fig.\ \ref{rect180fig}.  The film (width $2a$) is centered on the $x$ axis, but a slot (width $2h$) with sharp 90-degree corners has been cut out of the center for $x < 0$. The lower arm of width $W = a-h$ carries a uniform sheet-current density $\bm K = \hat x K_I$  until it reaches the turnaround. The current then turns around and finally flows in the upper arm with current density  $\bm K = -\hat x K_I$.  We wish to calculate the critical current at which the first vortex is nucleated at one of the sharp inner corners of the turnaround. 

The conformal mapping\cite{Kober57p160}
\begin{eqnarray}
\zeta'(w)&=&\frac{d\zeta(w)}{dw}= \frac{2a}{\pi}\frac{\sqrt{w^2-\alpha^2}}{w^2-1},
\label{zetaprimerect180}\\
\zeta(w)&=& \Big(\frac{2a}{\pi}\Big)\Big[\cosh^{-1}\Big(\frac{w}{\alpha}\Big)-i\pi/2 \nonumber \\
&-&\frac{\sqrt{1-\alpha^2}}{2}\cosh^{-1}\Big(\frac{w^2(2-\alpha^2)-\alpha^2}{\alpha^2(w^2-1)}\Big)\Big],
\label{zetarect180}
\end{eqnarray}
where $\alpha = \sqrt{2(h/a)-(h/a)^2}$,
maps points in the upper half $w$-plane ($w = u + i v$)  into the area $-a \le y \le a$ for $x > 0$ or $h < |y| < a$ for $x < 0$ in the $\zeta$-plane  ($\zeta = x + i y$) as shown in Fig.\ \ref{rect180fig}.  The inverse mapping $w(\zeta)$ can be obtained numerically.  

The complex potential ${\cal G}_\zeta(\zeta)$ is given by Eq.\ (\ref{calGzeta}) using the inverse mapping $w(\zeta)$ with $I=K_I W$, 
and the corresponding complex sheet current ${\cal K}_\zeta(\zeta)=d{\cal G}_\zeta(\zeta)/d\zeta = K_x(x,y)-iK_y(x,y)$ is
\begin{equation}
{\cal K}_\zeta(\zeta)=\frac{K_I W}{a \sqrt{w^2(\zeta)-\alpha^2}}.
\label{calKrect180}
\end{equation}
For  $-a < y < -h$, $K_x(x,y) \to K_I$ as $x \to -\infty$, and  for  $h < y < a$, $K_x(x,y) \to -K_I$ as $x \to -\infty$.  The streamlines of the  sheet current $\bm K = \hat K_x + \hat y K_y$  are obtained as contours of the stream function $S(x,y) = \Im{\cal G}_\zeta(x+iy)$, the imaginary part of ${\cal G}_\zeta(\zeta)$, shown in Fig.\ \ref{rect180fig}(a).  

The critical current of the rectangular 180$^\circ$ turnaround is reached when a vortex can be nucleated at one of the sharp inner corners shown in Fig.\ \ref{rect180fig}(a) or (b).
The lower right corner at $\zeta = -ih$ corresponds to $w = \alpha$.   
Expanding Eq.\ (\ref{zetaprime180}) for $\bm r_v=(\delta/\sqrt{2},h-\delta/\sqrt{2})$ or $\zeta_v =-ih+ \delta/\sqrt{2}-i\delta/\sqrt{2}$ along the diagonal extending away from the lower right inner corner of Fig.\ \ref{rect180fig}(a), we obtain
\begin{equation}
w(\zeta_v) = \alpha + \frac{i}{(2\alpha)^{1/3}}\Big(\frac{W}{a}\Big)^{4/3}\Big(\frac{3\pi\delta}{4a}\Big)^{2/3}
\end{equation}
to lowest order, where $\delta \ll a$. 
From Eqs.\ (\ref{calGzeta}), (\ref{Icirc}), and (\ref{DeltaI}) we obtain 
the Gibbs free energy,
\begin{equation}
G = \frac{\phi_0^2}{2\pi \mu_0 \Lambda} \ln \Big(\frac{3\delta}{\xi}\Big)-\phi_0 K_I\Big(\frac{W}{\pi \alpha}\Big)^{1/3}\Big(\frac{3\delta}{2}\Big)^{2/3}.
\label{Gdeltarect180}
\end{equation}
Following the steps that led to Eq.\ (\ref{Kcstrip}), we obtain  with $K_I = K_c$,
\begin{eqnarray}
\Big(\frac{3\delta_b}{2}\Big)^{2/3} \!\!\!\!\!&\!\!\!\!=\!\!& \!\!\!\!\frac{3\phi_0 \alpha^{1/3}}{4\pi^{2/3} \mu_0 \Lambda K_I W^{1/3}},\;
\delta_c\!=\!\frac{e^{3/2}\xi}{3}\!=\!1.49\xi,\\
K_c \!\!\!&= &\!\!\frac{\phi_0}{e \pi\mu_0\xi\Lambda}R,\;{\rm where }\;
R=\frac{3}{2}\Big(\frac{\pi\alpha \xi}{2W}\Big)^{1/3}
\label{Rrect180}
\end{eqnarray}
is the reduction factor due to current crowding at one of the sharp inner corners, and (because $h = a-W$) $\alpha = \sqrt{1-(W/a)^2}$.  
The barrier height for $K_I < K_c$ is
\begin{equation}
G_b =  \frac{3\phi_0^2}{4\pi\mu_0\Lambda}\ln\Big(\frac{K_c}{K_I}\Big).
\label{Gbrect180}
\end{equation}
Note that the prefactor is larger  than that in Eq.\ (\ref{Gbstrip}) by a factor of 3/2, which arises from the term proportional to $\delta^{2/3}$ in  Eq.\ (\ref{Gdeltarect180}).

In the limit when $W \ll a$,  $\alpha \to 1$, in which case Eq.\ (\ref{Rrect180}) yields $R \approx (3/2)(\pi\xi/2W)^{1/3}$, a result nearly the same as that in Eq.\ (\ref{Rrtangle}) with $a = W$ but larger by a factor of $2^{1/3} = 1.26$ because of the geometry differences.

The steps leading to Eq.\ (\ref{Rrect180}) should be valid so long as $h =(a-W)\ll \xi$, but the above approximations fail for very small gap widths $g = 2h \sim \xi$.  In this case, $\alpha \sim (\xi/W)^{1/2}$, and Eq.\ (\ref{Rrect180}) yields $R \sim (\xi/W)^{1/2}$.  In the limit as $h \to 0$, the value of $R$ is given by Eq.\ (\ref{R180}).

\section{90-degree turns\label{rightangle}}

In this section we first calculate how much the critical current is reduced by a sharp 90-degree turnaround, and we then describe the shape of an optimally rounded 90-degree turn that should exhibit no critical-current reduction.

\subsection{Sharp 90$^\circ$ turn\label{sharp90}}

\begin{figure}
\includegraphics[width = 6cm]{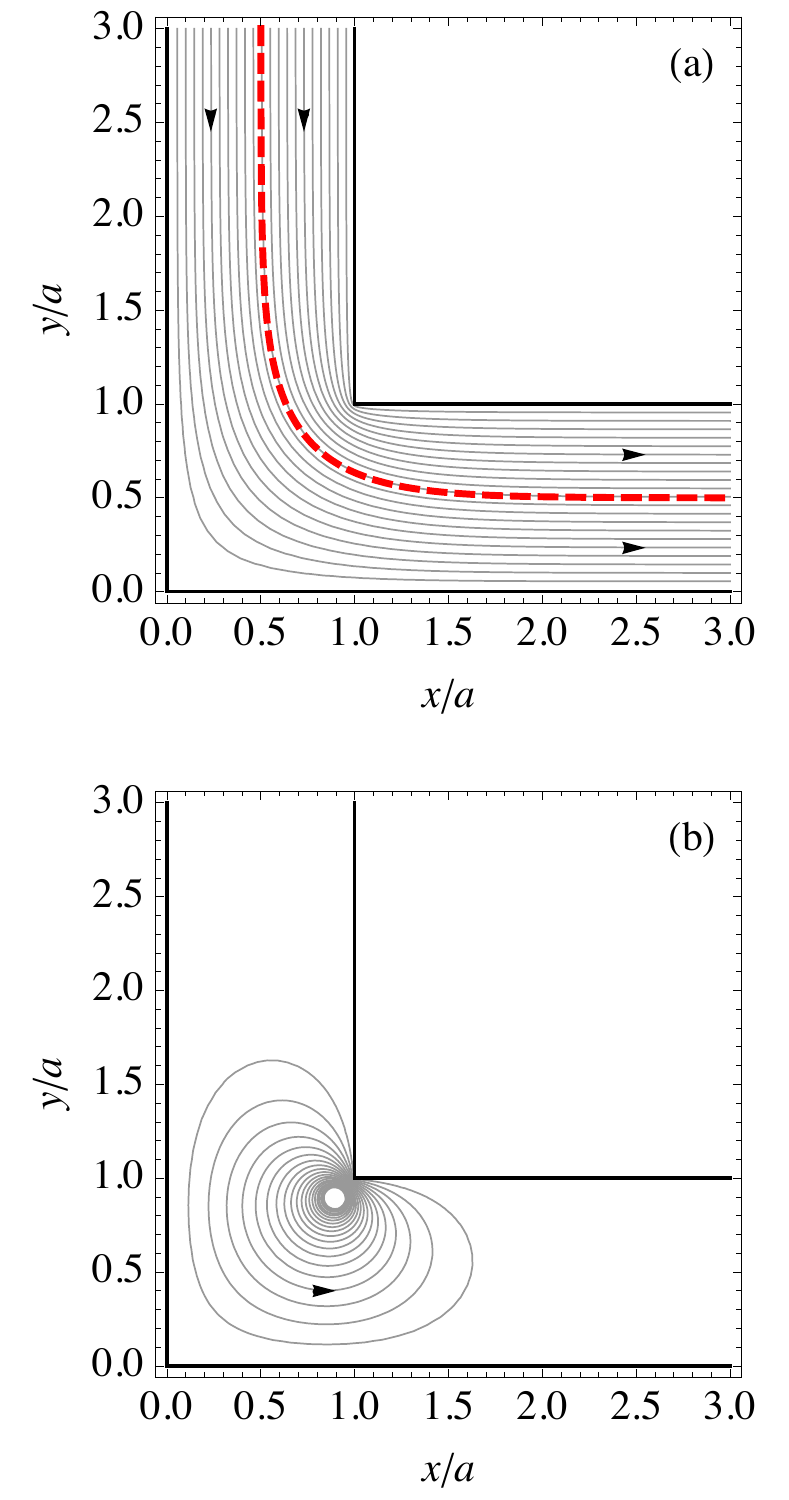}
\caption{(a) Current flow in a strip carrying current $K_I a$ around a 90$^\circ$ turn, shown by the contour plot of the stream function $S(x,y) = \Im {\cal G}(x+iy)$ [Eq.\ (\ref{Grtangle})], which has the values $S=0$ along the inner boundary ($y = a$ for $x \ge a$ and $x = a$ for $y \ge a$) and $S=-K_I a$ along the outer boundary ($y = 0$ for $x \ge 0$ and $x = 0$ for $y \ge 0$).   The contours correspond to streamlines of the sheet-current density $\bm K$, and the arrows show the direction of the current.  The dashed curve, which corresponds to $S = -K_I a/2$, separates the current-crowding region near the inner corner from the current-expanding region near the outer corner.  The magnitude of $\bm K$ is constant ($\bm K=K_I$) along the dashed curve. (b) Vortex-generated current flow, shown by the contour plot of the stream function $S_v(x_v,y_v;x,y)$, shown here for $(x_v,y_v) = (0.9 a,0.9 a)$, which has the values $S_v=0$ for $(x,y)$  along the inner and outer boundaries. The contours correspond to streamlines of the vortex-generated sheet-current density $\bm K_v$.  }
\label{rtangleplot}
\end{figure} 

Consider the current flow in a strip of width $a$ with a right-angle turn as  shown in Fig.\ \ref{rtangleplot}.  The complex potential that describes the flow of current is  
\begin{equation}
{\cal G}_\zeta(\zeta)=-\frac{K_I a}{\pi} \ln[w(\zeta)-1],
\label{Grtangle}
\end{equation}
where the conformal mapping from the $w$ plane ($w  = u +i v$) to the $\zeta$ plane ($\zeta = x + i y$) is\cite{Kober57p157} 
\begin{eqnarray}
\zeta'(w)&=&\frac{d\zeta(w)}{dw}=\frac{a\sqrt{2-w}}{\pi(1-w)\sqrt{w}},\label{zetaprime}\\
\zeta(w)&=&\frac{a}{\pi}\Big[\cos^{-1}(1-w)+\cosh^{-1}\Big(\frac{1}{1-w}\Big)\Big],
\end{eqnarray}
and $w(\zeta)$ is the inverse of $\zeta(w)$, which is readily obtained numerically.
The imaginary part of ${\cal G}$ is the stream function, $S(x,y) = \Im {\cal G}(x+iy).$ Shown in Fig.\ \ref{rtangleplot}(a) is a contour plot of $S(x,y)$ whose  contours correspond to streamlines of the sheet-current density $\bm K$.

To calculate the critical current at which a vortex is nucleated from the  inner corner, note that the corner at $\zeta = a + ia$ corresponds to $w = 2$.  Expanding Eq.\ (\ref{zetaprime}) about this point with $\zeta_v = a+ia + \delta e^{-3\pi/4}$ yields
\begin{equation}
w(\zeta_v) = 2 -i(3\pi \delta/a\sqrt{2})^{2/3}.
\end{equation}
We can use this expansion following the procedure of Sec.\ \ref{Kcstripsection} to obtain 
the Gibbs free energy,
\begin{equation}
G = \frac{\phi_0^2}{2\pi \mu_0 \Lambda} \ln \Big(\frac{3\delta}{\xi}\Big)-\phi_0 K_I \Big(\frac{a}{2\pi}\Big)^{1/3}(3\delta)^{2/3}.
\label{Gdeltartangle}
\end{equation}
Following the steps that led to Eq.\ (\ref{Kcstrip}), we obtain  with $K_I = K_c$,
\begin{eqnarray}
\delta_b^{2/3} &=& \frac{\phi_0}{2\pi \mu_0 \Lambda K_I}\Big(\frac{3\pi}{4a}\Big)^{1/3},\;\delta_c=\frac{e^{3/2}\xi}{3}=1.49\xi,\\
K_c &= &\frac{\phi_0}{e \pi\mu_0\xi\Lambda}R,\;{\rm where }\;
R=\frac{3}{2}\Big(\frac{\pi\xi}{4a}\Big)^{1/3}
\label{Rrtangle}
\end{eqnarray}
is the reduction factor due to current crowding at the sharp inner corner. For example, $R= 0.64$ when $\xi = 0.1 a$ and $R= 0.30$ when $\xi = 0.01 a$. 
The barrier height for $K_I < K_c$ is
\begin{equation}
G_b =  \frac{3\phi_0^2}{4\pi\mu_0\Lambda}\ln\Big(\frac{K_c}{K_I}\Big).
\label{Gbrtangle}
\end{equation}
Note that the prefactor is larger  than that in Eq.\ (\ref{Gbstrip}) by a factor of 3/2, which arises from the term proportional to $\delta^{2/3}$ in  Eq.\ (\ref{Gdeltartangle}).

\subsection{Optimally rounded 90$^\circ$ turn\label{90optimal}}

For the case of the right-angle turn shown in Fig.\ \ref{rtangleplot}(a), the complex sheet-current density in the $\zeta$ plane obtained from Eq.\  (\ref{Grtangle}) via $K_\zeta(\zeta)=d{\cal G}_\zeta(\zeta)/d\zeta$ is
\begin{equation}
{\cal K}_\zeta(\zeta) = K_x -i K_y = \frac{K_I\sqrt{w(\zeta)}}{\sqrt{2-w(\zeta)}},
\label{Krtangle}
\end{equation} 
where $K$, the magnitude of ${\cal K}_\zeta$, is also the magnitude of $\bm K$; i.e., $K = |{\cal K}_\zeta| = |\bm K|=\sqrt{K_x^2+K_y^2}$, and $K$ varies along all of the contours of Fig.\ \ref{rtangleplot}(a) except one.  For contours close to the inner corner, $K$ increases upon approaching the corner, which we call current crowding; for the contour along the inner boundary, $K$ even diverges at the corner.  As shown in Sec.\ \ref{sharp90}, this current crowding leads to a significant reduction in the critical current, because vortices  nucleate preferentially at the sharp inside corner.  On the other hand, for contours close to the outer corner, $K$ decreases upon approaching the corner, which we call current expansion.  

However, as first discovered by Hagedorn and Hall,\cite{Hagedorn63} there is one special contour near the middle for which 
$K$ remains constant with the value $K_I$; this contour, which we call the optimal contour, is shown as the dashed curve in Fig.\ \ref{rtangleplot}(a).  From Eq.\ (\ref{Krtangle}) we find that in the $w$ plane the optimal contour corresponds to $w = 1+iv$, where $v>0$, such that the optimal contour in the $\zeta$ plane, $\zeta_o(v) = x_o(v)+iy_o(v)$, is given by the parametric equation
\begin{equation}
\zeta_o = \frac{a}{2}\Big\{\Big[1+\frac{2}{\pi}\sinh^{-1}\Big(\frac{1}{v}\Big)\Big]+i\Big[1+\frac{2}{\pi}\sinh^{-1}v\Big]\Big\}.
\label{zetaortangle}
\end{equation} 
Alternatively, $v$ can be eliminated to obtain $y_o$ as a function of $x > a/2$ or $x_o$ as a function of $y > a/2$:
\begin{eqnarray}
y_o(x)&=&\!\frac{a}{2}\Big\{1\!+\!\frac{2}{\pi}\sinh^{-1}\!\Big[\frac{1}{\sinh[(\pi/a)(x-a/2)}\Big]\Big\},\label{yortangle}\\
x_o(y)&=&\!\frac{a}{2}\Big\{1\!+\!\frac{2}{\pi}\sinh^{-1}\!\Big[\frac{1}{\sinh[(\pi/a)(y-a/2)}\Big]\Big\}.\label{xortangle}
\end{eqnarray}

The above results have important consequences, because they can tell us the optimal film design that will prevent any significant reduction of the critical current due to current crowding at a right-angle turn.  Consider a long superconducting strip of width $W$ and critical sheet current given by Eq.\ (\ref{Kcstrip}).  If we wish the direction of the current to change by 90$^\circ$ at the corner of a strip for which the outer boundary is a right angle, as shown in Fig.\ \ref{rtangleplot}(a), the inner boundary of the strip should be chosen to be the smooth curve given by Eqs.\ (\ref{zetaortangle})-(\ref{xortangle}) but with $a/2=W$.  The minimum radius of curvature of this curve is $4W/\pi = 1.27 W$, which occurs at $x_o = y_o = W[1+(2/\pi)\sinh^{-1}(1)]=1.56W$.  As discussed in Secs.\ \ref{curve} and \ref{rhocsec}, so long as $\xi$ is much smaller than the minimum radius of curvature, the self-energy $E_{self}$ of a nucleating vortex for small $\delta$ is, to excellent approximation, the same as in Eq.\ (\ref{Gdelta}).  Moreover, since the sheet-current density along the entire inner boundary is constant with the value $K_I$, the work term $W_I$  and hence the entire Gibbs free energy for small $\delta$ are very nearly the same as in Eq.\ (\ref{Gdelta}).  Therefore, the critical current for the strip with a corner of the above-described design, i.e., the area between the dashed curve and the outer boundary in Fig.\ \ref{rtangleplot}(a), should be the same as that of a very long strip of constant width $W$ [see Eq.\ (\ref{Kcstrip})], so long as $\xi \ll W$.

\subsection{Rounded 90-degree turnaround at the end of straight strips\label{Cockroft90Section}}

\begin{figure}
\includegraphics[width = 7cm]{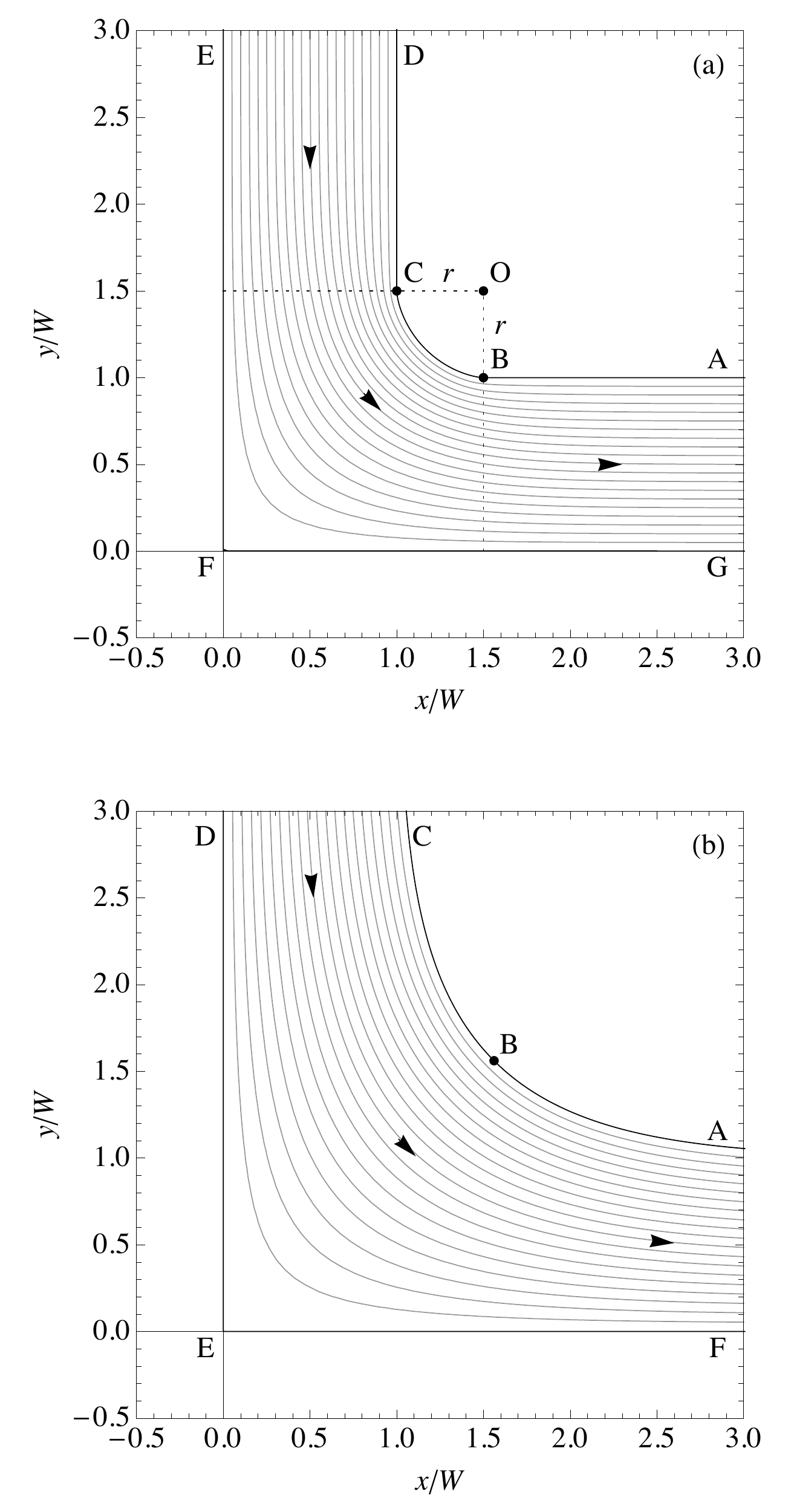}
\caption{(a) Current flow calculated in Sec.\  \ref{C90General} for a film carrying current $K_I W$ from a straight strip of width $W$ around a rounded 90-degree turn into another straight strip of width $W$, shown by the contour plot of the stream function $S(x,y)= \Im{\cal G}_\zeta(x+iy)$, which has the values $S=0$ along the outer boundaries of the film and $S=K_I W$ along the inner boundary.  The contours correspond to streamlines of the sheet-current density $\bm K$, and the arrows show the direction of the current. Here $r/W= 1/2$, and current crowding along the arc BC results in a critical current reduction factor $R = 0.654$. (b) Current flow calculated in Sec.\ \ref{C90Optimal}, where the connection is infinite in length and $r/W \to \infty$, a geometry for which there is no current crowding along the arc ABC and no critical-current reduction. }
\label{C90fig}
\end{figure}

We now use Cockroft's method\cite{Cockroft28} to calculate the current-crowding critical-current reduction factor $R$ for the case that two straight strips with constant width $W$ are connected by a 90-degree rounded corner, as shown in Fig.\ \ref{C90fig}(a). The results depend upon the parameters chosen.  In Sec.\  \ref{C90General} we discuss the behavior for which the ``radius'' $r$  shown in Fig.\ \ref{C90fig}(a)  is finite, and in Sec.\ \ref{C90Optimal} we treat the limiting case for    $r/W \to \infty$, which produces a curving inner boundary corresponding to optimal rounding discussed in Sec.\ \ref{90optimal}.

\subsubsection{General case\label{C90General}}
The conformal mapping\cite{Cockroft28}
\begin{eqnarray}
\zeta'(w)&=&\frac{d\zeta(w)}{dw}= A\frac{\sqrt{w+1}+\sqrt{\frac{b+1}{b-1}}\sqrt{w-1}}{(w-a)\sqrt{w-b}},
\label{zetaprimeC90}\\
A&=&W/g,\;a=(b^2+1)/2b,\label{A&a}\\
\zeta(w)&=&  2A\Big[\ln\Big(\frac{\sqrt{w+1}+\sqrt{w-b}}{\sqrt{b+1}}\Big)\nonumber \\
&&+\sqrt{\frac{b+1}{b-1}}\tan^{-1}\sqrt{\frac{(b+1)(w-b)}{(b-1)(w+1)}}\nonumber \\
&&+\sqrt{\frac{b+1}{b-1}}\ln\Big(\frac{\sqrt{w-1}+\sqrt{w-b}}{\sqrt{b-1}}\Big)\nonumber \\
&&+\tan^{-1}\sqrt{\frac{(b-1)(w-b)}{(b+1)(w-1)}}\Big],
\label{zetaC90} \\
g&=&\pi\Big(1+\sqrt{\frac{b+1}{b-1}}\Big),\label{g}
\end{eqnarray}
maps points in the upper half $w$-plane ($w = u + i v$)  into the area ABCDEFG in the upper half $\zeta$-plane  ($\zeta = x + i y$) shown in Fig.\ \ref{C90fig}(a).  The inverse mapping $w(\zeta)$ can be obtained numerically. 
Special points in the $\zeta$ and $w$ planes are related as follows: A, $w = -\infty$; B, $w = -1$; C, $w = +1$; D, $w=a-\epsilon$; E, $w = a+\epsilon$; F, $w = b$; and G, $w = +\infty$; where $1 < a < b$ and $\epsilon$ is a positive infinitessimal.  The ``radius'' $r$ shown in Fig.\ \ref{C180fig}(a) is determined from
\begin{eqnarray}
r &=& A\Big[\sqrt{\frac{b+1}{b-1}}\ln\Big(\frac{1+\sqrt{2/(b+1)}}{1-\sqrt{2/(b+1)}}\Big) \nonumber \\
&&-2\tan^{-1}\sqrt{\frac{2}{b-1}}.\label{r}
\end{eqnarray}

The complex potential describing the current flow within the area ABCDEFG shown in Fig.\ \ref{C90fig} is 
\begin{equation}
{\cal G}_\zeta(\zeta)=\frac{I}{\pi}\ln[w(\zeta)-a].
\label{calGzetaC90}
\end{equation}
The imaginary part of this is the stream function $S$, whose contours, shown in  Fig.\ \ref{C90fig}, are the streamlines of the sheet-current density $\bm K = \hat x K_x + \hat y K_y$.  The  corresponding complex sheet current, ${\cal K}_\zeta(\zeta)=d{\cal G}_\zeta(\zeta)/d\zeta = K_x(x,y)-iK_y(x,y)$, is
\begin{equation}
{\cal K}_\zeta(\zeta) = K_I\frac{(\sqrt{b+1}+\sqrt{b-1})\sqrt{w-b}}{\sqrt{(b+1)(w-1)}+\sqrt{(b-1)(w+1)}},
\end{equation}
where $K_I = I/W$.
For $0 < y < W$ and $x \gg W$, which corresponds to $w \to \infty,$ we have to good approximation $K_x = K_I$.  However, around the arc BC, which corresponds to $w = u$, where $-1 \le u \le 1$, we find that $K_{\rm BC} = |{\cal K}|$ is given by 
\begin{equation}
\frac{K_{\rm BC}}{K_I}=\frac{\sqrt{b+1}+\sqrt{b-1}}{\sqrt{2}}.
\end{equation}
Since $b > 1$ except in the special limit $b \to 1$, this equation tells us that current-crowding  ($K_{\rm BC} > K_I$) almost always occurs.  As discussed in Secs.\ \ref{curve} and  \ref{rhocsec}, so long as $\xi \ll r$, this current-crowding therefore leads to a critical sheet-current density of the form 
\begin{equation}
K_c = \frac{\phi_0}{e \pi\mu_0\xi\Lambda}R,\;{\rm where }\;
R=\frac{\sqrt{2}}{\sqrt{b+1}+\sqrt{b-1}}
\label{RBC90}
\end{equation}
is the reduction factor due to current crowding at the curving inner boundary BC, relative to the $K_c$ for a long straight strip [Eq.\ (\ref{Kcstrip})].  However, to treat the case for which $r' \le \xi$, an approach similar to that in Sec.\ \ref{rhocsec} would be required.

\begin{figure}
\includegraphics[width = 6cm]{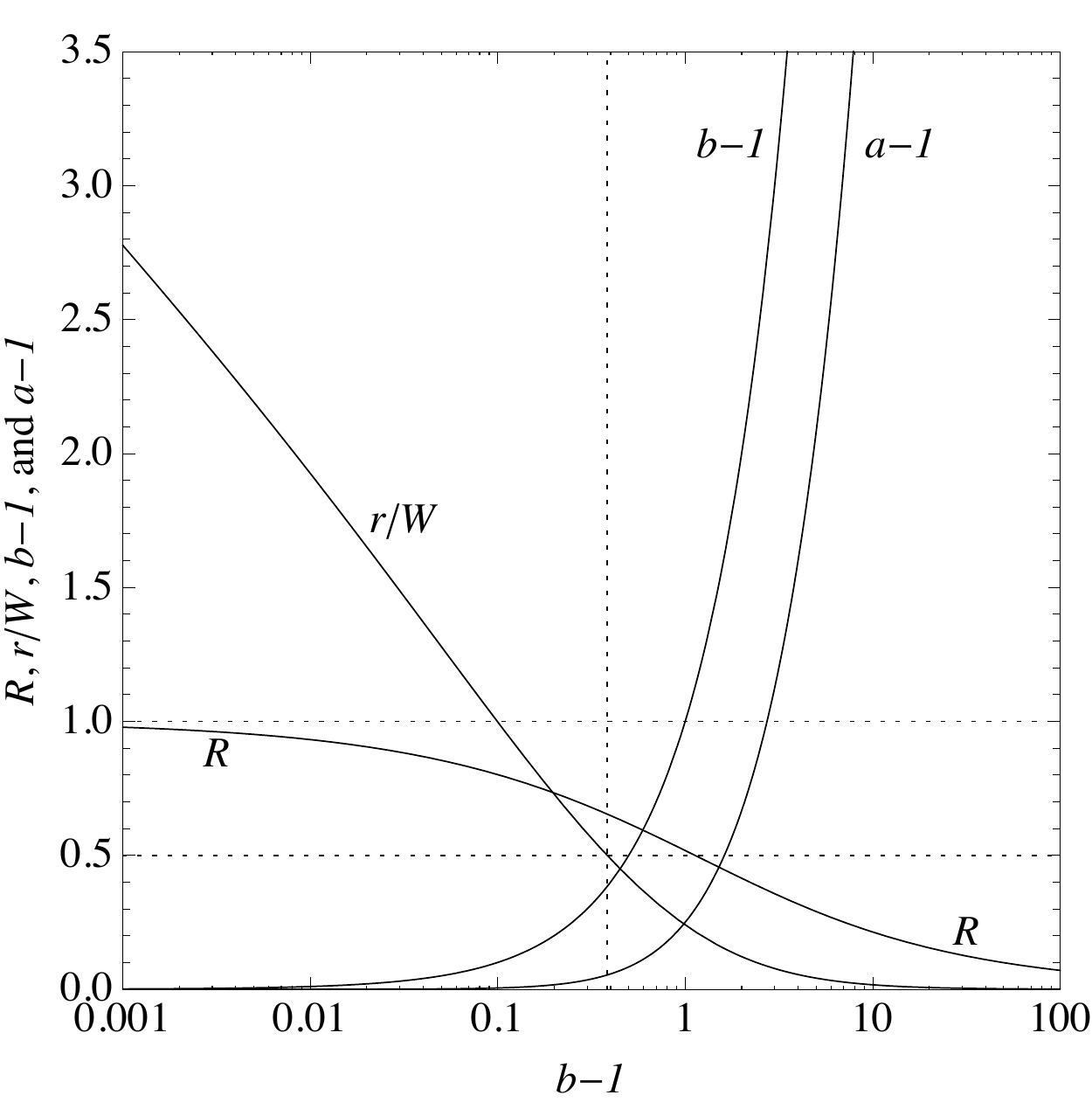}
\caption{Plots of the current-crowding critical-current reduction factor $R$ [Eq.\ (\ref{RBC90})], $r/W$ [Eq.\ (\ref{r})], $b-1$, and $a-1$ vs $b-1$ for the 90-degree turn discussed in Sec.\ \ref{C90General}.  The vertical dotted line marks the parameters for Fig.\ \ref{C90fig}(a) ($R = 0.654$, $r/W = 0.5$, $b = 1.383$, and $a=1.053$). }
\label{C90Retcfig}
\end{figure} 

Figure \ref{C90Retcfig} shows plots of $R$, $r/W$, $b-1$, and $a-1$ vs $b-1$.
While the plot shows $R$ approaching zero for $r/W \to 0,$ bear in mind that Eq.\ (\ref{RBC90}) is valid only for $r \gg \xi$.  Note that to achieve the goal of minimizing current crowding and maximizing the critical current, one should choose the shape of the 90-degree turn make $r/W$ as large as possible.  To achieve $R = 1$ requires $r/W \to \infty$ and $b \to 1$.  We discuss this limit in the following section.

\subsubsection{Optimal rounding when  $r/W \to \infty$ and  $b \to 1$ \label{C90Optimal}}

When $r/W \to\infty$ and $b \to 1$, the dependence of $\zeta(w)$ becomes dominated by values of $w$ very close to 1 and $b$, and it is appropriate to introduce the variable $\omega = (w-1)/(b-1)$ in Eqs.\ (\ref{zetaprimeC90})-(\ref{r}) and to take the limit as $b \to 1$.  This leads to the conformal mapping 
\begin{eqnarray}
\zeta'(\omega)&=&\frac{d\zeta(\omega)}{d\omega}= \frac{W(1+\sqrt{\omega})}{\omega\sqrt{\omega-1}},
\label{zetaprimeofomega90}\\
\zeta(\omega)&=&  \frac{2W}{\pi}\Big(\cos^{-1}\frac{1}{\sqrt{\omega}}+\cosh^{-1}\sqrt{\omega}\Big),
\label{zetaofomega90}
\end{eqnarray}
which maps points in the upper half $\omega$-plane  into the area ABCDEF in the upper half $\zeta$-plane  ($\zeta = x + i y$) shown in Fig.\ \ref{C90fig}(b). 
Special points in the $\zeta$ and $\omega$ planes are related as follows:  A, $\omega = -\infty$; B, $\omega = -1$; C, $\omega = -\epsilon$; D, $\omega = +\epsilon$; E, $\omega = 1$; and F, $w = +\infty$, where $\epsilon$ is a positive infinitessimal. The inverse mapping $\omega(\zeta)$ can be obtained numerically.  The inner boundary curve ABC for $\omega < 0$ is given by Eqs.\ (\ref{zetaortangle})-(\ref{xortangle}) with $v$ replaced by $1/\sqrt{|\omega|}$ and $a$ replaced by $2W$.  
This curve corresponds exactly to the optimally rounded inner boundary of the 90-degree turn discussed in Sec.\ \ref{90optimal}. 

The complex potential describing the current flow within the area ABCDEF shown in Fig.\ \ref{C90fig}(b) is 
\begin{equation}
{\cal G}_\zeta(\zeta)=\frac{I}{\pi}\ln\omega(\zeta).
\label{calGzetaC90opt}
\end{equation}
The imaginary part of this is the stream function $S$, whose contours, shown in  Fig.\ \ref{C90fig}(b), are the streamlines of the sheet-current density $\bm K = \hat x K_x + \hat y K_y$.  The  corresponding complex sheet current, ${\cal K}_\zeta(\zeta)=d{\cal G}_\zeta(\zeta)/d\zeta = K_x(x,y)-iK_y(x,y)$, is
\begin{equation}
{\cal K}_\zeta(\zeta) = \frac{I}{W}\frac{\sqrt{\omega(\zeta)-1}}{1+\sqrt{\omega(\zeta)}}.
\end{equation}
For $y < W$ and $x \gg W$, which corresponds to $\omega(\zeta) \to \pm\infty,$ we obtain $K_x = K_I$, where $K_I = I/W$.  Along the arc ABC, which corresponds to $\omega = u$, where $-\infty \le u \le 0$, we find that $K_{\rm BC} = |{\cal K}| = K_I$.  In other words, there is no current crowding, and the critical current is predicted to be exactly the same as for a long straight strip [Eq.\ (\ref{Kcstrip})].
As discussed above, however, this conclusion is based on the assumption that the coherence length $\xi$ is much smaller than the radius of curvature at the origin $\rho_c = 2W/\pi$. To treat the case for which $\xi \ge \rho_c$, an approach similar to that in Sec.\ \ref{rhocsec} would be required.

\section{T intersection\label{Tsec}}

To measure the critical current of a narrow strip, it is often the case that the current is fed in and taken out using wide contact pads at the ends of the strip, and the voltage is measured using sidebar contacts that intersect the strip in T intersections.  We defer to the next section the question of how the wide contact pads at the ends affect the measurement, and in the present section we first examine the extent to which the geometry of the T intersection can reduce the critical current because of current crowding at the sharp corners.  We then calculate the boundaries of a T intersection with rounded corners that should prevent any reduction of the critical current.

\subsection{Sharp corners\label{sharpT}}
\begin{figure}
\includegraphics[width = 6cm]{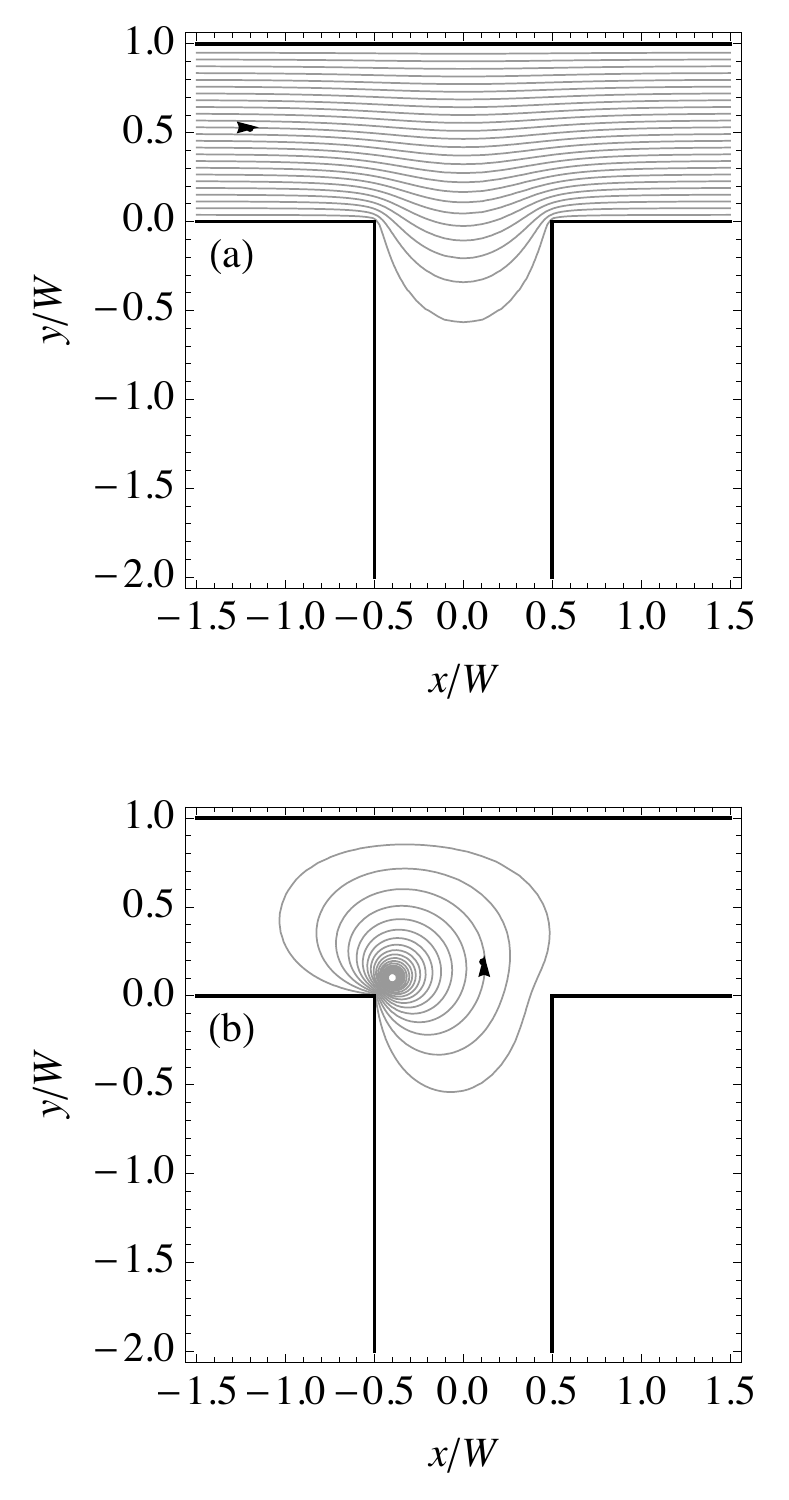}
\caption{(a) Current flow in a strip carrying total current $K_I W$ across the top of a T intersection, shown by the contour plot of the stream function $S(x,y)= \Im{\cal G}_\zeta(x+iy)$, which has the values $S=K_I W$ along the top of the T ($y = W$) and $S=0$ along the underside  ($|x|\ge b$, $y =0$) and  the sides  ($|x|=b$, $y \le 0$).  The contours correspond to streamlines of the sheet-current density $\bm K$, and the arrow shows the current direction.  (b) Vortex-generated current flow, shown by the contour plot of the stream function $S_v(x_v,y_v;x,y)$, which has the values $S_v=0$  along the boundaries. The contours, shown here for $(x_v,y_v) = (-b+0.1W,0.1W)$, correspond to streamlines of the vortex-generated sheet-current density $\bm K_v$, and the arrow shows the direction of the current. The plots show the behavior when $b=W/2$.}
\label{Tfig}
\end{figure}

The current flow in a strip of width $W$ across the top of a T intersection with a voltage-contact sidebar of width $2b$ is shown in Fig.\ \ref{Tfig}. 
The conformal mapping\cite{Kober57p158}
\begin{eqnarray}
\zeta'(w)&=&\frac{d\zeta(w)}{dw}= \frac{2b}{\pi}\frac{\sqrt{\beta^2-w^2}}{w^2-1},
\label{zetaprimeT}\\
\zeta(w)&=& i W -\frac{2b}{\pi}\sin^{-1}(w/\beta) \nonumber \\
&-&\frac{2W}{\pi}\tanh^{-1}\Big(\frac{w\sqrt{\beta^2-1}}{\sqrt{\beta^2-w^2}}\Big),
\label{zetaT}
\end{eqnarray}
where $\beta=\sqrt{1+W^2/b^2}$, maps points in the upper half $w$-plane ($w = u + i v$)  into the T-shaped region in the $\zeta$-plane  ($\zeta = x + i y$) shown in Fig.\ \ref{Tfig}.  The inverse mapping $w(\zeta)$ must be obtained numerically.

The complex potential ${\cal G}_\zeta(\zeta)$ is given by Eq.\ (\ref{calGzeta}) with $I=K_I W$ and the inverse mapping $w(\zeta)$.  The corresponding complex sheet current ${\cal K}_\zeta(\zeta)=d{\cal G}_\zeta(\zeta)/d\zeta = K_x(x,y)-iK_y(x,y)$ is
\begin{equation}
{\cal K}_\zeta(\zeta)=\frac{K_I\sqrt{\beta^2-1}}{\sqrt{\beta^2-w^2(\zeta)}}.
\label{calKT}
\end{equation}
For  $0 < y < W$, $K_x(x,y) \to K_I$ as $|x| \to \infty$. The streamlines of the  sheet current $\bm K = \hat X K_x + \hat y K_y$  are obtained as contours of the stream function $S(x,y) = \Im{\cal G}_\zeta(x+iy)$, the imaginary part of ${\cal G}_\zeta(\zeta)$, shown in Fig.\ \ref{Tfig}(a).  Note the current crowding at the inner corners of the T intersection at $(x,y) = (\pm b, 0)$, where $K = |\cal {\cal K}_\zeta(\zeta)|$ diverges.   

The critical current of the T intersection is reached when a vortex can be nucleated at the sharp corners in Fig.\ \ref{Tfig}(a) or (b) at  $(x,y) = (\pm b, 0)$.
 Expanding Eq.\ (\ref{zetaprimeT}) about $w = \beta$ (which corresponds to $\zeta = -b$) yields for $\zeta_v = -b + \delta e^{i\pi/4}$, $w(\zeta_v)=\beta +\delta w$, where
\begin{equation}
\delta w = \frac{i}{(2\beta)^{1/3}}\Big(\frac{3\pi(\beta^2-1)\delta}{4b}\Big)^{2/3}.
\label{deltawT}
\end{equation}
Thus for  $\bm r_v=(-b+\delta/\sqrt{2},\delta/\sqrt{2})$, where $\delta \ll b$, we can follow the procedure of Sec.\ \ref{Kcstripsection} to obtain
the Gibbs free energy,
\begin{equation}
G = \frac{\phi_0^2}{2\pi \mu_0 \Lambda} \ln \Big(\frac{3\delta}{\xi}\Big)-\phi_0 K_I\Big(\frac{W}{\pi \beta}\Big)^{1/3}\Big(\frac{3\delta}{2}\Big)^{2/3}.
\label{GdeltaT}
\end{equation}
Following the steps that led to Eq.\ (\ref{Kcstrip}), we obtain  with $K_I = K_c$,
\begin{eqnarray}
\Big(\frac{3\delta_b}{2}\Big)^{2/3}\!\! \!\!\!\!\!&=& \!\!\!\frac{3\phi_0 \beta^{1/3}}{4\pi^{2/3} \mu_0 \Lambda K_I W^{1/3}},\;
\delta_c\!=\!\frac{e^{3/2}\xi}{3}\!=\!1.49\xi,\\
K_c &= &\frac{\phi_0}{e \pi\mu_0\xi\Lambda}R,\;{\rm where }\;
R=\frac{3}{2}\Big(\frac{\pi\beta \xi}{2W}\Big)^{1/3}
\label{RT}
\end{eqnarray}
is the reduction factor due to current crowding at one of the sharp inner corners, and $\beta = \sqrt{1+(W/b)^2}$.  
The barrier height for $K_I < K_c$ is
\begin{equation}
G_b =  \frac{3\phi_0^2}{4\pi\mu_0\Lambda}\ln\Big(\frac{K_c}{K_I}\Big).
\label{GbT}
\end{equation}
Note that the prefactor is larger  than that in Eq.\ (\ref{Gbstrip}) by a factor of 3/2, which arises from the term proportional to $\delta^{2/3}$ in  Eq.\ (\ref{GdeltaT}).

In the limit when $W \ll b$,  $\beta \to 1$, in which case Eq.\ (\ref{RT}) yields $R \approx (3/2)(\pi\xi/2W)^{1/3}$, the same as that in Eq.\ (\ref{Rrect180}) in the limit  when $W \ll a$ and $\alpha \to 1$.

In the opposite limit when $W \gg b$, we have $\beta \to W/b$, in which case Eq.\ (\ref{RT}) yields $R \approx (3/2)(\pi\xi/2b)^{1/3}$.  The steps leading to Eq.\ (\ref{Rrect180}) should be valid when $b \gg \xi$, but the above approximations fail for very small contact widths $2b \sim \xi$; in this case,  Eq.\ (\ref{Rrect180}) yields $R \approx 1$.  When  $2b \ll \xi$, current-flow perturbation by the contact lead (the bottom of the T)  is negligibly small, the current-crowding effect essentially disappears, and the critical sheet current is practically the same as in a  straight long strip [Eq.\ (\ref{Kcstrip})].

\subsection{Rounded corners\label{roundedT}}

Based upon our findings in Secs.\ \ref{180optimal} and \ref{90optimal} for 180-degree turnarounds and right-angle turns, it should be possible to design T intersections with rounded corners and slightly widened strips near the intersection such that the critical current is not determined by vortex nucleation at the corners but rather is the same as for a long, straight strip. The mathematical form for a candidate T intersection with rounded inner corners can be derived as follows.  

\begin{figure}
\includegraphics[width = 6cm]{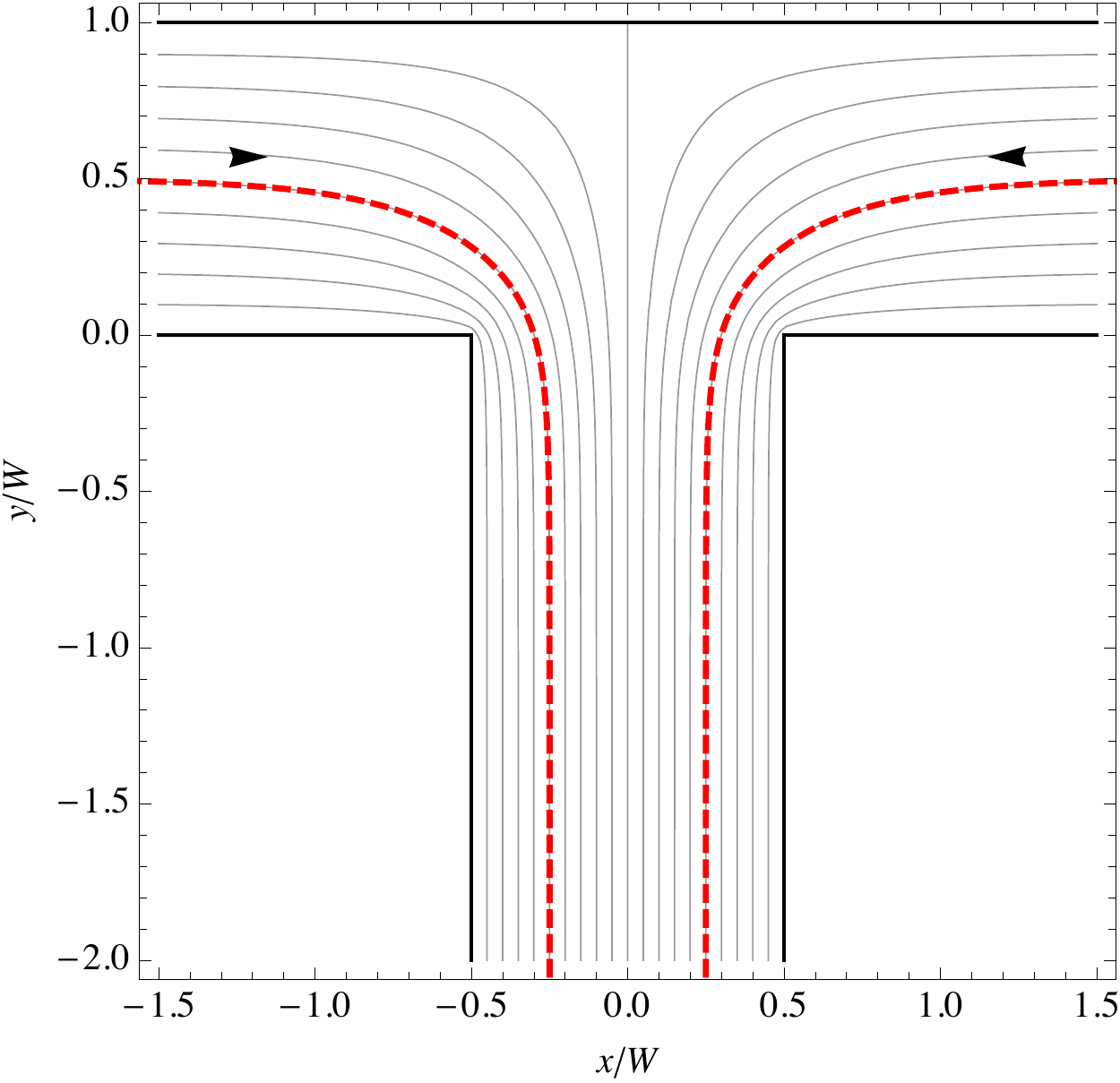}
\caption{Current flow in a strip carrying currents $K_I W$ in opposite directions merging at the top of a T intersection, shown by the contour plot of the stream function $S(x,y)= \Im{\cal G}_\zeta(x+iy)$ [Eq.\ (\ref{GroundedT})], which has the values  $S=2 K_I W$ along the right boundary  ($x > b$, $y =0$ and $x=b$, $y \le 0$),  $S=K_I W$ along the top of the T ($y = W$), and $S=0$ along the left boundary   ($x < - b$, $y =0$ and $x = -b$, $y \le 0$).  The contours correspond to streamlines of the sheet-current density $\bm K$, and the arrows show the current direction.  The right dashed curve  shows $S = K_I W/2$ and the left one shows $S = 3K_I W/2$.  Current crowding occurs only for contours near the sharp corners under the dashed curves.  The plot shows the behavior when $b=W/2$.}
\label{RoundedTfig}
\end{figure}

Let us  use the same conformal mapping as in Eqs.\ (\ref{zetaprimeT}) and (\ref{zetaT}) but consider equal current flow from the right and left ends of the top of the T (width $W$) at current density $K_I$ into the bottom of the T (width $2b$) at current density $K_I W/b$, as shown in Fig.\ \ref{RoundedTfig}.  The complex potential describing this current flow is
\begin{equation}
{\cal G}_\zeta(\zeta)=-\frac{K_I W}{\pi} \ln[w^2(\zeta)-1].
\label{GroundedT}
\end{equation}
The corresponding complex sheet current ${\cal K}_\zeta(\zeta)=d{\cal G}_\zeta(\zeta)/d\zeta = K_x(x,y)-iK_y(x,y)$ is
\begin{equation}
{\cal K}_\zeta(\zeta)=\frac{K_I W w(\zeta)}{b\sqrt{\beta^2-w^2(\zeta)}}.
\label{calKroundedT}
\end{equation}
For  $0 < y < W$, $K_x(x,y) \to \mp K_I$ as $x \to \pm \infty$, and for $|x| < b$, $K_y(x,y) \to - K_I W/b$ as $y \to -\infty$. The streamlines of the  sheet current $\bm K = \hat K_x + \hat y K_y$  are obtained as contours of the stream function $S(x,y) = \Im{\cal G}_\zeta(x+iy)$, the imaginary part of ${\cal G}_\zeta(\zeta)$, shown in Fig.\ \ref{RoundedTfig}.  Note the current crowding at the inner corners of the T intersection at $(x,y) = (\pm b, 0)$, where $K = |\cal {\cal K}_\zeta(\zeta)|$ diverges. 

The dashed contours in Fig.\ \ref{RoundedTfig} for $S = K_I W/2$ and $3K_I W/2$ correspond to the optimally rounded contours shown in Figs.\ \ref{180fig} and \ref{rtangleplot}.   As one moves along any contour 
under a dashed curve in Fig.\ \ref{RoundedTfig}, $K =|\bm K|$ has a maximum  near a sharp corner, but $K$ varies monotonically when one moves along one of the dashed curves.  Thus, for the type of current flow shown in Fig.\ \ref{RoundedTfig}, a patterned film in the shape of a T with rounded corners consisting of the region between the dashed curves and the straight line across the top of the T has the optimum shape.  When $\xi \ll W$ or $b$, the critical current will be unaffected by the bend and will be the same as for a long, straight film.  
To calculate the coordinates  $x_o$ and $y_o$ of the optimally rounded dashed curves, note that in the $w$-plane the corresponding contours are defined by $w_o(\eta) = u_o + iv_o = \cosh \eta + i \sinh \eta = \sqrt{2}\sin(\pi/4+i\eta)$ for $S = K_I W/2$ and  $w_o(\eta) = u_o + iv_o  = -\cosh \eta + i \sinh \eta = \sqrt{2}\sin(-\pi/4+i\eta)$ for $S = 3K_I W/2$, where in both cases $\eta \ge 0$.  Thus the coordinates of the dashed curves in the $\zeta$-plane can be expressed with the help of Eq.\ (\ref{zetaT}) via the parametric equations
\begin{eqnarray}
\zeta_o(\eta) &=& x_o(\eta)+i y_o(\eta) = \zeta(w_o(\eta)).
\end{eqnarray}

A T intersection with rounded corners as described above  would a good choice for a patterned structure to avoid the problem of critical-current reduction due to current crowding discussed in Sec.\ \ref{sharpT}.

\section{Wide end pad\label{WideEndsec}}

To supply current to a narrow strip, it is common to use wide contact pads at the ends of the strip.  In Sec.\ \ref{SharpEnd}  we show that with such a geometry the sharp corners where the strip connects to the end pad are weak points where vortex nucleation caused by current crowding  occurs at a current less than the critical current of the long strip alone.  However, in Sec.\ \ref{SmoothedEnd} we calculate smooth boundaries of the connection between the strip and the end pad for which there should be no reduction of the critical current due to current crowding.

\subsection{Sharp corners\label{SharpEnd}}

\begin{figure}
\includegraphics[width = 8cm]{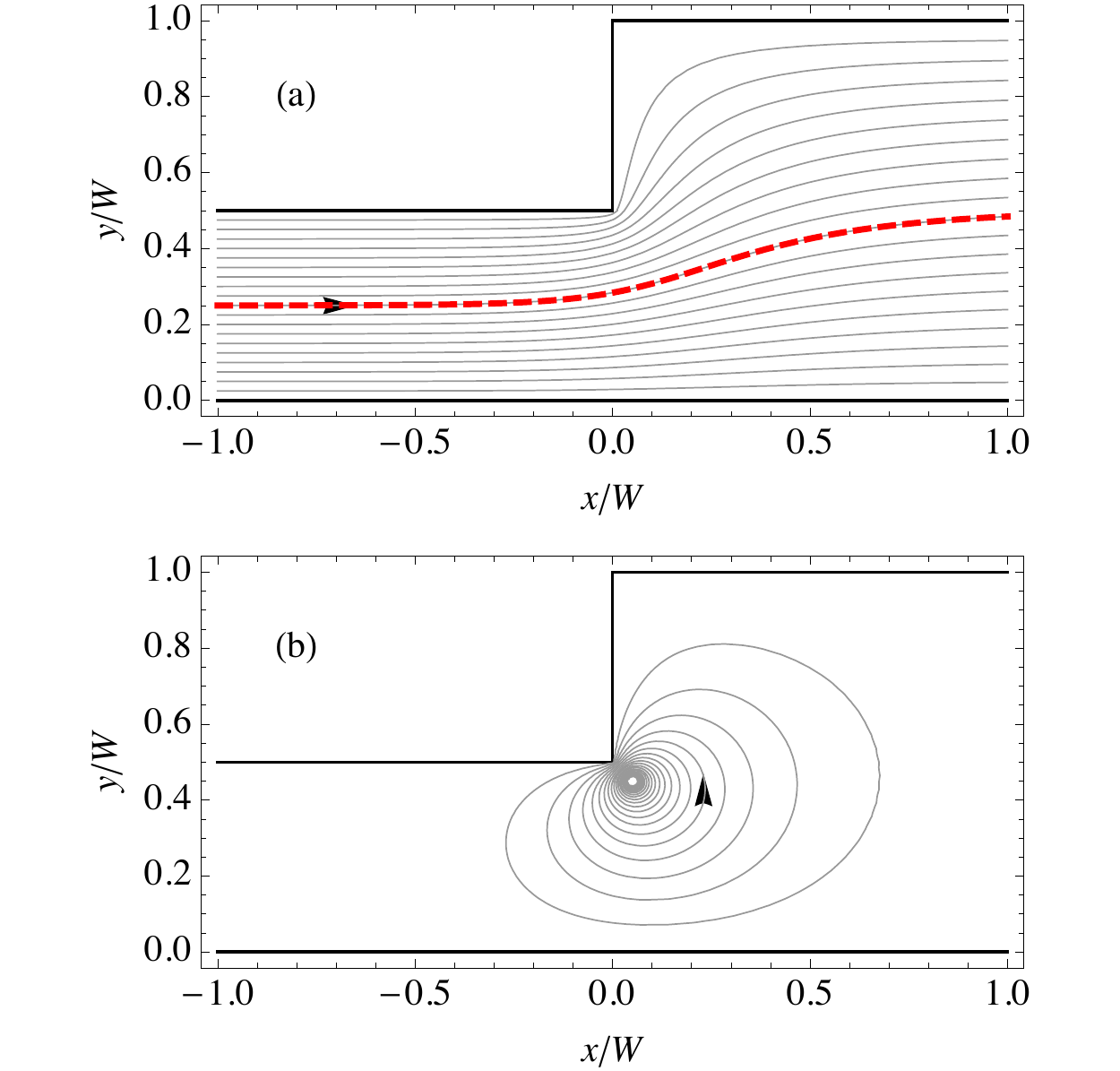}
\caption{(a) Current flow in the upper half of a strip of total width $W$ carrying total current $K_I W$ into a wide contact strip of total width $a$, shown by the contour plot of the stream function $S(x,y)= \Im{\cal G}_\zeta(x+iy)$, which has the values $S=0$ along the upper boundary and $S=-K_I W/2$ along the $x$ axis.  The contours correspond to streamlines of the sheet-current density $\bm K$, and the arrow shows the current direction.  Current crowding leading to $K>K_I$ occurs for all contours above the dashed contour for $S = -K_I W/4$.  (b) Vortex-generated current flow, shown by the contour plot of the stream function $S_v(x_v,y_v;x,y)$, which has the values $S_v=0$  along the boundaries. The contours, shown here for $(x_v,y_v) = (0.05W,0.45W)$, correspond to streamlines of the vortex-generated sheet-current density $\bm K_v$, and the arrow shows the direction of the current. The plots show the behavior when $a = 2W.$}
\label{WideEndfig}
\end{figure}

Consider the current flow in a strip of width $W$ ($x < 0$, $-W/2 < y < W/2$) connected at $x = 0$ to a wide contact pad of width $a>W$ ($x > 0$, $-a/2 < y < a/2$). The sheet-current density is $K_I$ for $x \ll -W$ and $K_I W/a$ for $x \gg a$, and to describe the overall current flow  we use the conformal mapping\cite{Churchill48b}
\begin{eqnarray}
\zeta'(w)&=&\frac{d\zeta(w)}{dw}= \frac{W\sqrt{\gamma+1}}{\pi(1-w^2)\sqrt{\gamma-w}},
\label{zetaprimeWideEnd}\\
\zeta(w)&=& \frac{i}{\pi} \Big[W \tan^{-1}\Big(\frac{\sqrt{w-\gamma}}{\sqrt{\gamma+1}}\Big)  \nonumber \\
&&\;\;\;\;+a \tan^{-1}\Big(\frac{\sqrt{\gamma-1}}{\sqrt{w-\gamma}}\Big)\Big] ,
\label{zetaWideEnd}
\end{eqnarray}
where $\gamma=(a^2+W^2)/(a^2-W^2)$, maps points in the upper half $w$-plane ($w = u + i v$)  into the region of the $\zeta$-plane  ($\zeta = x + i y$) defined by $0 < y < W/2$ for $x < 0$ and  $0 < y < a/2$ for $x > 0$.  The inverse mapping $w(\zeta)$ must be obtained numerically. The mapping of Eq.\ (\ref{zetaWideEnd}), shown  for $a = 2W$ and $\gamma = 5/3$ in Fig.\ \ref{WideEndfig}, corresponds to just the upper half of the strip and its contact pad, since the sheet-current density $\bm K(x,y)$ has mirror symmetry about $y = 0$.     

The complex potential ${\cal G}_\zeta(\zeta)$ describing total current $K_I W$ in the $x$ direction is given by Eq.\ (\ref{calGzeta}) with $I=-K_I W/2$ and the inverse mapping $w(\zeta)$.  The corresponding complex sheet current ${\cal K}_\zeta(\zeta)=d{\cal G}_\zeta(\zeta)/d\zeta = K_x(x,y)-iK_y(x,y)$ is
\begin{equation}
{\cal K}_\zeta(\zeta)=\frac{K_I\sqrt{\gamma-w(\zeta)}}{\sqrt{\gamma+1}}.
\label{calKWideEnd}
\end{equation}
For $x < 0$ and  $0 < y < W/2$, we obtain $K_x(x,y) \to K_I$ as $w(\zeta) \to -1$ and $x \to -\infty$, and for $x > 0$ and $0 < y < a/2$, we obtain  $K_x(x,y) \to K_I/2$ as $w(\zeta) \to +1$ and $x \to +\infty$. The magnitude of the sheet current, $K = |{\cal K}_\zeta(\zeta)|$, is zero at 
$(x,y)=(0,a/2)$, where $w(\zeta)=\gamma$, but diverges at $(x,y)=(0,W/2)$, where $w(\zeta) \to \infty.$  The streamlines of the  sheet current $\bm K = \hat K_x + \hat y K_y$  are obtained as contours of the stream function $S(x,y) = \Im{\cal G}_\zeta(x+iy)$, the imaginary part of ${\cal G}_\zeta(\zeta)$, shown in Fig.\ \ref{WideEndfig}(a).  Note the current crowding near the sharp corner at $(x,y) = (0,W/2)$, where $K = |\cal {\cal K}_\zeta(\zeta)|$ diverges.   

The critical current is reached when a vortex can be nucleated at  $(x,y) = (0,W/2)$ in Fig.\ \ref{WideEndfig}(a) or (b).
To calculate it, we first need the work term $W_I(\bm r_v) = \phi_0 \Delta I(\bm r_v)=\phi_0 [S(0,W/2)-S(x_v,y_v)]$.
However, we need this only very close to the corner.  Expanding Eq.\ (\ref{zetaprimeWideEnd}) about $w = \pm \infty$  yields, for $\zeta = iW/2 + \delta/\sqrt{2}-i \delta/\sqrt{2}$ and $\delta \ll W$,
\begin{equation}
\frac{1}{w} = -i\Big(\frac{3\pi\delta}{4W\sqrt{\gamma+1}}\Big)^{2/3}.
\label{wWideEnd}
\end{equation}
Thus for  $\bm r_v=(\delta/\sqrt{2},W/2-\delta/\sqrt{2})$, where $\delta \ll W$, we can follow the procedure of Sec.\ \ref{Kcstripsection} and use Eq.\ (\ref{wWideEnd}) to obtain 
the Gibbs free energy,
\begin{equation}
G = \frac{\phi_0^2}{2\pi \mu_0 \Lambda} \ln \Big(\frac{3\delta}{\xi}\Big)-\phi_0 K_I\Big(\frac{W}{\pi (\gamma+1)}\Big)^{1/3}\Big(\frac{3\delta}{2}\Big)^{2/3}.
\label{GdeltaWideEnd}
\end{equation}
Following the steps that led to Eq.\ (\ref{Kcstrip}), we obtain  with $K_I = K_c$,
\begin{eqnarray}
\Big(\frac{3\delta_b}{2}\Big)^{2/3} \!\!\!\!\!\!&=& \!\!\frac{3\phi_0 (\gamma+1)^{1/3}}{4\pi^{2/3} \mu_0 \Lambda K_I W^{1/3}},\;
\delta_c\!=\!\frac{e^{3/2}\xi}{3}\!=\!1.49\xi,\\
K_c\! &= &\!\!\frac{\phi_0}{e \pi\mu_0\xi\Lambda}R,\;{\rm where }\;
R\!=\frac{3}{2}\Big(\frac{\pi(\gamma+1)\xi}{2W}\Big)^{1/3}.
\label{RWideEnd}
\end{eqnarray}
is the reduction factor due to current crowding at the sharp corner $(x,y)=(0,W/2)$, and $\gamma+1=2a^2/(a^2-W^2)$.  The barrier height for $K_I < K_c$ is
\begin{equation}
G_b =  \frac{3\phi_0^2}{4\pi\mu_0\Lambda}\ln\Big(\frac{K_c}{K_I}\Big).
\label{GbWideEnd}
\end{equation}
Note that the prefactor is larger  than that in Eq.\ (\ref{Gbstrip}) by a factor of 3/2, which arises from the term proportional to $\delta^{2/3}$ in  Eq.\ (\ref{GdeltaWideEnd}).

In the limit when $a/W\to \infty$,  $\gamma+1 \to 2$, in which case Eq.\ (\ref{RWideEnd}) yields $R \approx (3/2)(\pi\xi/W)^{1/3}$, the same as that in Eq.\ (\ref{Rrect180}) in the limit $\alpha \to 1$ for a strip of width $W/2$.

The steps leading to Eq.\ (\ref{RWideEnd}) should be valid when $(a-W) \gg \xi$, but the assumptions fail when $(a-W) \sim \xi$, in which case $(\gamma +1) \sim W/\xi \gg 1$, and $R \sim 1.$   When  $(a-W) \ll \xi$, current-flow perturbation at the transition to the wider end  is negligibly small,  and the critical sheet current is practically the same as in a  straight long strip [Eq.\ (\ref{Kcstrip})].

\subsection{Rounded corners\label{SmoothedEnd}}

We now seek the mathematical form of optimally rounded boundaries of a transition region from a narrow strip to a wide strip, such that the critical current is not reduced by vortex nucleation at the corners but instead is the same as that of a long, straight, narrow strip. 
We already have determined in Secs.\ \ref{180optimal} and \ref{90optimal} the shapes of optimally rounded boundaries for which there is no critical-current reduction due to current crowding around 180-degree turnarounds or right-angle turns (see the dashed curves in Figs.\ \ref{180fig} and \ref{rtangleplot}).  Here we  use a similar approach.  

A careful analysis of the contours for $-K_I W/4 < S \le 0$ in Fig.\ \ref{WideEndfig}(a) reveals that $K$, the magnitude of the current density, has a maximum greater than $K_I$ as one moves along the contour from left to right.  However, for $S = -K_I W/4$, the maximum  occurs at $x = -\infty$, where $K= K_I.$  For contours with $-K_I W/2 \le S < -K_I W/4,$ $K$ decreases monotonically below $K_I$ as one moves along the contour from left to right. The shape of the optimal contour therefore can be obtained from $S = -K_I W/4$.  In the $w$-plane this corresponds to a circular arc of radius 1 centered at the origin, 
 $w_o(\eta) = u_o + iv_o = e^{i\eta}=\cos \eta + i \sin \eta$, where $0 \le \eta \le \pi$.  Thus the coordinates  $(x_o(\eta),y_o(\eta))$ of the dashed curve in the $\zeta$-plane, the upper boundary of the optimally rounded transition region, can be expressed with the help of Eq.\ (\ref{zetaWideEnd}) via the parametric equations
\begin{eqnarray}
\zeta_o(\eta) &=& x_o(\eta)+i y_o(\eta) = \zeta(w_o(\eta)).
\end{eqnarray}
Because of the mirror symmetry about the $x$ axis, the lower boundary of the optimally rounded transition region is defined by the coordinates $(x_o(\eta),-y_o(\eta))$.

Note that since the rounded transition region is essentially carved out of the middle of a long strip of width $W$ and an end pad of width $a$, far from the joint the widths are $W/2$ for the narrow strip and $a/2$ for the end pad.

\section{Edge defects\label{EdgeDefects}}

In the above calculations, we have calculated the critical currents of long, straight thin strips of uniform thickness or strips with straight or smooth edges interrupted by corners or turns.  We now calculate the critical-current reduction when the edges of the strips are not straight or smooth but have defects.  

\subsection{Semicircular notch}

Consider  a long, straight strip of constant width $W$ with a defect along the edge, modeled as a semicircular notch of radius $a$ at the edge, where $a \ll W$. The conformal mapping\cite{Churchill48c} 
\begin{eqnarray}
\zeta'(w)\!\!&=&\!\frac{1}{2}+\frac{w}{2\sqrt{w^2-(2a)^2}}, \label{zetaPrimeSemiNotch}\\
\zeta(w)\!\!&=& \!\!\frac{w+\sqrt{w^2-(2a)^2}}{2}, \label{zetaSemiNotch}\\
w(\zeta)\!\!&=& \!\!\zeta+a^2/\zeta,
\label{wSemiNotch}
\end{eqnarray}
maps points in the upper half $w$-plane ($w = u + i v$)  into the region of the $\zeta$-plane  ($\zeta = x + i y$) above  the $x$ axis for $|x| \ge a$ and above the semicircle for $|x| < a$. 

\begin{figure}
\includegraphics[width = 7cm]{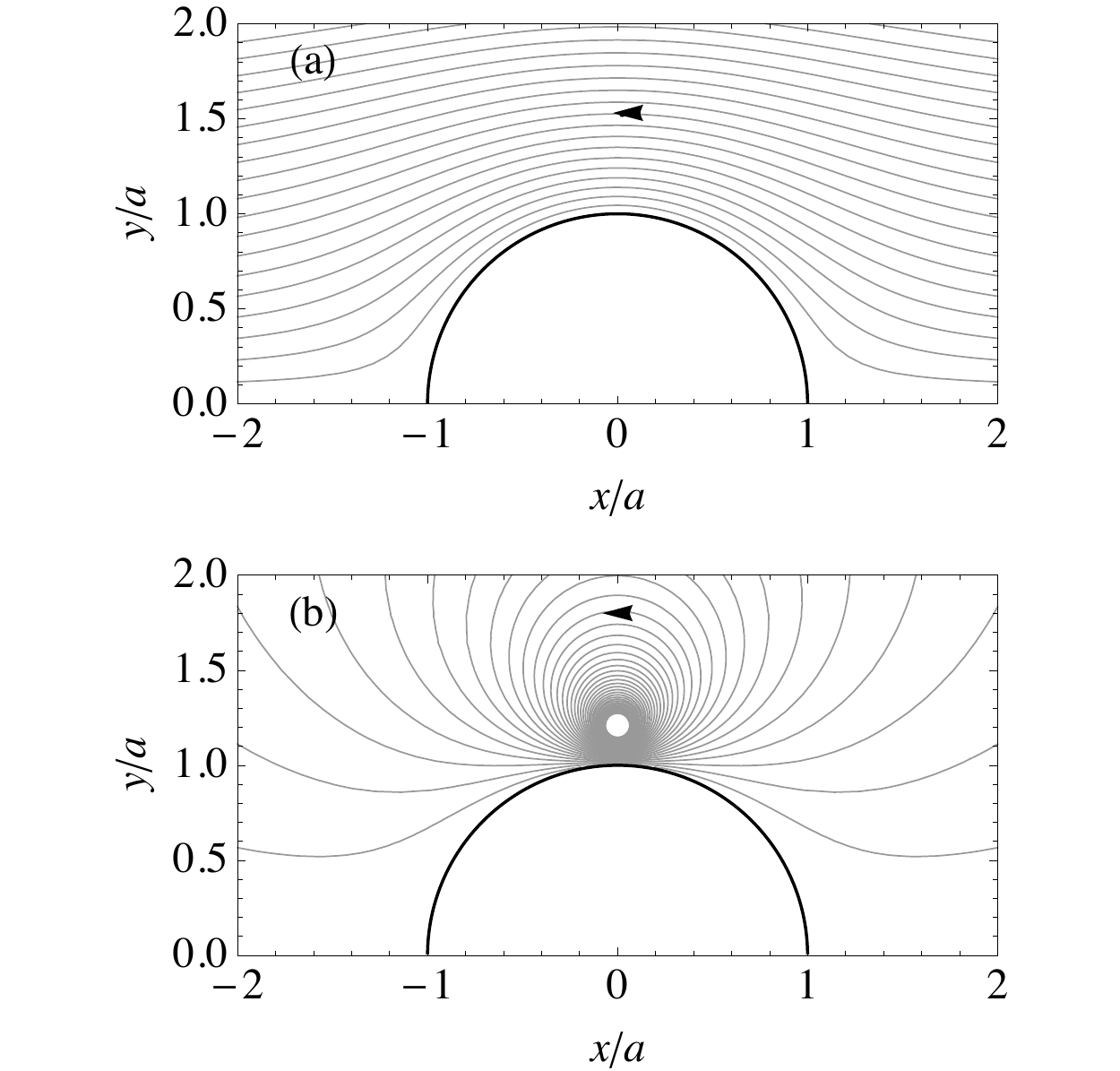}
\caption{(a) Current flow around a semicircular notch of radius $a$ near the edge of a strip, shown by the contour plot of the stream function $S(x,y)= \Im{\cal G}_\zeta(x+iy)$, Eq.\ (\ref{calGSemiNotch}), which has the values $S=0$ along the boundary.  The contours correspond to streamlines of the sheet-current density $\bm K$, and the arrow shows the current direction.  (b) Vortex-generated current flow, shown by the contour plot of the stream function $S_v(x_v,y_v;x,y)$, which has the values $S_v=0$  along the boundaries. The contours, shown here for $(x_v,y_v) = (0,1.2a)$, correspond to streamlines of the vortex-generated sheet-current density $\bm K_v$, and the arrow shows the direction of the current. }
\label{SemiNotchSfig}
\end{figure}

For a sheet-current density with the value $K_x = -K_I$ at large distances above the notch, the complex potential is 
\begin{equation}
{\cal G}_\zeta(\zeta)=-K_I w(\zeta), 
\label{calGSemiNotch}
\end{equation}
and its imaginary part is the stream function, shown in Fig.\ \ref{SemiNotchSfig}(a). Similarly, when a vortex is at the position $\zeta_v = i y_v$, the stream function $S_v$ can be calculated from Eq.\ (\ref{Gvzeta}) with $w(\zeta)$ obtained from Eq.\ (\ref{wSemiNotch}).  The stream function $S_v$, the imaginary part of ${\cal G}_{v\zeta}(\zeta_v,\zeta)$, is shown in Fig.\ \ref{SemiNotchSfig}(b). 

\begin{figure}
\includegraphics[width = 7cm]{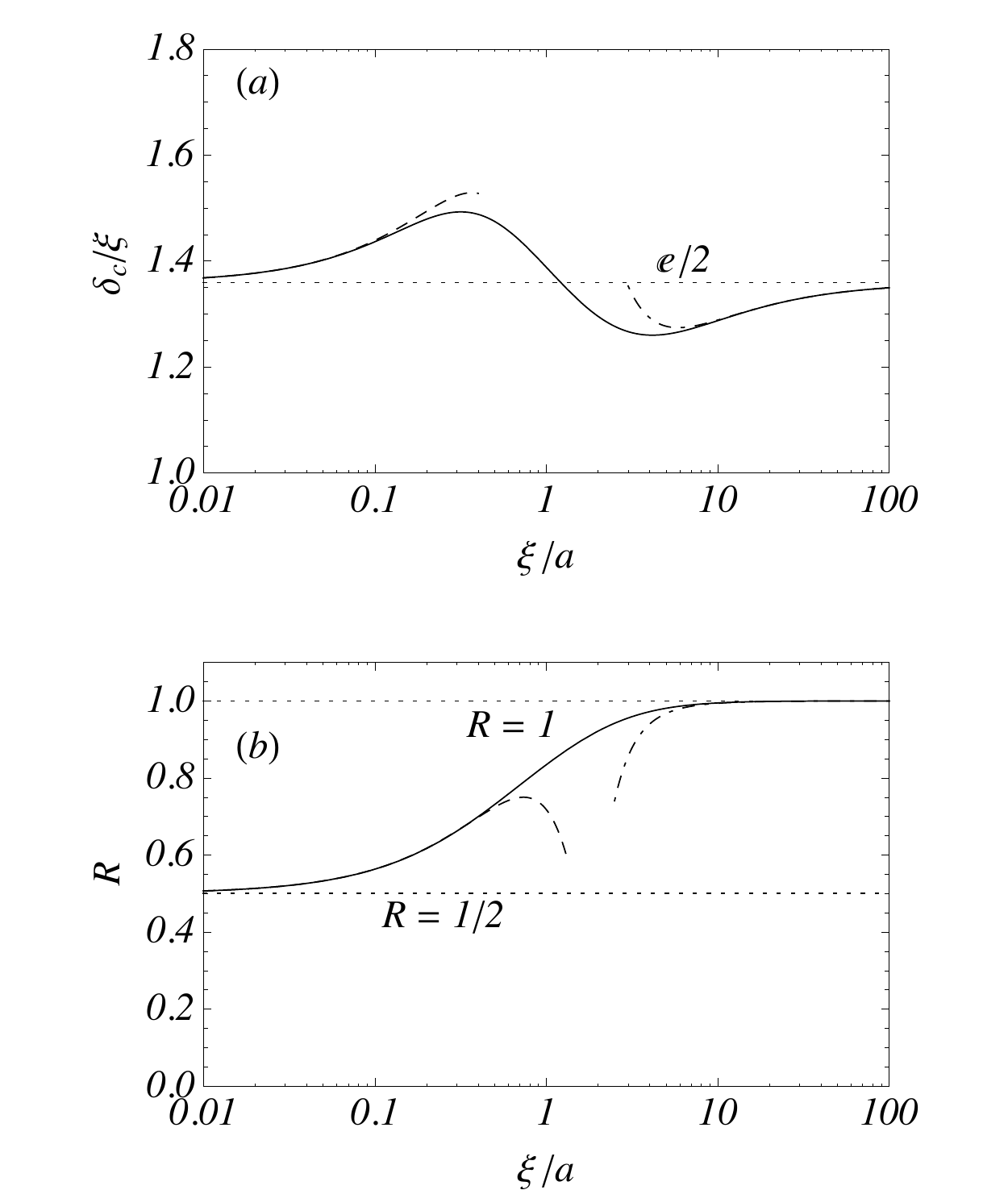}
\caption{Numerical results for (a) $\delta_c/\xi$ and (b) $R$ as functions of the ratio of the coherence length $\xi$ to the radius $a$ of the semicircular notch. Expansions in powers of $\xi/a$ are shown as dashed curves for $\delta_c$ [Eq.\ (\ref{deltacsmxi})] and $R$ [Eq.\ (\ref{Rsmxi})].  Expansions in powers of $a/\xi$ are shown as dot-dashed curves for $\delta_c$ [Eq.\ (\ref{deltaclgxi})] and $K_{0c}$ [Eq.\ (\ref{Rlgxi})].}
\label{SemiNotchfig}
\end{figure}

When a vortex is at the position $\zeta_v = iy_v$ ($y_v > a$)
on the $y$ axis, we obtain from Eqs.\ (\ref{Icirc}) and (\ref{DeltaI})
\begin{equation}
G = \frac{\phi_0^2}{2\pi \mu_0 \Lambda} \ln \Big[\frac{2y_v(y_v^2-a^2)}{\xi (y_v^2+a^2)}\Big]-\phi_0 K_I\Big(y_v-\frac{a^2}{y_v}\Big).
\label{GSemiNotch}
\end{equation}
Setting $\delta = y_v-a$ and following the steps outlined in Sec.\ \ref{Kcstripsection} for arbitrary values of $\xi/a$, we obtain $\delta_c$, the position for which $G = 0$, as shown in Fig.\ \ref{SemiNotchfig}(a).  The corresponding sheet-current density is
\begin{equation}
K_c= \frac{\phi_0}{e \pi\mu_0\xi\Lambda}R,
\label{RSemiNotch}
\end{equation}
where $R$, the reduction factor due to current crowding near the top of the notch, is shown in Fig.\ \ref{SemiNotchfig}(b).
 Analytic expansions (including only the first few terms in the series) for $\xi/a \ll 1$ are [dashed curves in Fig.\ \ref{SemiNotchfig}]
\begin{eqnarray}
\frac{\delta_c}{\xi}&=&\frac{e}{2}+\frac{e^2}{8}\Big(\frac{\xi}{a}\Big)-\frac{e^3}{16}\Big(\frac{\xi}{a}\Big)^{2}, \label{deltacsmxi}\\
R &= &\frac{1}{2}+\frac{e}{4}\Big(\frac{\xi}{a}\Big)-\frac{e^2}{16}\Big(\frac{\xi}{a}\Big)^2,
\label{Rsmxi}
\end{eqnarray}
and  corresponding expansions for $\xi/a \gg 1$ are [dot-dashed curves in Fig.\ \ref{SemiNotchfig}]
\begin{eqnarray}
\frac{\delta_c}{\xi}&=&\frac{e}{2}-\Big(\frac{a}{\xi}\Big)+\frac{8}{e}\Big(\frac{a}{\xi}\Big)^{2}, \label{deltaclgxi}\\
R&=&1-\frac{4}{e^2}\Big(\frac{a}{\xi}\Big)^2-\frac{368}{e^4}\Big(\frac{a}{\xi}\Big)^4.
\label{Rlgxi}
\end{eqnarray}

Regardless of the size of $\xi$ relative to $a$, vortex nucleation always occurs when the barrier height is reduced to zero at a distance $\delta_c$ of the order of $\xi$ from the semicircular notch.  When $\xi \ll a$, the critical-current reduction factor is $R = 1/2$, which arises from current crowding at the top of the notch, where the sheet-current density is a factor of two larger than far away from the notch.  However, because of the radius-of-curvature effect discussed in Sec.\ \ref{rhocsec}, $R$ approaches 1 as $\xi/a$ increases to large values.  This behavior of $R$ as a function of $\xi/a$ appears to be a general feature of edge defects in thin films:  When the linear dimensions of the edge defect are much smaller than $\xi$, the suppression of the critical current is negligible ($R \approx 1$). 

\subsection{Triangular notch}

Consider  a long, straight strip of constant width $W$ with a defect along the edge, modeled as a triangular notch in the shape of an isosceles triangle with base $2b$ along the edge, height  $a$,  two equal sides of length $c = \sqrt{a^2+b^2}$, and vertex angle $\theta_0= 2\tan^{-1}(b/a)$, where $a \ll W$.  We define $\mu = 1/(2-\theta_0/\pi)$, which varies between 1/2 (when $\theta_0 = 0$) and 1 (when $\theta_0 = \pi$).  
The conformal mapping\cite{Spiegel64} 
\begin{eqnarray}
\zeta'(w)\!\!&=&\!\frac{d\zeta(w)}{dw}= K(\mu)\frac{w^{\frac{1}{\mu}-1}}{(1-w^2)^{\frac{1}{2\mu}-\frac{1}{2}}}, \label{zetaprimeNotch}\\
K(\mu)\!\!&=&\!c \exp\Big[-i\frac{\pi}{2}\Big(\frac{1}{\mu}-1\Big)\Big]g(\mu),\label{KNotch}\\
g(\mu)\!\!&=&\!\!\frac{\sqrt{\pi}}{\Gamma(\frac{1}{2\mu})\Gamma(\frac{3}{2}-\frac{1}{2\mu})}, \label{gNotch}\\
\zeta(w)\!\!&=& \!\!K\mu w^\frac{1}{\mu} \; _2F_1\Big(\frac{1}{2\mu},\frac{1}{2\mu}-\frac{1}{2};\frac{1}{2\mu}+1;w^2) \nonumber \\
&&\!\!+i c \sin\Big[\frac{\pi}{2}\Big(\frac{1}{\mu}-1\Big)\Big],
\label{zetaNotch}
\end{eqnarray}
and $_2F_1(\alpha,\beta;\gamma;z)$ is the hypergeometric function, maps points in the upper half $w$-plane ($w = u + i v$)  into the region of the $\zeta$-plane  ($\zeta = x + i y$) above  the $x$ axis for $|x| \ge b$ and above the notch for $|x| < b$.  The inverse mapping $w(\zeta)$ must be obtained numerically. $g(\mu) = 1$ at $\mu = 1/2$ and 1, and rises smoothly to a maximum of $g = \sqrt{\pi}/\Gamma^2(3/4) = 1.180$ at $\mu = 2/3$.
\begin{figure}
\includegraphics[width = 5cm]{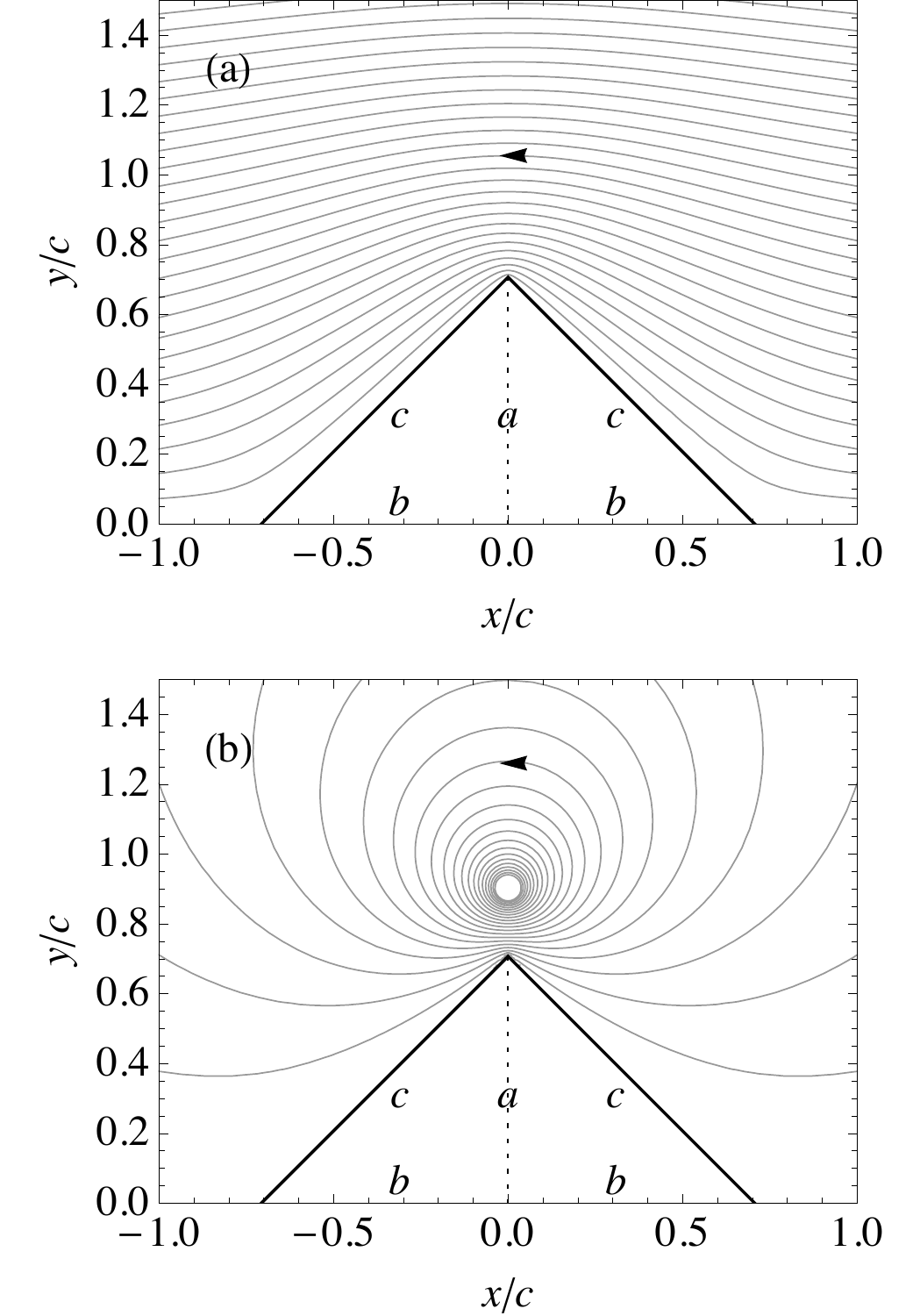}
\caption{(a) Current flow around a isosceles triangular notch (height $a$, base $2b$, and sides $c$) near the edge of a strip, shown by the contour plot of the stream function $S(x,y)= \Im{\cal G}_\zeta(x+iy)$, Eq.\ (\ref{calGNotch}), which has the values $S=0$ along the boundary.  The contours correspond to streamlines of the sheet-current density $\bm K$, and the arrow shows the current direction.  (b) Vortex-generated current flow, shown by the contour plot of the stream function $S_v(x_v,y_v;x,y)$, which has the values $S_v=0$  along the boundaries. The contours, shown here for $(x_v,y_v) = (0,0.9c)$, correspond to streamlines of the vortex-generated sheet-current density $\bm K_v$, and the arrow shows the direction of the current. The plots show the behavior when  $\mu = 2/3$ and the vertex angle is $\theta_0 = \pi/2$.}
\label{Notchfig}
\end{figure} 

For a sheet-current density with the value $K_x = -K_I$ at large distances above the notch, the 
  complex potential is 
\begin{equation}
{\cal G}_\zeta(\zeta)=-K_I g c w(\zeta), 
\label{calGNotch}
\end{equation}
and its imaginary part is the stream function, shown in Fig.\ \ref{Notchfig}(a). Similarly, when a vortex is at the position $\zeta_v = i y_v$, the stream function $S_v$ can be calculated from Eq.\ (\ref{Gvzeta}) with $w(\zeta)$ obtained as the inverse of $\zeta(w)$, Eq.\ (\ref{zetaNotch}).  The stream function $S_v$, the imaginary part of ${\cal G}_{v\zeta}(\zeta_v,\zeta)$, is shown in Fig.\ \ref{Notchfig}(b). 

\begin{figure}
\includegraphics[width = 7cm]{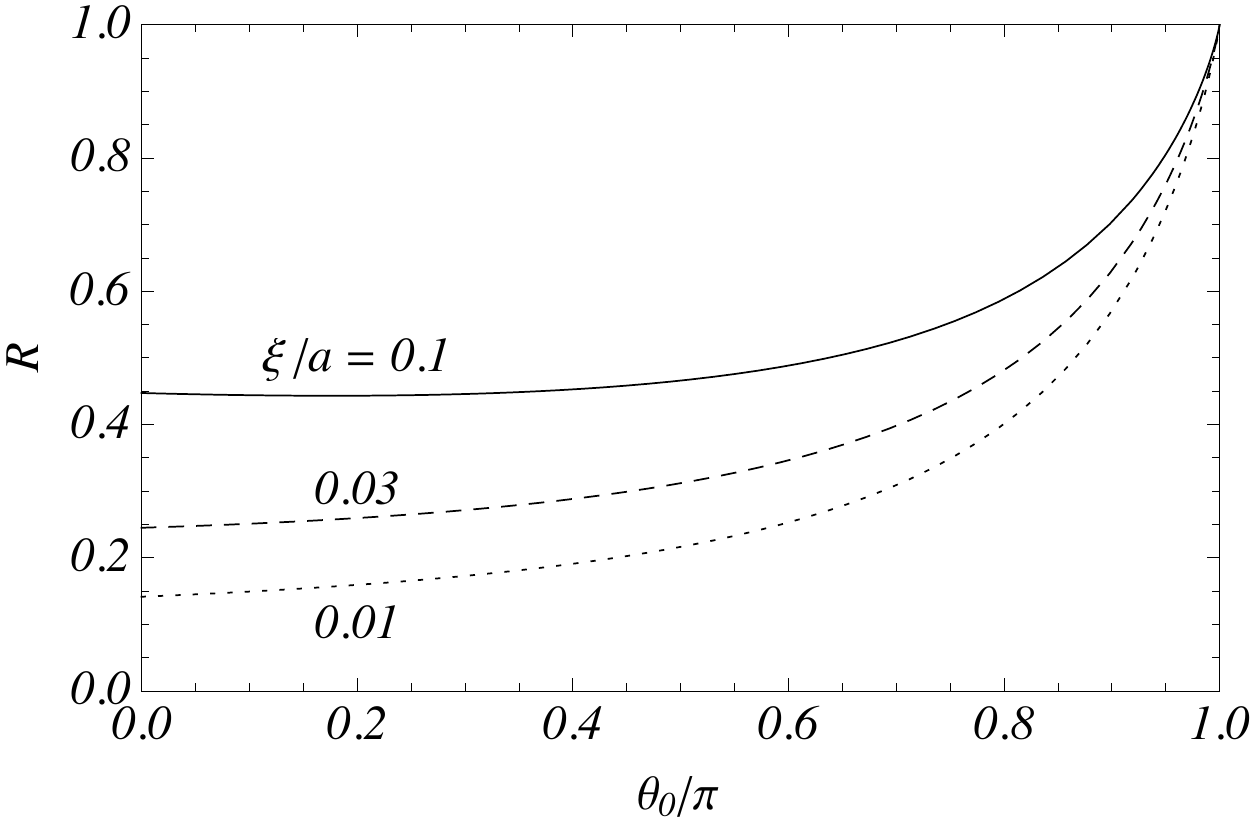}
\caption{The notch's critical-current reduction factor $R$, Eq.\ (\ref{RNotch}), versus the vertex angle $\theta_0$ for $\xi/a$ = 0.01, 0.03, and 0.1, where $a =c\cos(\frac{\theta_0}{2})= c\sin[\frac{\pi}{2}(\frac{1}{\mu}-1)].$}
\label{NotchRafig}
\end{figure}

At distances along the $y$ axis a short distance $\delta$ above the peak at $z = x + i y = ia$, which corresponds to $w = 0$, one can show from Eqs.\ (\ref{zetaprimeNotch}) and (\ref{KNotch}) that 
$w = iv$, where $v=(\delta/c\mu g)^\mu$.  Following the procedure of Sec.\ \ref{Kcstripsection}, we obtain
\begin{equation}
G = \frac{\phi_0^2}{2\pi \mu_0 \Lambda} \ln \Big(\frac{2\delta}{\mu\xi}\Big)-\phi_0 K_Igc\Big(\frac{\delta}{\mu gc}\Big)^{\mu}.
\label{GNotch}
\end{equation}
Following the steps that led to Eq.\ (\ref{Kcstrip}), we obtain  with $K_I = K_c$,
\begin{eqnarray}
\Big(\frac{\delta_b}{\mu gc}\Big)^{\mu}\!\!\!\!\!&=& \!\!\!\frac{\phi_0 }{2\pi \mu_0 \Lambda K_I \mu gc},\;
\delta_c\!=\!\frac{\mu e^{1/\mu}\xi}{2}\!,\\
K_c &= &\frac{\phi_0}{e \pi\mu_0\xi\Lambda}R,\;{\rm where }\;
R=\frac{1}{\mu}\Big(\frac{\xi}{2gc}\Big)^{1-\mu}
\label{RNotch}
\end{eqnarray}
is the reduction factor due to current crowding near the top of the notch, valid only for small values of $\xi/c$.
Figure \ref{NotchRafig} exhibits plots of $R$ vs $\theta_0/\pi$, where $\theta_0$ is the notch's vertex angle, and Table \ref{table1} exhibits the functional dependence of $R$ upon $\theta_0$ and $\mu$.
The result in Eq.\ (\ref{RNotch}) is analogous to those found in Refs.\ \onlinecite{Soininen94}  and \onlinecite{Aladyshkin01} as the solutions of related problems.
The barrier height for $K_I < K_c$ is
\begin{equation}
G_b =  \frac{\phi_0^2}{2\pi \mu \mu_0\Lambda}\ln\Big(\frac{K_c}{K_I}\Big).
\label{GbNotch}
\end{equation}
Note that the prefactor is larger  than that in Eq.\ (\ref{Gbstrip}) by a factor of $1/\mu$, which arises from the term proportional to $\delta^{\mu}$ in  Eq.\ (\ref{GNotch}).

\begin{table}
\centering
\caption{\label{table1}The notch's critical-current reduction factor $R$, Eq.\ (\ref{RNotch}), as a function of the vertex angle $\theta_0$ and $\mu = 1/(2-\theta_0/\pi)$. }
\begin{ruledtabular}
\begin{tabular}{ccc}
$\theta_0$ & $\mu$ & $R$\\
\hline
0 & 1/2 & $1.414(\xi/a)^{1/2}$\\
$\pi/4$ & 4/7 & $1.191(\xi/a)^{3/7}$\\
$\pi/2$ & 2/3 & $1.004 (\xi/a)^{1/3}$\\
$3\pi/4$ & 4/5 & $0.876 (\xi/a)^{1/5}$\\
$\pi$ & 1 & 1
\end{tabular}
\end{ruledtabular}
\end{table}

\section{Thermal excitation over the Gibbs free-energy barrier\label{BarrierClimbing}}

In the above sections we have calculated the critical current $I_c = K_c W$ by defining it as the current $I = K_I W$ at which the Gibbs free-energy barrier is reduced to zero. 
This is  the critical current that would be measured in the limit of zero temperature, $ T = 0$.  However,  because the Gibbs free-energy barrier height $G_b$ is proportional to $\ln(I_c/I)$, as shown in Eqs.\ (\ref{Gbstrip}), (\ref{Gbarc}), (\ref{Gb180}), (\ref{Gbrect180}), (\ref{Gbrtangle}), (\ref{GbT}), and (\ref{GbWideEnd}), experiments performed at finite temperature may observe switches into a state of finite voltage at currents less than the zero-Gibbs-free-energy-barrier critical current $I_c$.  Such switches can occur because nascent vortices may be thermally excited to the top of the free-energy barrier during the time of the experiment.  Because this is a statistical process, the switching currents can be described by a probability distribution that depends upon temperature, sample properties, and experimental conditions. 

We next derive the switching probability distribution applicable to experiments in which the current $I$ through the superconducting device is ramped from zero to some maximum value at a rate $r_I = dI/dt$, where $t$ is the time, such that $I = r_I t$.   Let $P_u(t)$ denote the probability that the device is still unswitched (in the zero-voltage state) at time $t$.  We assume that the probability per unit time of a switch into a finite-voltage state (resulting from a successful transit of a nascent vortex over the Gibbs free-energy barrier) is 
\begin{eqnarray}
\Omega &=& \omega \exp(-G_b/k_BT)=\omega (I/I_c)^N,\; I \le I_c, \label{OmegaEq}\\
&=& \omega,\; I > I_c,
\end{eqnarray}
where $\omega$ corresponds to the sample-dependent attempt frequency in Hz and $N$ is the ratio of  $G_b$'s energy scale  to $k_BT$: $N = \phi_0^2/2\pi\mu_0\Lambda k_B T$ for a straight strip or a gentle curve [Eqs.\ (\ref{Gbstrip}) and (\ref{Gbarc})],  $N = \phi_0^2/\pi\mu_0\Lambda k_B T$ for a sharp 180-degree turnaround [Eq.\ (\ref{Gb180})], or $N = 3\phi_0^2/4\pi\mu_0\Lambda k_B T$ for sharp right-angle inner corners [Eqs.\ (\ref{Gbrtangle}), (\ref{GbT}), and (\ref{GbWideEnd})]. For the cases of interest here, $N\gg 1$.  The rate of decrease of $P_u(t)$ is given by the equation
\begin{equation}
dP_u/dt = - \Omega P_u.
\label{dPudtEq}
\end{equation}
The solution of Eqs.\ (\ref{OmegaEq})-(\ref{dPudtEq}), expressed in terms of $I$, is
\begin{eqnarray}
\!\!\!\!\!P_u \!\!&=&\!\!\exp\!\Big(\!\!-\!\frac{\omega I^{N+1}}{(N\!+\!1)r_I I_c^{N}}\Big),\; I \le I_c, \label{PuEqless}\\
&=& \!\!\exp\!\Big(\!\!-\!\frac{\omega I_c}{(N\!+\!1)r_I}\Big)\!\exp\!\Big(\!\!-\!\frac{\omega (I\!-\!I_c)}{r_I}\Big),\; I \ge I_c.
\end{eqnarray}
The probability that the device has switched by the time the current reaches $I$ is $P_{sw} = 1-P_u$, and the switching probability distribution  $P'_{sw}=dP_{sw}/dI=-dP_u/dI$ is 
\begin{eqnarray}
\!\!\!\!\!\!\!\!\!\!P'_{sw} \!\!&=&\!\!\frac{\omega}{r_I}\Big(\frac{I}{I_c}\Big)^N\exp\!\Big(\!\!-\!\frac{\omega I^{N+1}}{(N\!+\!1)r_I I_c^{N}}\Big),\; I \le I_c, \label{PpswEqless}\\
&=& \!\!\!\frac{\omega}{r_I}\!\exp\!\Big(\!\!-\!\!\frac{\omega I_c}{(N\!+\!1)r_I}\!\Big)\!\exp\!\Big(\!\!\!-\!\!\frac{\omega (I\!-\!I_c)}{r_I}\!\Big),\; I \ge I_c.
\!\label{PpswEqmore}
\end{eqnarray}

\begin{figure}
\includegraphics[width = 8cm]{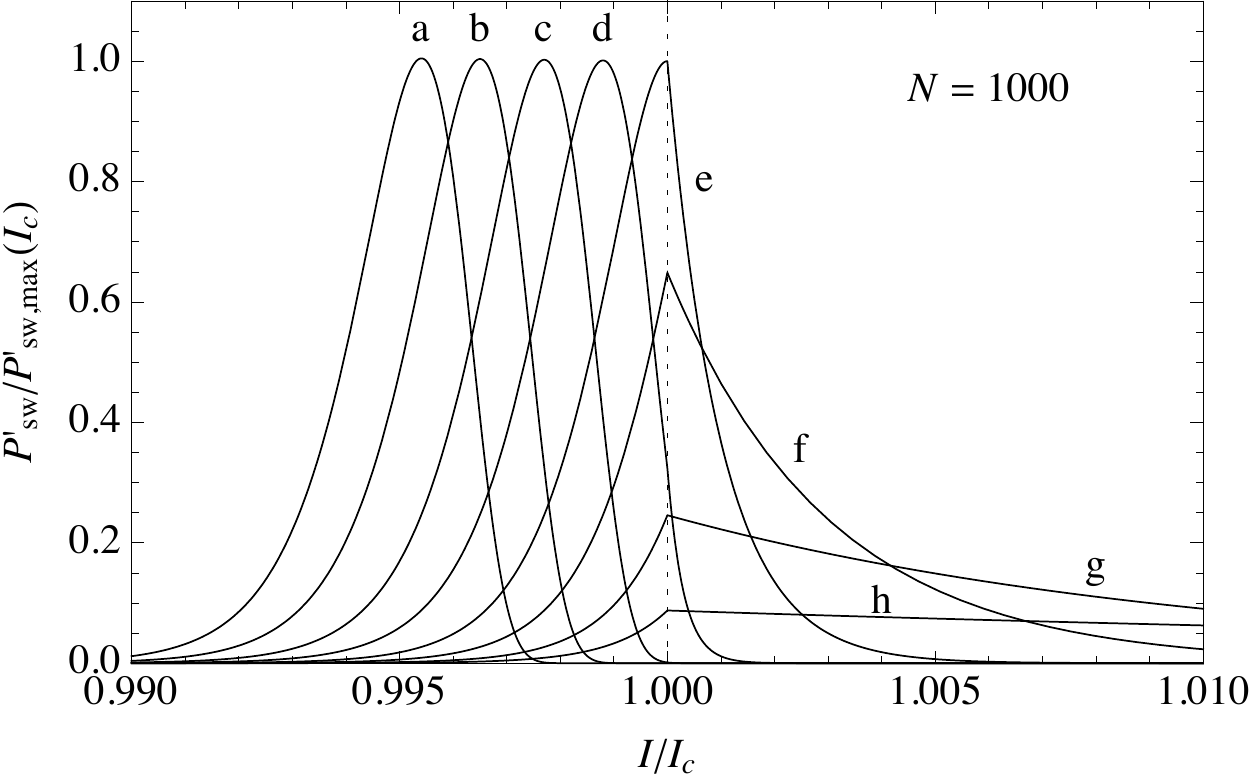}
\caption{Switching probability distribution due to thermally activated barrier-climbing, $P'_{sw}/P'_{sw,max}(I_c)$ vs $I/I_c$ [Eqs.\ (\ref{PpswEqless}) and (\ref{PpswEqmore})], where $P'_{sw,max}(I_c)=(N/I_c)\exp[-N/(N+1)],$ shown for $N = 1000$ and $r_I/\omega I_c$ = (a) $10^{-5}$, (b) $3 \times 10^{-5}$, (c) $10^{-4}$, (d) $3 \times 10^{-4}$, (e) $10^{-3}$ ($I_{max} = I_c$), (f) $3 \times 10^{-3}$, (g) $10^{-2}$, and (h) $3 \times 10^{-2}$.}
\label{Ppswfig}
\end{figure}

Figure \ref{Ppswfig} shows normalized plots of $P'_{sw}$ vs $I/I_c$ for $N = 1000$ and values of $r_I/\omega I_c$ spanning over three orders of magnitude.  

When $Nr_I/\omega I_c \le 1$, the switching probability distribution $P'_{sw}$ has its maximum $P'_{sw,max}$ at $I = I_{max}$ , where
\begin{equation}
I_{max} = I_c(Nr_I/\omega I_c)^{1/(N+1)}
\end{equation}
and
\begin{equation}
P'_{sw,max}=(N/I_{max})\exp[-N/(N+1)].
\end{equation}
At very low temperatures $N \gg 1$, and for a wide range of values of $r_I/\omega I_c$ we find that $I_{max}\approx I_c$ and $P'_{sw,max}\approx N/eI_c$.  Since $\int_0^\infty P'_{sw} dI = 1$, the width $\Delta I$ of the probability distribution is approximately $eI_c/N$.  Note that  $I_{max}= I_c$ when $Nr_I/\omega I_c = 1$. 
 
When  $Nr_I/\omega I_c > 1$, the maximum of $P'_{sw}$ becomes cusplike and remains at $I = I_c$. For high ramp rates $r_I \gg \omega I_c/N$, the probability of switching at a reduced current $I < I_c$ becomes very small, and the switching occurs chiefly for $I > I_c$ on a time scale of order $1/\omega$.    
At zero temperature, which corresponds to $N = \infty$, Eqs.\ (\ref{PpswEqless}) and (\ref{PpswEqmore}) show that no switching occurs for $I < I_c$, $P'_{sw,max}=\omega/r_I$ at $I = I_c$, and  $P'_{sw}=(\omega/r_I)\exp[-(\omega/r_I)(I-I_c)],$ with a width $\Delta I$ of order $r_I/\omega$.

\section{Comparison with experiment using sharp rectangular 180-degree turnarounds\label{ExpSec}}

In 2009, Yang et al.\cite{Yang09} observed that narrow hairpin turns 
in superconducting thin films exhibited reduced critical 
currents and hypothesized that the current-crowding 
effect described above might have explained their results, but at the time the authors were not aware of the work of Hagedorn and Hall, and further did not attempt a careful analysis.\cite{Hagedorn63} 
Yang et al.\ \cite{Yang09} measured the critical currents of superconducting meander structures with rectangular 180-degree turnarounds like that shown in Fig.\ \ref{rect180fig}. 
Combining displaced mirror images of rectangular turnarounds in a two-dimensional layout (a ``boustrophedonic'' pattern) results in a fill factor $f= W/p$, where $p = W +g$ is the pattern period (pitch) and $W$ and $g$ are the strip and gap widths far from the turnarounds. 

The results of Sec.\ \ref{rect180section} can be used to obtain a simple theoretical prediction for the dependence of the ratio of the critical current $I_c(f)$ at fill factor $f$ to $I_c(0)$, the  critical current  in the limit of infinite gap width (zero filling factor).  Since in the latter case $h/a \to 1$, the desired critical current ratio obtained from by Eq.\  (\ref{Rrect180}) is  $I_c(f)/I_c(0) = \alpha^{1/3}$, where $\alpha = \sqrt{2(h/a)-(h/a)^2}$. However, in terms of the fill factor $f$ we have $h/a = (1-f)/(1+f)$.  The theory therefore predicts
\begin{equation}
\frac{I_c(f)}{I_c(0)}=\frac{(1+2f-3f^2)^{1/6}}{(1+f)^{1/3}},
\label{Icf}
\end{equation}   
except for a crossover at $f \approx 1-\xi/W$ into the limit $f\to 1$, where Secs.\ \ref{180sharp} and \ref{rect180section} predict  
\begin{equation}
\frac{I_c(1)}{I_c(0)}=\frac{2^{4/3}}{3}\Big(\frac{\pi\xi}{W}\Big)^{1/6}.
\label{Icf1}
\end{equation}

\begin{figure}
\includegraphics[width = 8cm]{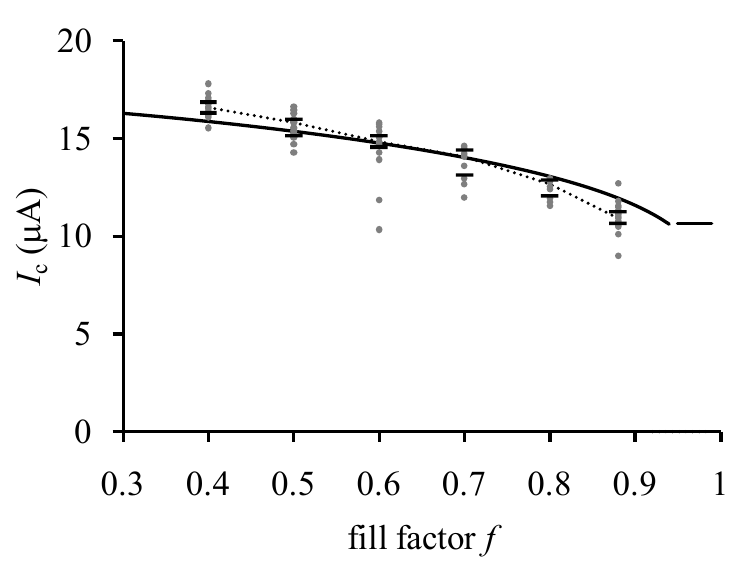}
\caption{Comparison between experimental results for sharp rectangular 180-degree turnarounds  \cite{Yang09} and the corresponding theoretical predictions of Eqs.\ (\ref{Icf}) and (\ref{Icf1}).}
\label{KarlFig}
\end{figure}

Figure \ref{KarlFig} shows a comparison between the experimental results and Eqs.\ (\ref{Icf}) and (\ref{Icf1}). The data are the same as reported in Ref.\ \onlinecite{Yang09}, while the theoretical fit includes a single free parameter, the critical current $I_c(0)$ in the limit of zero fill factor ($f = 0$), which was not measured. The fitted value of $I_c(0)$ = 17 $\mu$A was obtained by varying it until the fit appeared acceptable by eye. A few-percent variation in this fitting 
parameter resulted in a markedly unacceptable fit. 

According to Eq.\ (\ref{Rrect180}), since a fill factor of $f$ = 0 corresponds to the value $\alpha = 1$, this critical current (17 $\mu$A) is smaller by the factor $R = (3/2)(\pi \xi/2 W)^{1/3}$ than the ideal critical current of a long strip of width $W$.  With the values $\xi$ = 7 nm and $W$ = 90 nm\cite{Yang09}, $R$ = 0.744, suggesting that the critical current of a long strip of width 90 nm at the same temperature should be 23 $\mu$A.

\section{Discussion\label{Discussion}}

In this paper we have developed a relatively simple, general method for systematically estimating the critical current of narrow, thin-film superconducting strip patterns with various layouts including turns and turnarounds.   We have shown that if the latter have sharp inner corners, current crowding has the effect of reducing the critical current below that of a long, narrow superconducting strip of constant width.

Our results also have important consequences for thin-film superconducting single-photon detectors.\cite{Goltsman01,Engel06,Bell07,Kitaygorsky07,Bulaevskii11,Bulaevskii11a,Yamashita11}   These detectors carry currents close to the critical current while waiting for photons to arrive, but spurious dark counts arise when when  thermally excited vortices climb over the Gibbs free energy barrier  and  cross the strip, thereby producing a voltage pulse.  Our results suggest that the the frequency of dark counts in single-photon detectors is increased at sharp corners, where the Gibbs free energy barrier is reduced.

We have made a number of simplifications to obtain our results, and following are various extensions and improvements that could be made:

1.\  Numerical solutions of the time-dependent Ginzburg-Landau (TDGL)  equations in two dimensions, allowing for the growth of fluctuations and instabilities leading to vortex nucleation at the corners and subsequent propagation across the current-carrying strip, would provide more accurate values for the non-thermally activated critical currents.  The geometries of interest here, for which the strip widths are much less than the Pearl length $\Lambda = 2 \lambda^2/d$, provide the opportunity for considerable simplifications in the TDGL calculations, since the spatial variation of the order parameter and the current density can be calculated ignoring the effect of self-fields. 

2.\  The present calculations, which assume that the coherence length $\xi$ is much less than the strip widths, could be extended by relaxing this assumption.  The price to be paid, however, is that the results for the critical current would become much more complicated.  

3.\  In calculating the self-energy, we have accounted only for the kinetic energy of the supercurrent circulating around the vortex outside the vortex core.  (The magnetic field energy contribution is negligibly small when $\Lambda$ is much larger than the strip width.)  The accuracy of the self-energy calculation could be improved by using a variational method to include the contributions to the vortex energy inside the vortex core arising from loss of condensation energy and the kinetic-energy cost of bending the magnitude of the order parameter.\cite{Clem75}

4.\  The calculations in this paper make use of the London-model assumption that the superconducting order parameter is not suppressed by the current density.  The accuracy of the present results could be improved by accounting for the current-induced suppression of the order parameter, as done in Refs.\ \onlinecite{Bulaevskii11} and \onlinecite{Bulaevskii11a}.

5.\  In this paper we have calculated the critical current as that for which the Gibbs free-energy barrier for nucleation of a vortex is reduced to zero.  We believe that extensions of our theory to calculate the rate of 
 thermal excitation of vortices over the Gibbs free energy barrier at slightly lower currents
would confirm that the frequency of dark counts\cite{Engel06,Bell07,Kitaygorsky07,Bulaevskii11,Bulaevskii11a,Yamashita11}  (if time-resolved) and  the thermally activated resistance\cite{Bartolf10} (if time-averaged) are systematically increased at sharp corners as a result of current crowding.

It is important to note that in this paper we have assumed  that, in stark contrast to most earlier experiments measuring the critical current in type-II superconductors with much larger transverse dimensions, bulk pinning plays no role whatsoever in determining the critical current in  thin and narrow films under self-field conditions.\cite{Bartolf10}  In 2G coated conductors (thickness $\sim$ 1 $\mu$m and width $\sim 4$ mm) with strong pinning, what normally dominates the critical current is bulk pinning, \cite{BabaeiBrojeny05} and except in very rare cases\cite{Dinner11} the critical-current density due to bulk pinning is typically far below the Ginzburg-Landau depairing critical-current density.\cite{Blatter94}   However, edge pinning has been shown to play an increasingly dominant role in measurements using narrow strips or bridges (width $<$ 10 $\mu$m) to assure that the current supply is adequate to do the measurements.\cite{Elistratov02,Jones10}  The theoretical calculations in the present paper are intended to apply to even thinner and narrower superconducting strips (thickness $\sim$ 5 nm, width $\sim$ 50-200 nm).  In such films, vortices introduced into the strip by a large applied perpendicular magnetic field\cite{Kuit08} presumably can be pinned by bulk pinning sites.  However, if the applied field is removed, a high current can drive these vortices out of their pinning sites, causing the vortices to annihilate with their images upon exiting the strip.  New vortices cannot enter the strip until the current is high enough that the vortices can surmount the Gibbs free-energy barrier at the other edge of the strip. 

Mirror images of the 180-degree turnarounds 
discussed in Sec.\ \ref{180general} can be combined in a two-dimensional layout to produce periodic meanders with the filling factor $f= W/p$, where $p = W +g$ is the pattern period (pitch) and $W$ and $g$ are the strip  and gap widths far from the turnarounds.  In Sec.\ \ref{180optimal} we described how to pattern a 180-degree turnaround with an optimally rounded inner corner that would avoid current crowding and thereby maintain the critical current at the same level as that of an infinitely long straight strip.  However, this optimal design yields a filling factor of only $f= 1/3$, because here $W = a/2$ and $g = a$.  If it is desired to increase the filling factor above 1/3, our calculations indicate that this must come at the expense of the critical current, because, as shown in Sec.\ \ref{180nonoptimal}, even with rounded inner corners the critical current generally will be reduced, primarily because of unavoidable current crowding where the inner boundary has its minimum radius of curvature.  Figure \ref{Rvsgplot} and Eq.\ (\ref{Rrounded180}) present our calculations of the best one can do in alleviating critical-current reduction by rounding the inner corner of the 180-degree turnaround.  
Engineering considerations regarding the trade-offs between critical currents $I_c$ and filling factors $f$ will determine the  shape of the turnaround for a specific application.

The work we have done here in fitting this theory to prior work \cite{Yang09} suggests that the bias current in conventional SNSPDs may be limited by sharp corners. One is tempted to conclude from this result that appropriately designed devices (in which the inner corners are rounded to avoid current crowding) would be capable of far superior performance. However, this presumption would be slightly premature--every fabrication process is slightly different, and the process used in the Yang paper was unique in its details from other processes reported in the literature. More careful experiments are needed in this case. 

Although the focus in this paper has been on superconducting thin films, our results have relevance to the properties of normal-metal films.  The current flows shown in Figs.\ \ref{arcplot}, \ref{parabolicplot}(a), \ref{genhypplot}(a), \ref{180fig}(a), \ref{C180fig}, \ref{C180optfig}, \ref{rect180fig}(a), \ref{rtangleplot}(a), \ref{C90fig}, \ref{Tfig}(a), \ref{RoundedTfig}, \ref{WideEndfig}(a), \ref{SemiNotchSfig}(a), and \ref{Notchfig}(a)  apply equally well to normal films.  Current crowding at the inner corners of sharp bends leads to locally increased dissipation, increasing the electrical resistance of the strip, producing excess ohmic heating, and possibly increasing electromigration.  These are all undesirable properties that could be avoided by choosing film patterns that optimally round the inner corners as discussed above.

\acknowledgments

We thank Vikas Anant, Eric A. Dauler, Eduard F. C. Driessen, Hendrik L. Hortensius, Andrew J. Kerman, Teun M. Klapwijk, Vladimir G. Kogan, and Joel K. W. Yang for helpful discussions.  This research, supported in part by the U.S.\ Department of
Energy, Office of Basic Energy Science, Division of Materials
Sciences and Engineering, was performed at
the Ames Laboratory, which is operated for the U.S.\ Department
of Energy by Iowa State University under Contract No.
DE-AC02-07CH11358.  
This work also was supported in part by the 
Netherlands
Organization for Scientific Research.
\appendix

\section{Self-energy of a Pearl vortex in a sector of angle $\alpha$}

\begin{figure}
\includegraphics[width= 6 cm]{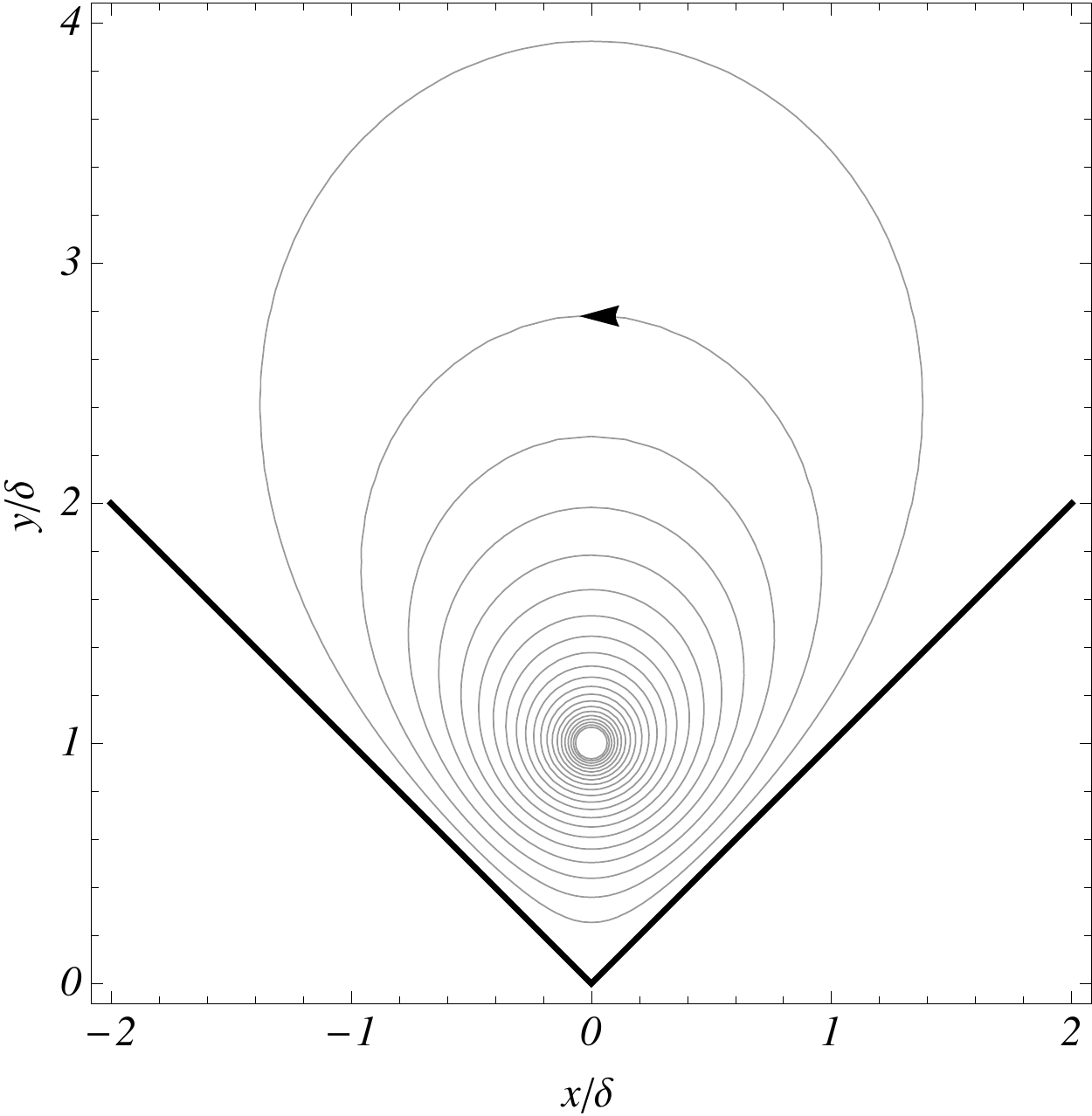}
\caption{Vortex-generated current flow in a thin-film angular sector of opening angle $\alpha$, shown by the contour plot of the stream function $S_v(x_v,y_v;x,y)$, which has the values $S_v=0$ for $(x,y)$ along the boundaries. The contours, shown here for $\bm r_v=(x_v,y_v) = (0,\delta)$ and $\alpha=\pi/2$, correspond to streamlines of the vortex-generated sheet-current density $\bm K_v$, and the arrow shows the direction of the current.}
\label{vortexsectorplot}
\end{figure}

Equation (\ref{Gvw}) gives the complex potential describing a vortex interacting with its image at the straight edge of a strip.  We have used this complex potential, combined with conformal mapping, to find the self-energy of a vortex in other strip geometries.  Using the conformal mapping,
\begin{eqnarray}
\zeta(w)&=& ie^{-i\alpha/2}w^{\alpha/\pi},\\
w(\zeta)&=&i e^{-i \pi^2/2\alpha}\zeta^{\pi/\alpha},
\end{eqnarray}
we can describe  the current flow surrounding the vortex in a thin-film sector of angular width $\alpha$ centered on the $y$ axis (see Fig.\ \ref{vortexsectorplot})       in terms of the complex potential in the $\zeta$ plane,
\begin{equation}
{\cal G}_{v\zeta}(\zeta_v;\zeta)=\frac{i\phi_0}{\pi \mu_0 \Lambda} \ln \Big(\frac{w(\zeta)-w^*(\zeta_v)}{w(\zeta)-w(\zeta_v)}\Big),
\label{Gvalpha}
\end{equation}
whose imaginary part is the stream function $S_v(x_v,y_v;x,y) = \Im{\cal G}_{v\zeta}(\zeta_v;\zeta)$.
When the vortex is at $(x_v,y_v) = (0,\delta)$ and $\xi \ll \delta$, the circulating current $I_{circ}(\bm r_v) = S_v(0,\delta;0,\delta+\xi)$ can be evaluated from Eq.\ (\ref{Gvalpha}) by replacing the numerator in the argument of the logarithm by $w(\zeta_v)-w^*(\zeta_v)$, where $w(\zeta_v) = i \delta^{\pi/\alpha}$, and the denominator by $\xi dw(\zeta_v)/d\zeta_v=(\pi \xi/\alpha)\delta^{\pi/\alpha-1}$.
The resulting self-energy $E_{self}(\bm r_v)= \phi_0 I_{circ}(\bm r_v)/2$ is  
\begin{equation}
E_{self}(\delta) = \frac{\phi_0^2}{2\pi\mu_0\Lambda}\ln\Big(\frac{2\alpha\delta}{\pi\xi}\Big).  
\label{Icircalpha}
\end{equation}
When the factor $2\alpha/\pi$ is an integer, it corresponds to the number of quadrants within which the vortex-generated current flows.\cite{Kogan11}  Note, for example, the factor 3 in the self-energy term of Eq.\ (\ref{Gdeltartangle}) [Fig.\ \ref{rtangleplot}(b)] and the factor 4 in Eq.\ (\ref{Gdelta180}) [Fig.\ \ref{180fig}(b)].


\begin{thebibliography}{99}
\bibitem{footnote1} Note that this is a change from the coordinate system used in Sec.\ \ref{uniformK}, where the sample boundaries were at $y = \pm W/2$.
\end{thebibliography}
\end{document}